\pdfoutput=1
\documentclass[a4paper,11pt]{article}

\newcommand\papertitle{Holographic RG Flows for Kondo-like Impurities}

\usepackage{jheppub}
\usepackage{xcolor}

\usepackage{tikz}
\usetikzlibrary{arrows, backgrounds, calc, decorations.pathreplacing, decorations.markings, fadings, quotes, positioning, shadings, intersections, angles}

\usepackage{amsmath,amsfonts,amssymb,mathrsfs}
\usepackage{bbm}
\usepackage[inline]{enumitem}
\usepackage{multirow}
\usepackage{subcaption}
\usepackage{nicefrac}
\usepackage{xspace}

\usepackage[colorlinks=true]{hyperref}
\hypersetup{
    bookmarks=true,         
    unicode=false,          
    pdftoolbar=true,        
    pdfmenubar=true,        
    pdffitwindow=false,     
    pdfstartview={FitH},    
    pdftitle={\papertitle}, 
    pdfauthor={Johanna Erdmenger, Charles M. Melby-Thompson, and Christian Northe},
    pdfnewwindow=true,      
    colorlinks=true,        
    linkcolor=blue,         
    citecolor=red,          
    filecolor=magenta,      
    urlcolor=cyan           
}

\makeatletter

\@addtoreset{equation}{section}
\makeatother

\newcommand{\p}	{\partial}

\newcommand{\R}	{\mathbb{R}}
\newcommand{\C}	{\mathbb{C}}
\newcommand{\Z}	{\mathbb{Z}}

\newcommand{\scD}{\mathscr{D}}

\newcommand{\cA}{\mathcal{A}}
\newcommand{\cB}{\mathcal{B}}
\newcommand{\cC}{\mathcal{C}}
\newcommand{\cD}{\mathcal{D}}
\newcommand{\cE}{\mathcal{E}}
\newcommand{\cF}{\mathcal{F}}

\newcommand{\cI}{\mathcal{I}}

\newcommand{\cL}{\mathcal{L}}
\newcommand{\cM}{\mathcal{M}}
\newcommand{\cN}{\mathcal{N}}
\newcommand{\cO}{\mathcal{O}}
\newcommand{\cP}{\mathcal{P}}
\newcommand{\cS}{\mathcal{S}}

\newcommand{\cV}{\mathcal{V}}
\newcommand{\cW}{\mathcal{W}}

\newcommand{\bx}{\mathbf{x}}
\newcommand{\bdr}{\mathbf{r}}
\newcommand{\bSigma}{\mathbf{\Sigma}}

\newcommand\affine[1] {\widehat{#1}}

\DeclareMathOperator{\sgn}{sgn}
\DeclareMathOperator{\Vol}{Vol}
\DeclareMathOperator{\Tr}{Tr}
\DeclareMathOperator{\tr}{\tr}

\newcommand{\dA}{A}

\newcommand{\braket}[2]	{\langle{#1}\vert{#2}\rangle}

\newcommand\id		{\mathbf{1}}

\newcommand\Algebra[1]	{\mathfrak{#1}}
\newcommand\Group[1]	{\mathrm{#1}}
\newcommand{\SO}	{\Group{SO}}

\newcommand{\SL}	{\Group{SL}}

\newcommand{\Spin}	{\Group{Spin}}
\newcommand{\SU}	{\Group{SU}}
\newcommand{\U}		{\Group{U}}

\newcommand{\su}	{\Algebra{su}}
\renewcommand{\u}	{\Algebra{u}}
\newcommand{\g}		{\Algebra{g}}

\newcommand\moduli{{\cM_{N_1N_5}}}

\newcommand\ads {\textsf{AdS}}
\newcommand\cft {\textsf{CFT}}
\newcommand{\AdS}	{\ads}

\renewcommand\S {\textsf{S}}

\newcommand{\dil}   {\phi}
\newcommand{\UV}    {\text{UV}}
\newcommand{\IR}    {\text{IR}}
\newcommand{\adsL}  {\mathsf{L}}
\newcommand{\adsR}  {\mathsf{R}}
\newcommand{\sfc}   {c}
\newcommand{\sfg}   {\mathsf{g}}
\newcommand{\ol}[1]	{\overline{#1}} 
\newcommand\cd{D}

\newcommand{\holA}{A} 
\newcommand{\hola}{a} 
\newcommand{\vaca}{\alpha} 

\newcommand{\holB}{B} 
\newcommand{\holb}{b} 
\newcommand{\vacb}{\beta} 

\newcommand{\holU}{U} 
\newcommand{\holu}{u} 
\newcommand{\vacu}{\eta} 

\newcommand{\holV}{V} 
\newcommand{\holv}{v} 
\newcommand{\vacv}{\nu} 

\newcommand{\qfive}     {q_{\scriptscriptstyle NS5}}
\newcommand{\qone}      {q_{\scriptscriptstyle F1}}
\newcommand{\pfive}     {q_{\scriptscriptstyle D5}}
\newcommand{\pone}      {q_{\scriptscriptstyle D1}}
\newcommand{\pthree}    {q_{\scriptscriptstyle D3}}

\newcommand{\pbare}     {\pone^\varnothing}

\newcommand{\Qfive}     {Q_{\scriptscriptstyle NS5}}
\newcommand{\Qone}      {Q_{\scriptscriptstyle F1}}
\newcommand{\Pfive}     {Q_{\scriptscriptstyle D5}}
\newcommand{\Pone}      {Q_{\scriptscriptstyle D1}}
\newcommand{\Pthree}    {Q_{\scriptscriptstyle D3}}

\newcommand{\defx}      {\xi}
\newcommand{\defxi}     {\psi_\defx}
\newcommand{\defc}      {\mathsf{c}}
\newcommand{\defw}      {\Xi}
\newcommand{\defrho}    {\psi_R}

\newcommand{\cc}        {\text{c.c}}

\newcommand\Secref[1]	{Section~\ref{#1}\xspace}

\newcommand\secref[1]	{section~\ref{#1}\xspace}
\newcommand\tableref[1]	{table~\ref{#1}\xspace}
\newcommand\figref[1]	{figure~\ref{#1}\xspace}
\newcommand\appref[1] {appendix~\ref{#1}\xspace}

\newcommand*\br{$\bullet$}
\newcommand*\nb {-}

\newcommand*\ti   {{\tilde\imath}}

\newcommand*\bq		{{\overline q}{}}

\newcommand*\bY		{{\overline Y}{}}

\newcommand*\btau	{{\overline\tau}{}}

\newcommand*\lr[1]{\overset{\leftrightarrow}{#1}}
\newcommand*\lrD{\lr{\cD}}

\newcommand*\onehalf {\nicefrac{1}{2}\xspace}
\newcommand*\res[2]  {{#1}\vert_{#2}}

\begin{document}
\title{\papertitle}

\author[a,b]{Johanna Erdmenger,}
\author[a]{Charles M. Melby-Thompson,}
\author[a,b]{and Christian Northe}

\affiliation[a]{Institut f\"{u}r Theoretische Physik und Astrophysik,\\Julius-Maximilians-Universit\"{a}t W\"{u}rzburg, Am Hubland, 97074 W\"{u}rzburg, Germany}

\affiliation[b]{W\"{u}rzburg-Dresden Cluster of Excellence ct.qmat,\\Julius-Maximilians-Universit\"{a}t W\"{u}rzburg, Am Hubland, 97074 W\"{u}rzburg, Germany}

\abstract{%
Boundary, defect, and interface RG flows, as exemplified by the famous Kondo model, play a significant role in the theory of quantum fields.
We study in detail the holographic dual of a non-conformal supersymmetric impurity in the D1/D5 CFT.
Its RG flow bears similarities to the Kondo model, although unlike the Kondo model the CFT is strongly coupled in the holographic regime.
The interface we study preserves $d = 1$ $\cN = 4$ supersymmetry and flows to conformal fixed points in both the UV and IR.
The interface's UV fixed point is described by $d = 1$ fermionic degrees of freedom, coupled to a gauge connection on the CFT target space that is induced by the ADHM construction.
We briefly discuss its field-theoretic properties before shifting our focus to its holographic dual.
We analyze the supergravity dual of this interface RG flow, first in the probe limit and then including gravitational backreaction.
In the probe limit, the flow is realized by the puffing up of probe branes on an internal $\S^3$ via the Myers effect.
We further identify the backreacted supergravity configurations dual to the interface fixed points.
These supergravity solutions provide a geometric realization of critical screening of the defect degrees of freedom.
This critical screening arises in a way similar to the original Kondo model.
We compute the $g$-factor both in the probe brane approximation and using backreacted supergravity solutions, and show that it decreases from the UV to the IR as required by the $g$-theorem.
}

\maketitle
\flushbottom

\section{Introduction}
\label{sec:introduction}

Boundaries, defects and interfaces play an important role in the world of quantum field theory (QFT).
One powerful example is the discovery of Dirichlet branes (D-branes) as the conformal boundary conditions of a string worldsheet, which revolutionized our understanding of many aspects of both quantum gravity and QFT.
Condensed matter theory presents us with many examples of both theoretical and practical importance.
For example, some gapped materials exhibit the fractional quantum Hall effect, resulting in edge currents localized on the boundary.
Among other applications, such materials are being studied for use as qubits in quantum computers \cite{Laughlin:1983fy, Kitaev:1997wr, 2008RvMP...80.1083N}.

Defects play a decisive role in the properties of materials via another famous phenomenon, the Kondo effect, which constitutes the primary motivation for the present paper.
Kondo's model \cite{Kondo} aimed to understand the anomalous logarithmic growth in resistivity exhibited by many metals at low temperatures.
The model is simplicity itself: free fermions coupled to a spin-\nicefrac{1}{2} impurity by a spin-spin interaction.
The study of the phenomenon underlying this coupling's temperature dependence was decisive in the development of the renormalization group (RG) \cite{Wilson:1974mb}.
The beta function of the impurity coupling is negative, and the model flows to a non-trivial fixed point in the IR.
As a result, the material's resistivity departs from the predictions of the Drude model due to the temperature dependence of the scattering amplitude of electrons by the impurities.
The effect therefore arises due to screening of the impurity by electrons.
Upon computing the RG flow of the impurity coupling, Kondo's model does indeed exhibit a logarithmic increase in resistivity at low temperatures.

The Kondo effect has been studied using many different techniques and generalized in a number of ways.
It has been solved using the Bethe ansatz \cite{Andrei:1980fv, Andrei:1982cr, WiegmannKondo, Wiegmann}, studied extensively using the tools of boundary conformal field theory (BCFT) \cite{Affleck:1990by, Affleck:1990iv, Affleck:1991tk}. More recently, it has been considered within the context of QCD \cite{Kimura:2018vxj} and quantum information \cite{Casini:2018cxg}.
There is a large-$N$ version of the Kondo model, in which the spin impurity lies in a totally antisymmetric representation of $\SU(N)$ \cite{Coleman, PhysRevB.35.5072, Bickers:1987zz, PhysRevB.58.3794}.
It is convenient in this family of models to write the spin impurity as a bilinear of `slave' Abrikosov fermions.
In this picture, the screening corresponds to the condensation of a bosonic operator built from the product of an electron and an Abrikosov fermion field.
The RG flow is triggered by a `double-trace' operator involving the product of this bosonic operator and its Hermitean conjugate.

It is natural for several reasons to complement these techniques with a study of analogues of the Kondo model realized using the anti-de Sitter/conformal field theory (AdS/CFT) correspondence.
On the one hand, both the simplicity and the applicability of Kondo's model make it an excellent prototype for the study of more complicated holographic defects, which already comprise a rich class of interesting objects.%
\footnote{Holographic realizations of defects of various types comprise a large literature. A few well-known and recent examples can be found in \cite{Bak:2003jk,Bobev:2013yra, Fujita:2016gmu, Melby-Thompson:2017aip, Karndumri:2017bqi, Fujita:2018dvl, Evans:2019pcs}.}
On the other hand, its importance means that any new generalizations or techniques provided by holography may lead to new insights, as has occurred in the study of QFT.

There exist in the literature a number of holographic constructions of Kondo-like models (we refer to these as ``holographic Kondo models''), some of which are to be found in \cite{Kachru:2009xf,Harrison:2011fs,Benincasa:2012wu}.
There are several important differences between the original model and its holographic versions.
One of these is that the free electron gas itself is replaced by a strongly coupled system.
A second is that the electron gas is a large-$N$ gauge theory, and can be described by pure geometry only in the absence of flavor symmetry.

A holographic model dual to a variant of the large-$N$ RG flow of \cite{Coleman, PhysRevB.35.5072, Bickers:1987zz, PhysRevB.58.3794} described above, also triggered by a double-trace operator and involving a condensation, was realized by one of the authors (JE) and collaborators in \cite{Erdmenger:2013dpa}.
This model is motivated by a top-down brane construction involving D5- and D7-brane probes in a D3-brane background.
While the D3-D5 brane and D3-D7 brane systems are separately supersymmetric, combining D5- and D7-brane probes breaks supersymmetry, resulting in the tachyon potential responsible for the condensation process.

Unfortunately, determining the exact form of the dual gravity action for the brane configuration described is a challenging task.
Therefore, in \cite{Erdmenger:2013dpa} a bottom-up model was considered that realizes the most salient features of the top-down model.
Essentially, this bottom-up model consists of an $\ads_2$ brane embedded in an $\ads_3$ BTZ background.
The brane possesses local fields dual to the condensing operator, the density of defect degrees of freedom, and the electron current.
The electrons of the dual field theory are chiral.
This model leads to  a number of interesting physical results.
The impurity screening is geometrically realized in the dual model as a decrease of the flux through the boundary of $\ads_2$ that is proportional to the number of defect degrees of freedom.
The resistivity does not have the $\ln T$ behaviour of the original Kondo model, but rather a polynomial behaviour with real exponent as expected for an impurity in a strongly coupled system \cite{Luttinger}.
Including the backreaction, field-theoretical results for the impurity entanglement entropy were reproduced in \cite{Erdmenger:2015spo}.
Two-point functions and spectral functions displaying Fano resonances were calculated in \cite{Erdmenger:2016vud,Erdmenger:2016jjg}.
Quenches of the holographic Kondo coupling were investigated in \cite{Erdmenger:2016msd}.
The case of two defects was studied in \cite{OBannon:2015cqy}.

In spite of these successes, it is desirable to construct a top-down model that realizes some of the features not present in the model described above.
It is desirable for such a model to have several properties: that its field theory description is known, that it possesses non-chiral fermions, and that it has a parameter corresponding roughly to a number of electron ``flavors''. Finally, if we are interested in making quantitiative comparisons between the gravitational and field theoretic descriptions, it is also desirable for the RG flow to preserve some fraction of supersymmetry.

\bigskip

The focus of this paper is the construction and study of a holographic Kondo model with these properties.
Our inspiration is drawn from the D-brane picture of the Kondo model \cite{Alekseev:2000fd, Fredenhagen:2000ei}, which essentially describes the screening of a Kondo impurity via the Myers effect \cite{Myers:1999ps} on $\S^3$: a stack of D0-branes, representing the UV impurity, condenses into a single D2-brane in the IR.

As our ambient model we choose the best-studied example of $\ads_2$/$\cft_3$ holography: the D1/D5 system, whose type IIB superstring dual lives on $\ads_3\,\times\,\S^3\,\times\,M_4$.
One immediate advantage over the model of \cite{Erdmenger:2013dpa} is that the Lagrangian of the ambient CFT is known.
Moreover, it naturally provides the $\S^3$ we need to embed the usual Kondo flow into a holographic setup.
The impurities, or interfaces%
\footnote{Interfaces are more general than defects, in that two distinct QFTs may live on either side of an interface.}, %
are realized in the gravitational dual by adding $p$ fundamental strings preserving half of the supersymmetries of the ambient D1/D5 CFT.

The CFT lives on the Higgs branch of a D1/D5 gauge theory compactified on a compact Calabi-Yau 4-manifold $M_4$ \cite{Seiberg:1999xz}, in which the D1-branes dissolve into the D5-branes and can be interpreted as gauge instantons on $M_4$.
The gauge theory Lagrangian of the interface itself is described in the language of the ADHM construction \cite{Atiyah:1978ri} via the Wilson line operator constructed in \cite{Tong:2014cha}, which naturally associates to each point of $M_4$ a connection on the target space of the CFT.
This connection naturally induces an interface joining distinct D1/D5 CFTs.
We also consider a generalization of these interfaces in which the fundamental strings are replaced by $(p,q)$-string bound states.

\medskip

On the gravity side these interfaces correspond to $(p,q)$-strings intersecting or ending on the D1/D5 system.
The holographic dual of this system is obtained by taking the near-horizon limit.
In the probe brane approximation, the near horizon geometry is $\ads_3\times \S^3\times M^4$, and the $(p,q)$-strings lie on an $\ads_2$ slice that is localized in $\S^3\times M^4$.
We show that these defects possess a marginally relevant operator, and that the RG flow it generates preserves four supersymmetries.
After deforming by this operator, the brane locus inside $\S^3$ puffs up via the Myers effect from a point into an $\S^2$.
Conflating the radial coordinate with the RG scale, this process is simply the sigma model description of the Kondo effect \cite{Affleck:1990zd,Affleck:1990by,Affleck:1990iv}.

While the probe brane computation is sufficient to determine the leading interaction between the interface and the CFT, it does not capture the effect of the interface on CFT observables.
These contributions are encoded in the gravitational backreaction of the interface.
While solving the backreacted geometry for the full interface flow is a difficult problem, we consider the simpler problem of obtaining the supergravity solutions dual to the interface fixed points.
These solutions can be obtained from the general ansatz of asymptotically $\ads_3\times \S^3\times M^4$ solutions to IIB supergravity of \cite{ChiodaroliOriginal} by relaxing the regularity conditions employed there.

In contrast to the probe brane description, where we are constrained to keep the F1 charge $p$ of the interface macroscopically small compared to the background's D1 and D5 charge, using our backreacted solutions we can dial it up to the order of the D5 charge.
In particular, when $p$ equals the D5 charge the D3 has slid down from the north to the south pole of the $\S^3$ and the interface vanishes.
An analogue of this scenario occurs in the original Kondo flow, where it is termed ``critical'' or ``exact'' screening.
Mathematically, the spin of the conduction electrons forms an $\SU(2)$ singlet with the spin of the impurity.
Physically, the conduction electrons absorb the impurity.

An important property of any boundary RG flow is that the ``number of boundary degrees of freedom'', as measured by the $g$-factor \cite{Affleck:1991tk}, does not increase from the UV to the IR \cite{Friedan:2003yc, Casini:2016fgb}.
The $g$-factor of a conformal boundary (or, by the folding trick, a conformal interface) may be defined as the overlap of the interface's boundary state with the vacuum, $g=\braket{0}{\cB}$.
It was shown in \cite{Calabrese:2009qy} that the boundary entropy can be extracted from entanglement entropy, which in holography is computed to leading order in $1/c$ by the Ryu-Takayanagi prescription \cite{Ryu:2006bv}.
Using the results of \cite{ChiodaroliEntropy} we compute the $g$-factor at the fixed points of our RG flow and confirm that these results satisfy the $g$-theorem.

\bigskip

The paper is organized as follows.
We begin with a review of the most important background material:
thus, \secref{sec:kondo defects} reviews relevant features of the Kondo model and sets the stage for the questions we address in the remainder of the paper, while \secref{sec:D1/D5} reviews the D1/D5 system and sketches pertinent features of its description in the UV as a gauge theory and in the IR as a sigma model.
\Secref{sec:interfaces} introduces the interface brane configurations and describes aspects of its gauge theory realization, followed by a detailed description of the interface flows in the probe brane approximation.
In \secref{sec:sugra} we turn to a fully backreacted supergravity analysis of the UV and IR fixed point interfaces of the RG flow based on the results of \cite{ChiodaroliOriginal, ChiodaroliJunctions, ChiodaroliEntropy}.
As one possible application of these results, we apply these solutions in \secref{sec:entropy} to compute the change in interface entropy along the RG flow, which we compare with expectations from the probe brane limit.
We summarize our conclusions in \secref{sec:conclusions} and suggest promising directions for future work.

\section{Kondo-like defects in AdS/CFT}
\label{sec:kondo defects}

\def\pr{\partial}
\def\N{{\mathcal{N}}}

The Kondo model \cite{Kondo} has played a prominent role in the development of modern quantum field theory, notably the renormalization group and boundary conformal field theory.
It was originally introduced to explain the anomalous logarithmic growth in resistivity in metals with magnetic impurities at low temperatures, and consisted of free spin-\onehalf electrons interacting via spin-spin coupling with a heavy spin-$s$ magnetic impurity.
The interaction is described by a delta-function potential,
\begin{equation}
  V \propto \delta(\vec r) \vec{J} \cdot \vec{S} \,,
\end{equation}
where $\vec J$ is the spin charge density of the electron field and $\vec S$ is the spin operator of the impurity.

Since electrons scatter in this model only in the $S$-wave channel, by restricting to this channel Affleck and Ludwig were able to reduce the electron-impurity interaction to a two-dimensional system with Hamiltonian \cite{Affleck:1990by}
\begin{equation} \label{KondoHamiltonian}
H = \frac{v_F}{2 \pi} i \psi^\dagger \pr_x \psi \, + \,
\frac{v_F}{2}  \lambda_K \delta(x) \vec{J} \cdot \vec{S} \,.
\end{equation}
Here, $\psi = (\psi_\uparrow,\psi_\downarrow)$ is the electron spin field doublet, $v_F$ is the Fermi velocity, and $\lambda_K$ is the Kondo coupling between electron and impurity spin.
The electron spin charge density operator $\vec J$ is given explicitly by
\begin{equation}
\vec{J} = \psi^\dagger \vec{\sigma} \psi \, ,
\end{equation}
where $\vec{\sigma}$ are the Pauli matrices.

The Kondo coupling possesses a negative beta function, which means that it flows from a trivial UV to an interacting IR fixed point.
Physically, an electron cloud forms around the impurity, screening it.
In the 1990's, the Kondo model and its generalizations to ``spin'' group $\SU(N)$ and $k$ electron flavors were studied extensively by Affleck and Ludwig within the framework of BCFT \cite{Affleck:1990by,Affleck:1990iv}.
Among other results, they used the BCFT fusion rules to derive the operator spectrum of the non-trivial IR BCFT.

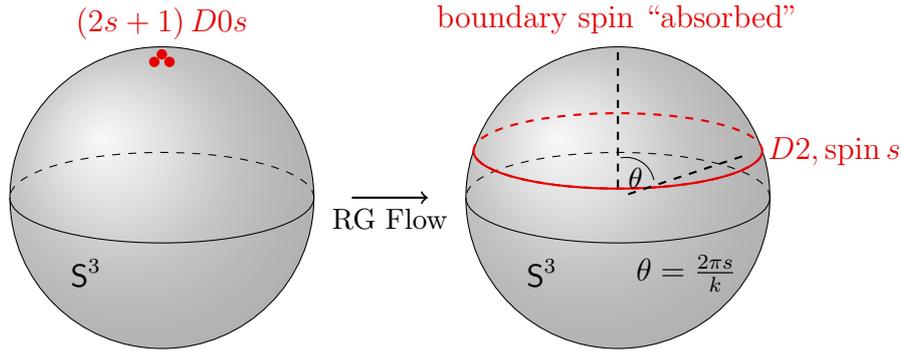
\begin{figure}[t]
\tikzstyle{D0}=[circle, draw=red!90!black, fill=red!90!black, minimum size=3.5pt, inner sep=0pt]
\begin{center}
 \begin{tikzpicture}[]
\def\shift{6cm }
\def\R{2cm}
  \shade[ball color = gray!80, opacity = 0.4] (0,0) circle (\R);
  \draw (0,0) circle (\R);
  \node at (-1,-1){\large $\S^3$};
  \node[D0] 		at (-.1cm,1.8cm)  {};
  \node[D0] 		at (.1cm,1.8cm) {};
  \node[D0] (TopD0)	at (0,1.9cm) {};
  \node[red!90!black, above] at (TopD0.north){\large $(2s+1)\, D0s$};
  \draw (-2,0) arc (180:360:2 and 0.6);
  \draw[dashed] (2,0) arc (0:180:2 and 0.6);

  \shade[ball color = gray!80, opacity = 0.4, xshift=\shift] (0,0) circle (\R);
  \coordinate (center) at (\shift,0);
%
  \node at (center){};
  \draw[xshift=\shift] (0,0) circle (\R);
  \node[xshift=\shift] at (-1,-1){\large $\S^3$};
  \node[xshift=\shift] at (.9,-1){\large $\theta=\frac{2\pi s}{k}$};
  \draw[xshift=\shift] (-2,0) arc (180:360:2 and 0.6);
  \draw[dashed, xshift=\shift] (2,0) arc (0:180:2 and 0.6);
  \draw[draw=red!90!black, thick, xshift=\shift] (-1.9,0.62) arc (180:360:1.9 and 0.5);
  \draw[draw=red!90!black, dashed, thick, xshift=\shift] (1.9,0.62) arc (0:360:1.9 and 0.5);
  \node[xshift=\shift] (D2ArcBegin)		at (1.85,0.62){};
  \node[red!90!black, right of=D2ArcBegin] {\large $D2,\textrm{spin}\, s$};
  \node (Northpole2)	at (\shift,1.9cm) {};
  \node[red!90!black, above] at (Northpole2.north){\large boundary spin ``absorbed''};

  \draw [->, thick] (2.5,0) -- (3.5,0) node [midway, below] {\normalsize RG Flow};
  \node[xshift=\shift]	(Center)	at (0,0){};
  \node[xshift=\shift]	(Northpole)	at (0,2.1){};
  \draw[dashed, thick] (Center) -- (Northpole);
  \draw[dashed, thick] (Center) -- (D2ArcBegin);
  \node[xshift=\shift, anchor=south west] at (0,0){$\theta$};
  \node[xshift=\shift]	(angleBottom)	at (.5,.1){};
  \node[xshift=\shift]	(angleTop)	at (-.1,.5){};
  \draw[bend right] (angleTop) to [bend left=45] (angleBottom);
\end{tikzpicture}\vfill
\caption{A stack of $2s+1$ D0 branes condense into a single brane of spin $s$ at fixed polar angle $\theta=\frac{2\pi s}{k}$. This process describes the absorption of an impurity by surrounding electrons in the Kondo model.}
\label{fig: BranePolarization}
\end{center}
\end{figure}

In this approach, the electron system is viewed as the $\affine{\su(2)}_k$ Wess-Zumino-Witten (WZW) CFT, and the Kondo effect is realized as an RG flow connecting two rational boundary conditions.
In the geometric language of string theory, the boundary conditions at the flow endpoints describe either D0-branes located at the poles, or D2-branes lying on $\S^2$s at the quantized polar angles $\theta=\frac{2\pi j}{k}$ inside $\S^3$.
A brane's \emph{spin} $j$ takes the half-integer values $0,1,\dots,k/2$, with $j=0,k/2$ corresponding to a D0-brane situated at the north or south pole, respectively.
In the UV, the Kondo impurity in the spin-$s$ representation of $\SU(2)$ is described by a stack of $(2s+1)$ D0-branes at the north pole.
Over the course of the RG flow, the stack of D0-branes condenses into a single D2-brane of spin $s$ \cite{Myers:1999ps, Fredenhagen:2000ei, Fredenhagen:2002qn}, which represents the IR impurity (\figref{fig: BranePolarization}).

We are only concerned with the case $2s\leq k$,%
\footnote{The case $2s>k$, under-screened impurities, has also been studied \cite{Affleck:1990by, 1997PhRvB..56..668L, 2004PhRvB..70p5112F}. Its IR fixed points are only reached for infinite coupling $\lambda_K$.}
whose IR fixed points lie at a finite coupling $\lambda_K$ and are described by interacting CFTs.
When $2s=k$, we speak of ``exact'' or ``critical'' screening.
In this case, while the UV consists of decoupled defect degrees of freedom together with the boundary condition that the outgoing mode equals the incoming mode, in the IR all the impurity degrees of freedom disappear, leaving behind the boundary condition that the outgoing wave is now the negative of the incoming wave.

There are different approaches for constructing gravity duals of spin impurity models.
In \cite{Kachru:2009xf,Harrison:2011fs}, a holographic dual of the IR fixed point associated with a spin impurity interacting with strongly coupled $\N=4$ Super Yang-Mills theory was constructed using D5-branes interacting with D3-branes at a point in the D3-brane worldvolume.
In the IR, the D5-branes dissolve and are replaced by a three-form flux. For general dimensions, a similar mechanism was established in \cite{Benincasa:2012wu}.

In  a series of papers \cite{Erdmenger:2013dpa,Erdmenger:2014xya,OBannon:2015cqy,Erdmenger:2015spo,Erdmenger:2016vud,Erdmenger:2016jjg,Erdmenger:2016msd}, one of the authors (JE) studied a holographic defect model where the defect RG flow linking the UV and IR fixed points could be written down explicity.
In this model, the flow is initiated by a marginally relevant double-trace operator.
Such an operator appeared long ago in the condensed matter literature in the context of the large-$N$ Kondo model \cite{ReadNewns,Coleman}.
This description is relevant when the electron has $N$ flavors with $\SU(N)$ flavor symmetry, and the impurity lies in a totally antisymmetric representation.
To reach this form the impurity is replaced with a new fermi field $\chi$, the Abrikosov fermion, in the fundamental representation.
The Hilbert space of such a fermi field is the sum of all anti-symmetric representations, and contains the original impurity representation.
On this extended Hilbert space, the spin operator takes the form
\begin{equation} \label{decomp}
  S^a = \chi^\dagger T^a \chi \, .
\end{equation}
If $N$ is large, at leading order in $1/N$ the $\SU(N)$ generalization
$J^aS^a$ of the interaction term in \eqref{KondoHamiltonian} reduces
to a product of two scalar operators, $\mathcal{O}
\mathcal{O}^\dagger$, with $\cO = \psi^\dagger\chi$. Due to its
product structure involving two gauge invariant objects, the operator
$\cO \mathcal{O}^\dagger$ is referred to as `double-trace'
operator. It is this operator that drives the RG flow to a non-trivial
IR fixed point.
In the IR, the operator $\mathcal{O} = \psi^\dagger \chi$ condenses, spontaneously breaking the spurious $U(1)$ symmetry introduced in \eqref{decomp}.
To recover the original model, the filling number $\chi^\dagger\chi$ is constrained to match the number of boxes $q$ in the Young tableau for the totally antisymmetric representation of $SU(N)$ under consideration.

The model of \cite{Erdmenger:2013dpa} was chosen because it shares many features with this construction.
It is inspired by a brane construction involving D7- and D5-brane probes in a D3-brane background, as shown in \tableref{tab: OldKondo}.
The strings stretching between the D3- and D7-branes give rise to a chiral fermionic field $\psi$ in the fundamental representation of $SU(N)$ living on their $(1+1)$-dimensional intersection.
The D3-D5 strings give rise to a fermion field $\chi$ localized along their $(0+1)$-dimensional intersection, also in the fundamental.
One notable difference from the above scenario is that the $\SU(N)$ group is gauged in the D-brane approach.

\begin{table}[hb]
\begin{center}
\begin{tabular}{|r|c|c|c|c|c|c|c|c|c|c|}
\hline
& 0 & 1 & 2 & 3 & 4 & 5 & 6 & 7 & 8 & 9 \\
\hline
$N$ D3 & \br & \br & \br & \br & \nb & \nb & \nb & \nb & \nb & \nb \\\hline
$1$ D7 & \br & \br & \nb & \nb & \br & \br & \br & \br & \br & \br \\\hline
$1$ D5 & \br & \nb & \nb & \nb & \br & \br & \br & \br & \br & \nb \\\hline
\end{tabular}
\caption{Brane configuration for the holographic Kondo model of
  \cite{Erdmenger:2013dpa}.  \label{tab: OldKondo}}
\end{center}
\end{table}

In the absence of D5-branes, the D3/D7 system has 8 ND directions and so preserves 8 supercharges, as does the D3/D5 system in the absence of D7-branes.
However, the D5- and D7-branes have 2 ND directions and so their interaction breaks all supersymmetries.
This interaction manifests itself as a complex tachyon field $\Phi$,
which is  identified as the gravity dual of the operator $\mathcal{O}
= \psi^\dagger \chi $. 

In the near-horizon limit, the $N$ D3-branes provide an $\ads_5\times \S^5$ supergravity background, on which live D5- and D7-brane probes wrapping $\ads_2\times \S^4$ and $\ads_3\times \S^5$, respectively.
The Dirac-Born-Infeld (DBI) action for the D5-brane contains a gauge field $a_\mu$ on the $\ads_2$ subspace spanned by the time and radial coordinate in the $\ads$ geometry.
The $a_t$ component of this gauge field is dual to the charge density of the Abrikosov fermions, $q = \chi^\dagger \chi$.
The D7-brane action contains a Chern-Simons (CS) term for a gauge field $A_\mu$ on $\ads_3$.
As noted before, the D5-D7 strings lead to a complex scalar tachyon field.

The holographic dictionary of this model is summarized in the following table:
\begin{center}
\begin{tabular}{|l|c|l|}
\hline
{Operator} & & {Gravity field} \\ \hline
Electron current $J^\mu  = \bar \psi \gamma^\mu \psi$ &$\Leftrightarrow$ & Chern-Simons gauge field $A$ in $\ads_3$ \\  \hline
Charge $q = \chi^\dagger \chi$   & $\Leftrightarrow$ & 2d gauge field $a$ in $\ads_2$
\\ \hline
Operator ${\cal O} =
\psi^\dagger \chi$  & $\Leftrightarrow$ & 2d complex scalar $\Phi$ in
                                          $\ads_2$  \\
\hline
\end{tabular}
\end{center}
Unfortunately, as of yet the precise form of the tachyon potential is unknown.
The approach of \cite{Erdmenger:2013dpa} was therefore to study a bottom-up holographic model incorporating the features shared by the top-down construction and the multi-channel Kondo model.
The action of this model contains a potential for the tachyonic scalar $\Phi$ responsible for the RG flow, which takes the simple form
\begin{equation} \label{potK}
V (\Phi^\dagger \Phi) = M^2 \Phi^\dagger \Phi \, .
\end{equation}
The parameter $M^2$ is chosen such that $\Phi^\dagger \Phi$ is a relevant operator in the UV limit.
It becomes marginally relevant when perturbing about the fixed point.
In this model the three-dimensional gauge field $A_\mu$ is responsible for a phase shift similar to the one observed in the field-theory Kondo model. Note however that it is non-dynamical.

This model provides a geometric realization of the spin screening
mechanism: The flux through the boundary of $\ads_2$ at the horizon, which counts the
number of defect degrees of freedom, decreases when the temperature is
lowered \cite{Erdmenger:2013dpa}. By adding the backreaction of the
defect brane on the ambient geometry, the impurity entanglement
entropy was calculated in \cite{Erdmenger:2015spo} and shown to agree with previous field
theory results subject to identifying a geometric scale with the Kondo
correlation length.

This model has clearly given rise to a number of successes, both at
the fundamental level and as far as applications to condensed matter
physics are concerned. For
further progress, however, a new model is desired that provides a
top-down string-theory construction for a defect CFT in 1+1
dimensions.

\section{D1/D5 system: gauge theory, sigma model, and holographic dual}
\label{sec:D1/D5}
While bottom-up holographic models have been found that share many characteristics with the Kondo effect, it is desirable to study holographic Kondo analogues in a context where field-theoretic and dual gravitational descriptions are both available.
Our primary interest, therefore, is to study defects with Kondo-like behavior within the context of a concrete top-down model whose CFT and gravitational descriptions are known and reasonably well understood.

The most accessible example of such $\ads_3/\cft_2$ pairs is provided by the low-energy limit of the D1/D5-brane system and its gravitational dual.
This duality is well-understood---having been identified already in Maldacena's original paper \cite{Maldacena1997}---and thus provides a natural starting point for studies of holographic defect RG flows.
These theories provide the ambient CFTs we will sew together with our interfaces.
(The objects studied in this paper are not strictly speaking defects, which inhabit a subspace of a single CFT, but are rather interfaces separating distinct members of a discrete family of CFTs.)

\smallskip

In this section we will briefly review those details of the D1/D5-brane system relevant to the work that follows.
We first review the brane configuration, its effective field theory (EFT) description as a gauge theory, and finally its IR realization as a non-linear sigma model (NLSM) on the moduli space of instantons.
We then briefly discuss the $\U(N_5)$ connection on it required to construct the interfaces we will study in \secref{sec:interfaces}.
Finally, we review the gravitational dual to the D1/D5 CFT.

\subsection{Brane configuration and effective field theory}
The construction of the D1/D5 CFT begins with the spacetime $\R^{1,1}\times M_4\times \R^4$, where the 4-manifold $M_4$ is either $T^4$ or K3.
For simplicity and concreteness we will focus on the $T^4$ case.
We introduce $N_1$ D1-branes and $N_5$ D5-branes extended along $\R^{1,1}$, with the remaining four directions of the D5-branes wrapped on $M_4$.
This system preserves eight supercharges, and as a 2d field theory has $(4,4)$ supersymmetry.

\begin{table}[h]
\centering
\setlength\tabcolsep{1em}
\begin{tabular}{|c|cc|cccc|cccc|}\hline
& \multicolumn{2}{c|}{CFT} & \multicolumn{4}{c|}{$M_4$} & \multicolumn{4}{c|}{$\R^4$}  \\
           &   0 &   1 &   2 &   3 &   4 &   5 &   6 &   7 &   8 &   9 \\\hline
D5 $(N_5)$ & \br & \br & \br & \br & \br & \br & \nb & \nb & \nb & \nb \\
D1 $(N_1)$ & \br & \br & \nb & \nb & \nb & \nb & \nb & \nb & \nb & \nb \\\hline
\end{tabular}
\caption{The D1/D5 brane configuration.}
\label{tab:D1/D5}
\end{table}

The brane's massless effective field content is organized into multiplets of $\cN=(4,4)$ supersymmetry  as follows.
The D1-D1 (D5-D5) strings each furnish a vector multiplet and a hypermultiplet, both in the adjoint of $\U(N_1)$ ($\U(N_5)$).
We have in addition the D1-D5 strings, which form a hypermultiplet in the $(N_1,\overline{N}_5)\oplus(\overline{N}_1,N_5)$ of $\U(N_1)\times\U(N_5)$.
The IR theory is described by the fluctuations around a scalar configuration with vanishing potential energy.
Unlike higher dimensions, where each such configuration would be a vacuum, IR divergences guarantee that these fluctuations probe the entire manifold of such configurations, and it is (a quantum-corrected version of) this manifold on which the IR CFT lives.
This ``moduli space'' of maximally supersymmetric configurations is separated broadly into Higgs and Coulomb branches (more generally there are mixed branches).

The branch of interest to us is the Higgs branch, where the hypermultiplet scalars have non-vanishing expectation---including the adjoint scalar expectations, which necessarily become non-commuting.
Geometrically, such configurations describe non-commutative D1-branes puffing up along the compact directions of the D5-branes.
From the point of view of the D5-brane gauge theory, they are interpreted as $N_1$ gauge instantons of 4d $\U(N_5)$ gauge theory compactified on $M_4$.

In the IR limit the system is an $\cN=(4,4)$ NLSM on the moduli space of solutions to the D-term equations.
This moduli space $\moduli$ is a hyperk\"ahler manifold, and the infrared theory is a $(4,4)$ SCFT with $\moduli$ as its target space.
Classically, this space is simply the moduli space of $N_1$ instantons in 4d YM on $M_4$, and the metric on $\moduli$ is that induced on the moduli space by the Yang-Mills action functional.
However, it receives large quantum corrections in the neighborhood of the singular point where the Higgs and Coulomb branches classically meet \cite{Seiberg:1999xz}.
More generally, $\moduli$ can be deformed in several ways, resolving this singularity classically and giving a positive binding energy to the D1/D5 system \cite{David:2002wn}.

\subsection{Gauge theory description}
We now discuss some pertinent details of the gauge theory description.
As discussed above, with respect to the preserved $\cN=(4,4)$ superalgebra the D1- and D5-branes each have a vector and a hypermultiplet in the adjoint of their respective gauge group.
The hypermultiplet of the D1 gauge theory has as its four scalars the coordinates on the torus, while those of the D5 gauge theory are the gauge field zero modes.
These two gauge theories interact purely via D1-D5 strings, which form a bifundamental hypermultiplet.
The 2d gauge theory thus takes the form
\begin{align}
  L &= L_{\U(N_1)}^\text{gauge} + L_{\U(N_5)}^\text{gauge}
  + L_\text{D1}^\text{hyper} + L_\text{D5}^\text{hyper}
  + L_\text{D1-D5}^\text{hyper} \,.
\end{align}

An $\cN=(4,4)$ vector multiplet is the dimensional reduction of a 6d vector multiplet and contains fields $(A_\mu, A_I, \lambda_{+i\alpha}, \lambda_{-i\dot\alpha})$, together with a symmetric doublet of auxiliary fields $D_{(ij)}$.%
\footnote{Additional details of the $\cN=(4,4)$ gauge multiplet can be found in \appref{app:vectorMultiplet}.}
Here, $\mu=(01)$, $I=(6789)$, $(\alpha,\dot\alpha)$ are doublet indices for the $\SU(2)_-\times\SU(2)_+$ R-symmetry, and $i$ is a doublet index for the $\SU(2)$ rotating the three complex structures on the hypermultiplet target space.
The basic bosonic Lagrangian is the dimensional reduction of six-dimensional
Yang-Mills theory,
\begin{align}
  L^\text{gauge} = \frac{1}{g^2} \Tr \biggl(
  & \frac{1}{2}(F_{01})^2
    + \frac{1}{2}\cD_+ A^I \cD_- A_I
    + \frac{1}{4}[A^I,A^J][A_I,A_J]
    + \frac{1}{4}D^{ij}D_{ij} \biggr)
  \,.
  \label{eq:basic action}
\end{align}
In our conventions $\cD_\pm=\cD_0\pm \cD_1$.
A hypermultiplet consists of a complex scalar doublet $q_i$ and two complex Weyl fermion doublets $(\psi_{-\alpha},\psi_{+\dot\alpha})$, all transforming in some representation of the gauge group.%
\footnote{When the hypermultiplet target space is a curved hyperk\"ahler manifold, $i$ is a local frame index on which $\SU(2)$ acts as a structure group. On $\R^4$ this gives rise to a global symmetry group, which is in any event broken by the periodicity conditions upon compactification to $T^4$.}
We set $\bq^i = (q_i)^\dagger$ and similarly with $\psi_\pm$.
Its bosonic Lagrangian is
\begin{align}
  L^\text{hyper} =
  & - \cD^\mu \bq^i \cD_\mu q_i
    + \bar q^i (D_i{}^j - A^I A_I \delta_i^j) q_j
    \,.
\end{align}
We denote the D1-D1 and D5-D5 hypermultiplets by $(Y_i, \eta_{-\alpha}, \eta_{+\dot\alpha})$.
For example, for the D1 multiplets $Y_i$ is a complex doublet built from the coordinates $Y_{I'}$ on $M_4$.
The D1-D5 hypermultiplets we call $(q_i, \psi_{-\alpha}, \psi_{+\dot\alpha})$, where $q_i$ is an $N_5\times N_1$ matrix transforming in the fundamental of $\U(N_5)$ and the antifundamental of $\U(N_1)$, while $\bq^i = (q_i)^\dagger$ is its hermitian conjugate.

\bigskip
Let us now return briefly to the properties of the classical moduli space $\moduli$.
The allowed configurations are those satisfying the commutation relations
\begin{align}
  [A^\text{D1}_I, A^\text{D1}_J] &= 0 &
  [A^\text{D5}_I, A^\text{D5}_J] &= 0 \\
  [A^I_\text{D1}, Y^i_{D1}] &= 0 &
  [A^I_\text{D5}, Y^i_{D5}] &= 0 \\
  A^I_\text{D5} q_i - q_i A^I_\text{D1} &= 0 \,,
\end{align}
together with the D-term equations
\begin{align}
  [\bY_\text{D1}^{(i}, Y^{j)}_\text{D1}] + \bq^{(i} q^{j)} + \frac{\zeta_\text{D1}^{ij}}{N_1} &= 0 &
  [\bY_\text{D5}^{(i}, Y^{j)}_\text{D5}] - q^{(i} \bq^{j)} + \frac{\zeta_\text{D5}^{ij}}{N_5} &= 0
  \,.
  \label{eq:D-terms}
\end{align}
We are interested in the Higgs branch, the branch of the solution space of the D-term equations for which $q_i$ has maximal rank.
The commutator equations then imply that $A_I$ vanishes for both branes, except for an overall center-of-mass coordinate,
\begin{align}
  A^\text{D1}_I &= a_I \id_{N_1} \,, &
  A^\text{D5}_I &= a_I \id_{N_5} \,.
  \label{eq:AI condition}
\end{align}
It is important for us that this coordinate may vary on $\R^{1,1}$.
This is intimately connected with another deformation: the $\theta$ parameter.

\medskip

To the action \eqref{eq:basic action} we can add two additional supersymmetric terms, the Fayet-Iliopoulos and the 2d $\theta$ terms:
\begin{equation}
  L^\text{gauge}{}' = L^\text{gauge} + L^\theta + L^\text{FI} \,.
\end{equation}
They take the form
\begin{align}
  L^\theta &= \frac{\theta}{2\pi} \Tr(F_{01}) \,, &
  L^\text{FI} &= \frac{1}{2g^2N} \zeta^{ij} \Tr(D_{ij}) \,.
\end{align}
From the supergravity point of view, $\theta_{N_1}$ arises from $C^{(0)}$ and $\theta_{N_5}$ arises from $\int_{M_4}C^{(4)}$.
These parameters are important in what follows: it is the value of $\theta_{N_5}$ that distinguishes the two CFTs separated by the interfaces we study.

It is well-known that such terms behave like the insertion of an electric charge $\frac{\theta}{2\pi}$ at a boundary \cite{Coleman,Witten:1995im}, leading to a non-zero electric field:
\begin{equation}
  \cE = \frac{g^2\theta}{2\pi} \,.
\end{equation}
This is of course expected: from the worldvolume point of view, our interfaces arise from string endpoints and thus carry electric charge.

It is important that, in the presence of a background electric field, some supersymmetries are broken.
This is not a problem for us:
To construct defect-type solutions preserving 4 superscharges, the ambient field theory need only preserve those same 4 supercharges.%
\footnote{Actually, the brane configurations preserve the same number of supersymmetries---it is merely that only half of these supersymmetries coincide with the supersymmetries preserved by the vanilla D1/D5 system.}
Such a solution is found easily.
Consider the supersymmetry variations $(\epsilon_+,\epsilon_-)$ satisfying
\begin{align}\label{defectSUSYs}
  \epsilon_{+i\alpha} &= \tau^9{}_\alpha{}^{\dot\alpha} \epsilon_{-i\dot\alpha} \,.
\end{align}
The unit quaternions $\tau^I{}_\alpha{}^{\dot\alpha}$ are sigma matrices for $\R^4$; detailed conventions can be found in \appref{app:vectorMultiplet}.
Let us consider configurations satisfying \eqref{eq:AI condition} that are invariant under these transformations, which boils down to demanding that the gaugino variations $\delta(\lambda_{+i\alpha}\pm\tau^9{}_\alpha{}^{\dot\alpha}\lambda_{-i\dot\alpha})$, whose building blocks are found at the end of \appref{app:vectorMultiplet}, vanish. We obtain the conditions
\begin{align}\label{vanishingGauginoVariations}
  F_{01} + \cD_1 A_9 &= 0 , &
  \cD_0 A_9 &= 0 , &
  \cD_\pm A_{6,7,8} &= 0 .
\end{align}
In an appropriate gauge, this is equivalent to setting
\begin{align}
  A_9 = A_0 = - \cE x \id
\end{align}
where $\cE$ is the background electric field.
(We also fix $A_{6,7,8}=0$.)
While the classical vacuum energy does not vanish in this case, it saturates a BPS condition with respect to a central charge equal to the fundamental string charge.
Note that on the Higgs branch of the D1/D5 system $A_9$ is equal for all branes, implying that the background gauge fields also coincide:
\begin{align}
  A_0^\text{D1} &= a_9(x) \id_{N_1} \,, &
  A_0^\text{D5} &= a_9(x) \id_{N_5} \,.
\end{align}

The gauge theory description of these solutions lacks manifest Lorentz invariance, but the brane construction of this phase (D1- and F1-strings dissolved in D5-branes) makes it clear that the theory is preserved by Lorentz transformations.
They are, however, not the same ones preserved by the pure D1-brane phase: the addition of the F-string charge causes the brane configuration to lie at a non-zero angle in the 1-9 plane.

\subsection{NLSM and the ADHM connection}\label{sec:ADHM}
The target space $\moduli$ was defined as the space of solutions to the D-term equations \eqref{eq:D-terms}, which we parametrize by coordinates $Z^A$.
In the sigma model limit the gauge kinetic term doesn't contribute, so that the gauge field equations of motion reduce classically to the constraints
\begin{align}
  \bY_\text{D1}^i \lrD_\mu Y^\text{D1}_i - q_i \lrD_\mu \bq^i  &= 0 &
  \bY_\text{D5}^i \lrD_\mu Y^\text{D5}_i + \bq^i \lrD_\mu q_i &= 0 \,.
  \label{eq:gauge constraint}
\end{align}
These equations are satisfied by the following ansatz.
Split the gauge connections $A_\mu$ into $A^{(0)}_\mu + a_\mu$, where $A^{(0)}_\mu$ is the background $\U(1)$ electric field.
As described above, on the Higgs branch of the background the abelian parts of the $\U(N_1)$ and $\U(N_5)$ gauge fields coincide in an appropriate gauge, and the charge assignments therefore imply that $A_\mu^{(0)}$ does not contribute to this constraint.

We now let $Z^A$ be coordinates on the moduli space, so that the non-linear sigma model has $Z^A(x^\mu)$ as its fields.
Restricting to field configurations of the form $\Phi(x^\mu) = \Phi(Z^A(x^\mu))$ for any field $\Phi$ in the gauge theory guarantees the D-term equations are satisfied everywhere on $\R^{1,1}$.
We take $a_\mu$ to have the form
\begin{align}
  a_\mu &= V_A \p_\mu Z^A \,,
\end{align}
where $V_A$ is a connection for $\U(N_1)\times\U(N_5)$ on $\moduli$ with covariant derivative
\begin{align}
  \delta_A &= \frac{\p}{\p Z^A} - i V_A \,.
\end{align}
Equation \eqref{eq:gauge constraint} is then guaranteed to hold provided $V_A$ satisfies the conditions
\begin{align}
  \bY_\text{D1}^i \lr{\delta}_{\!A} Y^\text{D1}_i + \bq^i \lr{\delta}_{\!A} q_i &= 0 \,, &
  \bY_\text{D5}^i \lr{\delta}_{\!A} Y^\text{D5}_i - q^i \lr{\delta}_{\!A} \bq_i &= 0
\end{align}
for all $Z^A$.
This is a linear equation for $V_A$ that can be solved at a generic point on the Higgs branch.
Restricted to these configurations, the bosonic part of the Lagrangian on the Higgs branch now takes the form
\begin{align}
  - \cD^\mu \bY_\text{D1}^i \cD_\mu Y^\text{D1}_i
  - \cD^\mu \bY_\text{D5}^i \cD_\mu Y^\text{D5}_i
  - \cD^\mu \bq^i \cD_\mu q_i
  &=
  - \frac{1}{2} g_{AB} \p^\mu Z^A \p_\mu Z^B \,,
\end{align}
where the metric on $\moduli$ is given by
\begin{align}
  \frac{1}{2}g_{AB} &=
  \Tr_{N_1} \bigl(
    \delta_{(A} \bY_\text{D1}^i \delta_{B)} Y^\text{D1}_i
    + \delta_{(A} \bq^i \delta_{B)} q_i
  \bigr)
  + \Tr_{N_5} \delta_{(A} \bY_\text{D5}^i \delta_{B)} Y^\text{D5}_i \,.
\end{align}

The connection $V_A$ also contributes via the $\theta$ term.
Dropping the contributions of the background $\U(1)$ gauge field, it simply takes the form
\begin{align}
  \frac{\theta}{2\pi} \int \Tr Z^*(\mathbb{F}) \,,
\end{align}
where $\mathbb{F}$ is the field strength of $V$.
This is a flat $B$-field on the target space $\moduli$, which for generic values of $\theta$ has a non-trivial effect on the CFT.

\medskip

In what follows, we will require a connection on $\moduli$ parametrized by $M_4$.
In fact, we will mostly discuss the familiar ADHM case $M_4=\R^4$.
The approach of \cite{Tong:2014cha} to the ADHM construction on $\R^4$ proceeded by viewing the ADHM equations as defining a $\U(N_5)$ connection $\Omega$ on $\R^4\times\moduli$.
The ADHM connection is then simply the restriction of $\Omega$ to $\R^4$ at fixed $Z\in\moduli$.
In \cite{Tong:2014cha} it was pointed out that likewise for any $y\in\R^4$ we obtain a $\U(N_5)$ connection $\Omega$ on $\moduli$.

The connection on $\R^4 \times\moduli$ is defined as follows.
Start with a $(N_5+2N_1)$-dimensional complex vector bundle $\widetilde{\cV}$ transforming in the $(1,N_5)\oplus 2(N_1,1)$ of $\U(N_1)\times\U(N_5)$, where the doublet index of the second factor is labeled by $\ti$.
The connection on this bundle is defined by $V_A$ along $\moduli$, and is trivial along  $\R^4$.

Above any point $(y,Z)\in\R^4\times\moduli$, pick an element $(v,w^{\ti})$ of $\widetilde{\cV}_{(y,Z)}$.
The ADHM map $\Delta^\dagger$ is a homomorphism from $\widetilde{\cV}$ to a vector bundle that is a doublet transforming in the $N_1$ of $\U(N_1)$, given by
\begin{align}\label{ADHMoperator}
  \left[\Delta^\dagger \begin{pmatrix} v \\ w^\ti \end{pmatrix} \right]^i
  &= \bq^i v + (Y^{I'} - y^{I'}) \tau_{I'}{}^{i}{}_{\ti} w^\ti \,.
\end{align}
The $\tau_{I'}$ here denote sigma matrices for the frame indices on $T^4$. $\Delta^\dagger$ generically has full rank, and so its kernel bundle is a complex vector bundle of dimension $N_5$.
We now choose over each $(y,Z)$ a Hermitian orthonormal basis $\{U^a\}$ for the kernel ($a=1,\ldots,N_5$), so that
\begin{align}\label{ADHMkernel}
  \Delta^\dagger U^a &= 0 &
  (U^a)^\dagger U^b &= \delta_a^b \,.
\end{align}
This condition gives a Hermitian vector bundle $\cV$ with structure group $\U(N_5)$.
It is equipped with the natural $\U(N_5)$ connection
\begin{align}
  \Omega_a{}^{b} &= i (U^a)^\dagger \scD U^b \,,
\end{align}
where $\scD$ is the gauge-covariant differential for $\widetilde{\cV}$ over $\R^4\times\moduli$; $\Omega$ is simply the restriction of $\scD$ to $\cV$.
With respect to the coordinates $(y^I, Z^A)$, $\scD$ decomposes into $(\p_I,\delta_A)$.
The components of $\Omega$ along $\R^4$ are the ADHM connection at fixed $Z\in\moduli$, while those along $\moduli$ define a $\U(N_5)$ connection on $\moduli$ for each $y\in \R^4$.%
\footnote{CMT would like to thank D.\ Gaiotto for explaining this connection and introducing him to reference \cite{Tong:2014cha}.}

When $M_4$ is compact (as in the case of interest) the map $\Delta$ must be modified.
We will not concern ourselves with these details here and simply assume an appropriate $\Delta$ exists.
Once $\Delta$ is in hand, the construction of $\Omega$ proceeds unchanged.

\subsection{Gravity dual}
Finally, we briefly review the holographic dual of the D1/D5 CFT.
The dual theory is obtained by taking the near-horizon limit of the D1/D5 black brane solution of Type IIB supergravity.
In string frame, the black brane solution takes the form
\begin{align}
ds^2 &= (Z_1 Z_5)^{-1/2}dx^2(\R^{1,1}) + (Z_1 Z_5)^{1/2}dx^2(\R^4) + \left(\frac{Z_1}{Z_5}\right)^{1/2} ds^2(M_4) \\
F^{(3)} &= 2r_1^2 g_s e^{-2\dil} *_6 \omega_{\S^3} + \frac{2r_5^2}{g_s}\omega_{\S^3} \\
e^{-2\dil} &= \frac{1}{g_s^2}\frac{Z_5}{Z_1}  \,,
\label{eq:D1/D5 sugra}
\end{align}
with $\omega$ the unit volume form on $\S^3$, $*{}_6$ the Hodge dual in the Euclidean $(\R^{1,1},\R^4)$ plane, and
\begin{align}
  Z_1 &= 1 + \frac{r_1^2}{r^2} & r_1^2 &= \frac{g_s N_1 \alpha'}{v_4}\\
  Z_5 &= 1 + \frac{r_5^2}{r^2} & r_5^2 &= g_s N_5\alpha'\,.
\end{align}
Here $r$ is the radius in the $\R^4$ plane, while $v_4$ is the volume of $M_4$ at $r=\infty$ in units of $(2\pi)^4\alpha'^2$.%
\footnote{We emphasize that, in our conventions, the string coupling $g_s$ is \emph{a parameter of the solution}, equal to the value of $e^\dil$ in the asymptotically flat region, and does not appear explicitly in either $\kappa_{10}^2$ or any D-brane action.
In particular, this means our RR fields are rescaled by a factor of $g_s$ relative to the most common convention.
In addition, the $p$-brane tension $T_p$ never includes the contribution from the asymptotic dilaton, meaning that $T_1 = (2\pi\alpha')^{-1}$ is the tension parameter used for both the D-string and the F-string.}
The IR brane dynamics are captured by the limit $r \ll r_1,r_5$.
The resulting near-horizon metric is
\begin{align}
ds^2 &= L^2(ds_{\ads_3}^2 + ds_{\S^3}^2) + \left(\frac{N_1}{v_4 N_5}\right)^{1/2} ds_{M^4}^2 \notag\\
F^{(3)} &= 2\alpha' N_5(\omega_{\ads_3}+\omega_{\S^3}) \notag\\
e^{-2\dil} &= \frac{1}{g_6^2}\frac{N_5}{N_1}
\label{eq:D1/D5 dual}
\end{align}
where $g_6^2=\frac{g_s^2}{v_4}$, $L^2 = r_1 r_5$.
Also, $ds_{\ads_3}^2$ and $ds_{\S^3}^2$ are unit radius metrics and $\omega_{\ads_3}$ and $\omega_{\S^3}$ are their volume forms.
For the supergravity approximation to be accurate, we must also have $r_{1,5}$ much larger than both the string and Planck lengths.

The holographic correspondence posits that Type IIB string theory on this background is equivalent to the D1/D5 CFT.
Therefore, this background and its deformation by interfaces forms the focus of this paper.

\section{Interfaces in the D1/D5 system}
\label{sec:interfaces}
We now add an interface located at $x^1=0$ to the CFT associated with the D1/D5-brane system. For this purpose, we consider an extended object localized at $x^1=0$ that intersects the D1/D5-brane system. This amounts to a UV brane/defect system whose IR limit will then be an interface CFT.

More specifically, we obtain the UV configuration in question by placing $p$ infinitely long fundamental strings at $x^1=0$, such that they intersect the D1/D5 system. These strings extend in the $09$ directions.
We can choose them to end freely on the D1/D5 system, thereby truncating them to semi-infinite strings.
Moreover, the strings can be given a finite length by introducing a D3-brane extended in the $0678$ directions, on which they are allowed to end.
The addition of this D3-branes does not break any additional symmetries.
The corresponding brane configuration is summarized in \tableref{tab:branes}.
\begin{table}[!h]
\centering
\setlength\tabcolsep{1em}
\begin{tabular}{|c|cc|cccc|cccc|}\hline
& \multicolumn{2}{c|}{CFT} & \multicolumn{4}{c|}{$M_4$} &&     &     &     \\
           &   0 &   1 &   2 &   3 &   4 &   5 &   6 &   7 &   8 &   9 \\\hline
D5 $(N_5)$ & \br & \br & \br & \br & \br & \br & \nb & \nb & \nb & \nb \\
D1 $(N_1)$ & \br & \br & \nb & \nb & \nb & \nb & \nb & \nb & \nb & \nb \\
F1 $(p)$   & \br & \nb & \nb & \nb & \nb & \nb & \nb & \nb & \nb & \br \\
D3 $(1)$   & \br & \nb & \nb & \nb & \nb & \nb & \br & \br & \br & \nb \\\hline
\end{tabular}
\caption{Brane configuration generating a class of defects.}
\label{tab:branes}
\end{table}

Na\"ively, the gauge-theory realization of a single string ending on
the D5-branes is simply given by a supersymmetric Wilson line in the fundamental representation,
\begin{equation}
  \cW = \Tr \cP\exp\biggl( i \int (A^\text{D5}_0-A_9^\text{D5}) dt \biggr) \,.
\end{equation}
However, due to the non-vanishing 1-5 hypermultiplet the long strings ending on the D5-branes mix with those ending on the D1-branes \cite{Tong:2014yna,Tong:2014cha}.
As we will review below, the net effect is that the coupling of these strings to the CFT is defined in terms of the $\U(N_5)$ connection on $\moduli$ reviewed in section \secref{sec:ADHM}.

The content of this section is organized as follows.
We first determine in \secref{sec:4.1} the basic properties of the above brane configuration, including the supersymmetries preserved by the interface and its description in the low-energy NLSM via the connection reviewed in \secref{sec:ADHM}.
\Secref{sec: pqStringProbeBrane} reviews the near-horizon limit of the brane configuration \tableref{tab:branes} in the S-dual frame, which we use in \secref{sec:4.3} to construct the holographic interface RG flows.
This is done in the following way.
When we applied S-duality to the brane configuration of \tableref{tab:branes}, the fundamental strings became D-strings.
Introducing a small non-abelian polarization near the boundary, these D-strings now flow in the infrared to D3-branes wrapped on $\S^2\subset\S^3$ via the Myers effect \cite{Myers:1999ps}.
Once the D3-branes have puffed up to well above the string scale, we may return to the D1/D5 frame, where the flow is now described by studying the D3-brane configuration as a function of the radial coordinate (which plays the role of the RG scale).
We are most interested in supersymmetric flows, which we derive by imposing kappa symmetry and solving the resulting first-order differential equation exactly.
Our solutions show that the D3 defects slide down on the $\S^3$ until they stabilize at a finite polar angle.
These results will be used to identify which flows join the fixed points described by the backreacted supergravity solutions of \secref{sec:sugra}.
%

\subsection{SUSY interfaces in the D1/D5 system}
\label{sec:4.1}
Consider a brane configuration as in \tableref{tab:branes}, with the D3-branes well separated from the D1/D5 branes in the 9 direction.
The Type IIB supersymmetries preserved by this configuration are those satisfying%
\footnote{For our conventions regarding IIB supergravity see \appref{app:IIB}.}
\begin{align}
  \Gamma_{01} J \epsilon_\text{IIB} &= \epsilon_\text{IIB} &
  \Gamma_{012345} J \epsilon_\text{IIB} &= \epsilon_\text{IIB} &
  \Gamma_{09} K \epsilon_\text{IIB} &= \epsilon_\text{IIB} \,,
\end{align}
where $J,K$ are matrices acting on the $\SL(2,\R)$ doublet index of the spinor $\epsilon$.
In terms of the $\N=1, d=6$ supersymmetry algebra along the $016789$ directions, this corresponds to the condition $\Gamma_{09}\epsilon_\text{6d} = \epsilon_\text{6d}$.
Upon dimensional reduction to 2d this becomes the condition
\begin{equation}
  \epsilon_{+i\alpha} = \tau^9{}_\alpha{}^{\dot\alpha} \epsilon_{-i\dot\alpha}
  \qquad\Leftrightarrow\qquad
  \epsilon_{-i\dot\alpha} = \btau^9{}_{\dot\alpha}{}^{\alpha} \epsilon_{+i\alpha} \,.
\end{equation}
Supersymmetry parameters satisfying this condition close on translations along $x^0$.

\medskip

Such strings can break on the D1/D5 system, allowing the two ends of the string to be separated in the $x^1$ direction.
The same supersymmetries are preserved regardless of whether the F1-strings are infinite or terminate; in the latter case, however, the string acts as an interface rather than a defect.
This is due to the fact that when $M_4$ is compact, the theta angle $\theta_\text{D5}$ of the D5 gauge theory jumps as the defect locus is crossed:
\begin{equation}
  \Delta\theta_\text{D5} = \frac{2\pi p}{N_5} \,.
\end{equation}
This relation is derived in the holographic limit from the supergravity equations of motion in \appref{app:theta}.

As was pointed out in \cite{Tong:2014cha}, it is not consistent with supersymmetry to introduce only 5-3 strings: it is also necessary to include 1-3 strings, which have non-trivial Yukawa couplings with the 1-5 and 5-3 strings.
On the Higgs branch of the D1/D5 system the Yukawa couplings cause the BPS 5-3 string to become mixed with 1-3 strings, while the remaining modes receive a further gap.
As a result, the modes remaining at low energy are an $N_5$-dimensional subbundle of the (1-3)$+$(5-3) vector bundle that naturally lives in the fundamental of a $\U(N_5)$ gauge group on $\moduli$.
This bundle is simply the bundle reviewed in \secref{sec:ADHM}.

Denote the BPS interface fermions which result from the mixing by $\eta_a$ and the $\U(N_5)$ connection on $\moduli$ by $\Omega_{Aa}{}^b$.
The coupling to the Wilson line is then \cite{Tong:2014cha}
\begin{align}
 S_\eta =  \Tr \int\!dt\, \bar\eta^a (i\delta_a^b \p_t + \dot Z^A \Omega_{Aa}{}^b) \eta_b \,.
  \label{eq:interface coupling}
\end{align}
This term drives the boundary RG flow. It corresponds to a marginally relevant
deformation. We note a formal similarity of the $(0+1)$-dimensional
fermionic interface degrees of freedom $\eta_a$ \footnote{We note that upon introducing the defect into the sigma
  model discussion of the preceding section, the $(0+1)$-dimensional fields $\eta_a$ arise form the fields and objects in \eqref{ADHMoperator} and \eqref{ADHMkernel} through
\begin{equation}
 \begin{pmatrix} v \\ w^\ti \end{pmatrix}=U_a\eta_a.
\end{equation}
While $v$ stems from the (5-3) strings and transforms in the
fundamental of $\U(N_5)$, the $w^\ti$ ($\ti=1,2$) stem from the (1-3)
strings and transform in the fundamental of $\U(N_1)$. The $U_a$ are
$(N_5+2N_1)$-dimensional vectors.} with the Abrikosov
fermions in the large $N$ Kondo model described in \eqref{decomp}. Note however that the
Wilson line operator \eqref{eq:interface coupling} is single-trace. It
thus differs in nature from
the double-trace operator driving the RG flow in the holographic
Kondo model of \cite{Erdmenger:2013dpa}, as described below
\eqref{decomp} above.  
We expect that switching on this Wilson line operator induces a RG flow analogous to the one discussed for
the D0/D2-brane case around Fig.~\ref{fig: BranePolarization}, with
the F1 string puffing up. We leave a field-theory analysis of this
mechanism to the future and turn to its gravity dual in the subsequent.

\bigskip
When the interface also has D1-brane charge its description is more involved.
A more thorough discussion of this case will be presented in future work \cite{next-paper}, but we make a few brief comments on this case here.

In this case, the value of $N_1$---and thus the rank of the gauge group---jumps across the interface.
Given a marked point $y$ on $M_4$ there is a natural embedding of $\moduli$ into $\cM_{(N_1+q),N_5}$ in which we allow $N_1$ general instantons and place $q$ small instantons at $y$.
By imposing that the fields of the $\cM_{(N_1+q),N_5}$ theory are restricted at $x^1=0$ to $\moduli$, we may further couple to the interface degrees of freedom via \eqref{eq:interface coupling}.

\subsection{Holographic realization}
\label{sec: pqStringProbeBrane}

The holographic description of these interfaces follows simply from the brane constructions: they correspond to $(p,q)$ strings on an $\ads_3\times \S^3\times M_4$ background.
When we discuss the brane polarizations that trip the interface flows, however, it is simplest to work in the $S$-dual frame.
The D1/D5 system interface system then becomes an F1/NS5 system with $(q,p)$ string interfaces.

There is also another reason to work in the dual frame.
The larger the D1-brane tension is in comparison to the fundamental string tension, the smaller the incidence angle in the brane configuration between the $(p,q)$-string and the D1/D5 system becomes.
At small string coupling, where we can trust the DBI-CS action, the D1-brane tension becomes very large, causing the $(p,q)$-string to merge with the CFT.
In order to prevent this, we have to make the string coupling large, invalidating the worldsheet description.
As a result, we don't expect the $(p,q)$-string worldsheet action to give accurate results at small coupling when $q\ne 0$.
The F1/NS5 description, on the other hand, is useful when the D1/D5-frame coupling is large, and this is the limit in which adding D1-brane charge has a small effect on the shape of the string.

We assume for simplicity that the axion $C^{(0)}$ vanishes in our background.
In this case, in string frame the S-dual configuration takes the form
\begin{align}
  \widehat{ds}{}^2 &= \alpha' N_5 (ds_{\ads_3}^2 + ds_{\S^3}^2) + \frac{1}{g_s} ds_{M^4}^2 &
  e^{-2\hat\dil} &= g_6^2 \frac{N_1}{N_5} \\
  \hat{H}{}^{(3)} &= 2\alpha' N_5(\omega_{\ads_3} + \omega_{\S^3}) = d\hat{B}{}^{(2)} &
  \hat{B}{}^{(2)} &= \hat{B}_{\ads_3} + \hat{B}_{\S^3} \\
  \hat{B}_{\ads_3} &= \alpha' N_5 (\psi + \tfrac{1}{2}\sinh 2\psi )\omega_{\ads_2} &
  \hat{B}_{\S^3} &= \alpha' N_5 (\theta-\tfrac{1}{2}\sin 2\theta) \omega_{\S^2} \,.\label{KalbRamondField}
\end{align}
(Note that $g_s$ and $g_6$ denote the values of these quantities in the D1/D5 frame.)
The $\ads$ radius in this frame is $\hat{L}^2 = \alpha' N_5$.

We now introduce $p$ D1-branes with $q$ units of fundamental string charge dissolved in them, and look for solutions preserved by the $d=1$ $\cN=4$ superconformal algebra.
These solutions appear widely in the literature (see e.g.\ \cite{Bachas:2000fr, Bachas:2001vj}), but we cover them here briefly.
We first consider the case where all $p$ D1-branes lie on the same locus.
In this case, a superconformal configuration lies on an $\ads_2$ slice.
Choosing an embedding of $d=1$ superconformal symmetry into that of $d=2$ induces a slicing of $\ads_3\times \S^3$ by $\ads_2\times \S^2$,
\begin{equation}
  \widehat{ds}_{6}^{2}=\alpha' N_5\left(d \psi^{2}+\cosh ^{2} \psi \, d s_{\ads_{2}}^{2}+d \theta^{2}+\sin ^{2} \theta \, d s_{\S^{2}}^{2}\right)
\end{equation}
Note that we have suppressed the metric of $M_4$, as it is irrelevant to what follows.
The $R$-symmetry of the $\cN=(4,4)$ algebra preserving this configuration is the $\SU(2)$ acting on the $S^2$.

The Lagrangian for the center of mass is
\begin{align}
  L_{(q,p)} &= - p\, T_1 e^{-\hat\dil} \sqrt{ -\det(\hat g + \hat B - 2 \pi\alpha' F) } \,,
\end{align}
where $\hat g$ and $\hat B$ are the pullbacks of these fields to the D1 worldsheet, and $F$ is the worldsheet gauge field strength.
If we are to preserve $\SO(2,1)\times\SU(2)$ symmetry, the interface must lie along a slice of $\ads_2$, and the field strength must be proportional to its volume form.
We thus write
\begin{equation}
  2\pi\alpha'F = \alpha' N_5 f\, \omega_{\ads_2} \,.
\end{equation}
On this configuration, the Lagrangian now takes the form
\begin{equation}
  L_{(q,p)} =
  - p\, T_1 e^{-\hat\dil} \sqrt{ g_{\ads_2} } \bigl( {\cosh}^4\psi - (\sinh\psi + \psi - f)^2 \bigr)^{1/2} \,.
\end{equation}
The value of $\psi$ is obtained by extremizing $L_{(q,p)}$, giving $\psi = f$.
Picking coordinates $(t,z)$ on $\ads_2$, the value of $f$ is determined by the string charge $q$ via the relation
\begin{align}
  q &= -\frac{\p L_{(q,p)}}{\p F_{tz}} = p\, T_1 e^{-\hat\dil} \frac{B_{tz} - 2\pi\alpha' F_{tz}}{\sqrt{-\det(\hat g + \hat{B} - 2\pi\alpha'F)}} \\
  &= p\,e^{-\hat\dil} \sinh\psi \in \Z \,.
\end{align}
On shell, the center of mass Lagrangian now takes the form
\begin{align}
  L_{(q,p)}^\text{o.s.} &= - p\, T_1 e^{-\hat\dil} \alpha' N_5 \sqrt{-g_{\ads_2}} \cosh\psi \\
  &= - \alpha' N_5 T_1 \sqrt{ q^2 + e^{-2\hat\dil}p^2 } \sqrt{-g_{\ads_2}} \,,
  \label{eq:string-on-shell}
\end{align}
which is simply the action of a brane of tension $T_{(q,p)}$ on an $\ads_2$ background of radius $\alpha' N_5$.

\subsection{Non-abelian description of the interface flow}
\label{sec:4.3}
Our next task is to deform the defect by a relevant operator and see where it flows.
The defect flow we are interested in is tripped by a non-abelian deformation of the defect embedding coordinates.
Since such a process is not easily described in the fundamental string language, we continue in the S-dual frame.
Note that in this section we set $q$ (the D1-brane charge in the original frame and the F-string charge in the S-dual frame) to zero.
In principle there is no problem with setting $q\ne 0$, but we note that in this case $p$ and $q$ must have a common divisor;
this is because the component of the gauge group that survives in the BPS vacuum is $\U(d)$, where $d$ is the greatest common divisor of $p$ and $q$ \cite{Witten:1995im}.

Before we discuss the RG flow, recall that the radial coordinate in $\ads$ has the interpretation of an energy scale in the field theory. Therefore, in our construction, it assumes the role of RG time.

We now wish to consider relevant deformations of our brane configuration.
Abelian deformations -- for example, a shift in the location on $M_4$ or $\S^3$ -- are all irrelevant operators.
To find a natural relevant deformation we must turn to non-abelian polarization of the defect.

This type of deformation is familiar from the $\SU(2)$ WZW model.
Starting with the BCFT corresponding to $p$ D0-branes on $\SU(2)\simeq \S^3$, there exists for $p>1$ a relevant boundary deformation.
The deformation involves a maximally non-abelian deformation of the $\S^3$ embedding coordinates.
Because the $H$ field is non-vanishing on $\S^3$, a set of branes so polarized becomes unstable toward flow to a single D2-brane wrapped stably on some $\S^2\subset \S^3$.

The $\S^3$ of our model is in fact described by just such a WZW sector.
What has changed is that the string worldsheet theory in the presence of a deformation must remain conformal, so that the RG flow in the WZW model is now be realized as a ``dynamical'' process evolving in the direction of increasing $z$. Our analysis parallels the exposition in appendix B of \cite{Camino:2001at}.

Let us examine briefly what happens when we first turn on the flow.
To simplify matters, we switch to stereographic coordinates on $\S^3$:
\begin{align}
  ds^2_{\S^3} &= \frac{(2\, d\vec{x})^2}{(1+r^2)^2} \,, \qquad\qquad \vec{x}\in\R^3 \,.
\end{align}
Stereographic coordinates are related to polar coordinates by $r=\tan\tfrac{\theta}{2}$.
The $B$-field \eqref{KalbRamondField} on $\S^3$ now takes the form
\begin{equation}
B_{\S^3}
= \hat{L}^2 b(\theta) \, \omega_{\S^2}
= \hat{L}^2 b(\theta) \frac{\epsilon_{ijk}x^i dx^j\wedge dx^k}{r^3} \,, \qquad\qquad
b(\theta) = \theta - \sin\theta\cos\theta \,.
\end{equation}
For convenience, we set $g(r)=\frac{4}{(1+r^2)^2}$ so that $g_{ij}(\S^3) = g(r)\delta_{ij}$.

As the brane's worldsheet coordinates we fix the $(t,z)$ directions, and pick the pole $\vec{x}=0$ to be the $\S^3$ location of the D1-branes in the UV.
We now study a deformation of the system in which the $\S^3$ embedding coordinate matrix $\vec\bx$ of the D1-branes in stereographic coordinates takes the form
\begin{align}\label{eq:x matrices}
  \bx^i &= \lambda f(z)\bSigma_i \,,
\end{align}
where the Hermitian matrices $\bSigma_i$ satisfy the $\su(2)$ commutation relations
\begin{align}
  [\bSigma_i,\bSigma_j] &= i\epsilon_{ijk}\bSigma_k \,.
\end{align}
We further assume that the fundamental of $\u(p)$ is irreducible under $\su(2)$,  making it the spin $\frac{p-1}{2}$ representation.
Then $\bdr^2=\vec\bx^2 = C_2(\Sigma)(\lambda f)^2\id$, where $C_2(\Sigma)=\frac{p^2-1}{4}$ is the quadratic Casimir of $\su(2)$ in the representation defined by $\vec\bSigma$, which turns $r=\sqrt{C_2}\lambda f\id$ into an abelian quantity.
We further assume that the brane has a fixed location in the $x^1$ and $M_4$ directions.
Moreover, we require $\psi$ to be an abelian constant, which ensures the preservation of the $(0+1)$-dimensional conformal group $\SO(2,1)$.

The non-abelian DBI Lagrangian takes the form \cite{Myers:1999ps}
\begin{align}\label{nonAbelianDBI}
  I_\text{DBI} &= -T_\text{D1}\Tr\left(
  	e^{-\dil}\sqrt{-\det(E_{ab}+E_{ai}(Q^{-1}-\delta)^{ij}E_{jb}+\lambda F_{ab})\det(Q^i{}_j)},
  \right)
\end{align}
where
\begin{align}
  E_{\mu\nu} &= \hat{g}_{\mu\nu} + \hat{B}_{\mu\nu}\,, &
  Q^i{}_j &= \delta^i{}_j - i\lambda[\bx^i,\bx^k]E_{kj} \,.
\end{align}
In this expression, $\xi^a=(t,z)$ denote the worldsheet variables, while $x^i$ denote the transverse variables.

The full form of \eqref{nonAbelianDBI} must be supplemented by terms involving higher powers in covariant derivatives.
However, it has been proven that, when commutator terms are grouped with covariant derivatives of $F$, that the totally symmetric prescription yields correct results when covariant derivative terms of $F$ are much smaller than $F$ itself \cite{Tseytlin:1997csa}.
Here, however, our aim is to show that flows corresponding to configurations \eqref{eq:x matrices} exist, and it is enough to find the leading contributing terms in the expansion of \eqref{nonAbelianDBI}.
In particular, to determine whether a relevant deformation exists, we need to know whether the potential is stable or unstable to perturbations away from $\theta=0$.

We are assuming that $q=0$, and therefore $\psi = F_{ab} = 0$.
The relevant components of $E_{\mu\nu}$ are then
\begin{equation}
E_{ij} = \hat{L}^2\biggl(g\,\delta^i\,_{j}
+ \frac{b(\theta(r))}{r^2} \epsilon_{ijk}\bSigma^k\biggr),
\quad
E_{ab} = g_{ab},
\quad
E_{ia}=0=E_{ai},
\end{equation}
and we also have%
\footnote{We refrain from matching index placement on both sides since raising and lowering involves $E^{ij}$.}
\begin{equation}\label{nonAbelianDBI Q}
Q^i\,_{j}=\biggl(1-\frac{2L^2b}{\lambda\sqrt{C_2}}\biggr)\delta^i\,_{j}\id+\frac{2L^2b}{\lambda (C_2)^{3/2}}\bSigma^j\bSigma^i+\frac{2L^2}{\lambda C_2}r^2\,g\,\epsilon_{ijk}\bSigma^k
\end{equation}
Then
\begin{equation}\label{nonAbelianDBIpullbackDeterminant}
-\det(E_{ab}+E_{ai}(Q^{-1}-\delta)^{ij}E_{jb})=\frac{L^2}{z^2}\biggl(\frac{L^2}{z^2}+\frac{(\p_z r)^2}{C_2}\bSigma^i \bigl(Q^{-1}\bigr)_{ij} \bSigma^j \biggr)
\end{equation}
where $Q^{ij}(Q^{-1}\bigr)_{jk}=\delta^i\,_{k}$ with $Q^{ij}=E^{ij}-i\lambda[\bSigma^i,\bSigma^j]$ and $E^{ij}E_{jk}=\delta^i\,_{k}$.

In order to extract the dimension of the perturbing operator, we need only contemplate the leading behavior of the potential generated by \eqref{nonAbelianDBI}.
Inspection of \eqref{nonAbelianDBIpullbackDeterminant} makes clear that this determinant contains no terms that are pure powers of $\theta$, but always feature derivatives in $z$.
Hence, we content ourselves with expanding its contributions to leading order, $\bigl(Q^{-1}\bigr)_{ij}=4L^2\delta^i\,_{k}+\cO(\theta)$ and $r=\theta/2+\cO(\theta^3)$.
We get
\begin{equation}
-\det(E_{ab}+E_{ai}(Q^{-1}-\delta)^{ij}E_{jb})=\frac{L^2}{z^2}\biggl(\frac{L^2}{z^2}+L^2 (\p_z\theta)^2 \biggr)+\dots
\end{equation}
The important terms must then come from the other determinant in \eqref{nonAbelianDBI}.
Indeed, we find terms of the correct orders when expanding \eqref{nonAbelianDBI Q} with $r^2\,g = \sin^2\theta$,
\begin{equation}
Q^i\,_{j}=\delta^i\,_{j}+\theta^2\frac{2L^2}{\lambda C_2}\epsilon_{ijk}\bSigma^k+\theta^3\frac{4}{3(C_2)^{3/2}\lambda}\bigl(\bSigma^j\bSigma^i-C_2\delta^i\,_{j}\bigr).
\end{equation}
This yields
\begin{equation}
\sqrt{Q^i\,_{j}}=\biggl(1-\theta^3\frac{4}{3(C_2)^{1/2}}\biggr)\id+\cO(\theta^4).
\end{equation}
Overall we then get
\begin{equation} \label{margrel}
e^{\dil}I_\text{DBI} =
  	\frac{L^2}{z^2}+\frac{L^2}{2}(\p_z\theta)^2-\frac{L^2}{z^2}\frac{4L^2}{3\lambda\sqrt{C_2}}\,\theta^3+\dots
  	\,.
\end{equation}
Here, the $\theta$ deformation has no mass term, but the leading (cubic) term is unstable.
The holographic dictionary thus implies that the perturbation is marginally relevant---just as the Kondo deformation is.
This confirms that flows corresponding to configurations \eqref{eq:x
  matrices} arise under deformation by a (marginally) relevant
operator.

It is natural to identify the field-theory operator dual to $\theta$
in \eqref{margrel} with the Wilson line operator in
\eqref{eq:interface coupling}. We leave a detailed analysis of this
duality to future work.

\subsection{D3-brane description of the interface flow}
\label{sec:D3 probe brane}
%
In this section, we demonstrate the existence of a non-trivial IR fixed point.
It is well known that, when D1-branes undergo a non-abelian polarization as above, they puff up into a fuzzy sphere.
When the size of the fuzzy sphere is well above the string scale, the system has a simple and convenient description in terms of the DBI-CS action of a single D3-brane \cite{Myers:1999ps}.
In this section we derive the explicit supersymmetric flow from a localized D1/F1 interface in the UV to a puffed up D3 interface in the infrared using the abelian D3-brane action.
These flows are closely related to the ``baryon vertex'' solutions first derived in \cite{Imamura:1998gk}, and for vanishing D1-brane charge is equivalent to the $p=5$ case studied in \cite{Camino:2001at}.

Note that in this section we return to the more familiar D1/D5 frame, where the background is supported by the RR 3-form field strength.
(The analysis is not, however, much different in the F1/NS5 frame.)
In this D1/D5 background, the D3-brane DBI-CS action takes the form
\begin{align}\label{D3 DBI CS action}
  \cL_\text{DBI} &= -T_\text{D3} e^{-\dil} \sqrt{D}
  + T_\text{D3} \int C^{(2)} \wedge \cF \,,
\end{align}
where
\begin{align}
  D &= -\det(\hat g + \cF) \,, &
  \cF &= 2\pi\alpha' F \,.
\end{align}
Since we are interested in D3-branes with manifest $\SU(2)$ symmetry, the
probe brane geometry takes the form of an $\S^2\subset \S^3$ fibered over a codimension~1 surface $\Sigma\subset\ads_3$.
We take $\xi=(t,z,\phi,\chi)$ as our worldvolume coordinates, with unit sphere metric
\begin{align}
  ds^2_{\S^3} &= d\theta^2 + {\sin}^2\theta\,(d\phi^2 + {\sin}^2\phi\,d\chi^2) \,,
\end{align}
Moreover, the D3-branes have $p$ units of F1 charge and $q$ units of D1 charge dissolved in them.
These charges are supported by gauge flux on the $\ads_2$ and the $\S^2$, respectively.
In terms of worldvolume fields,
\begin{align}
  p &= \int_{\S^2} \frac{\p\cL}{\p F_{tx}} \in \Z \,, &
  q &= \frac{1}{2\pi}\int_{\S^2} F_{\phi\chi} \in \Z \,.
\label{eq:D3 quantization}
\end{align}

Supersymmetric configurations are ones such that the effect of spacetime supersymmetry on the brane configuration are pure gauge, in the sense that they can be eliminated by acting with a $\kappa$-symmetry transformation.
$\kappa$-symmetry is a local fermionic symmetry on the brane worldvolume, and is the fermionic partner to worldvolume coordinate reparametrizations.
The action of $\kappa$-symmetry is determined by a fermionic parameter $\kappa$ that is a spacetime spinor.
Half of its components act trivially, while the non-trivial components are those satisfying $\Gamma_\kappa\kappa=\kappa$, where $\Gamma_\kappa$ is a brane configuration-dependent matrix whose form we will return to momentarily.
If $\zeta$ is a spacetime Killing spinor, then the condition that $\zeta$ is preserved by a given brane configuration is that $\Gamma_\kappa\zeta=\zeta$ \cite{Bergshoeff:1997kr}.

Consider momentarily the flat space configuration consisting of a $(p,q)$ string crossing the D1/D5 system and preserving maximal supersymmetry.
This system preserves the 4 supersymmetries whose variation parameters $\epsilon$ satisfy
\begin{equation}
\epsilon
= \Gamma_{01}J \epsilon
= \Gamma_{012345}J \epsilon
= \Gamma_{09}K \epsilon \,.
\label{eq:flow susies}
\end{equation}
While in the infrared the symmetry algebra is enhanced by superconformal generators, since we are interested in the interface flow itself,
we require only the supersymmetries \eqref{eq:flow susies} to be preserved.

If we now consider the backreacted D1/D5 geometry, the Killing spinors of this background differ only by a scale factor \cite{BLT}.
As a result, our goal is to find brane configurations such that $\Gamma_\kappa\epsilon=\epsilon$ for any Killing spinor satisfying \eqref{eq:flow susies}.
Since it is natural for us to work in polar coordinates on $\R^4$, the final constraint should be expressed in terms of $(r,\theta,\phi,\chi)$.
This yields the following constraints:
\begin{align}
  \Gamma_{\underline{tx}} \epsilon &= J\epsilon &
  \Gamma_{\underline{tr}} \epsilon &= e^{-\theta\Gamma_{\underline{r\theta}}} K \epsilon \\
  \Gamma_{\underline{t\theta}}\epsilon &= -\Gamma_{\underline{tr\phi\chi}} J \epsilon &
  \Gamma_{\underline{\phi\chi}}\epsilon &= -\Gamma_{\underline{r\theta}}J\epsilon \\
  \Gamma_{\underline{tr\phi\chi}} \epsilon &= -(\sin\theta+\Gamma_{\underline{r\theta}}\cos\theta)I\epsilon &
  \Gamma_{\underline{tx\phi\chi}} \epsilon &= -\Gamma_{\underline{r\theta}} \epsilon \\
  \Gamma_{\underline{tr}} \epsilon &= (\cos\theta - \Gamma_{\underline{r\theta}}\sin\theta)K\epsilon &
  \Gamma_{\underline{t\theta\phi\chi}} \epsilon &= -(\cos\theta-\Gamma_{\underline{r\theta}} \sin\theta) I \epsilon \,.
\end{align}
$\Gamma_\kappa$ is given by~\cite{Cederwall:1996pv,Cederwall:1996ri}
\begin{equation}
  d^4\xi \sqrt{-\det(\hat g + \cF)} \Gamma_\kappa
  = \gamma^{(4)} I + \gamma^{(2)}\wedge \cF\, J + F\wedge \cF\,I \,,
\end{equation}
where
\begin{equation}
  \gamma^{(k)} = \frac{1}{k!} \gamma_{m_1}\cdots\gamma_{m_k} d\xi^{m_1}\wedge\cdots\wedge d\xi^{m_k}
\end{equation}
and
\begin{equation}
  \gamma_m = \frac{\p x^\mu}{\p\xi^m} \Gamma_\mu
  \qquad\qquad
  \Gamma_\mu = e_{\mu}^{\underline{\mu}} \Gamma_{\underline{\mu}} \,.
\end{equation}
With $\cF_{tx}$ and $\cF_{\phi\chi}$ the only non-vanishing components of $\cF$, the constraint takes the form
\begin{align}
  (\gamma_{tx\phi\chi} + \cF_{tx} \cF_{\phi\chi}) I \epsilon + (\gamma_{tx} \cF_{\phi\chi} + \gamma_{\phi\chi}\cF_{tx}) J \epsilon &= \sqrt{-\det(\hat g + \cF)} \epsilon \,.
\end{align}
This needs to be solved for any $\epsilon$ satisfying the above relations.
Applying these relations allows us to reduce to the equations
\begin{align}
  U^{-1/2} \cF_{\phi\chi} + V \frac{d}{dx}(r\sin\theta) &= \sqrt{-\det(\hat g + \cF)} \label{eq: firstConstraint}\\
  \cF_{\phi\chi}(\cF_{tx} - \frac{d}{dx}(r\cos\theta) ) &= 0 \\
  V( \cF_{tx} -\frac{d}{dx} (r\cos\theta) ) &= 0 \\
  \cF_{\phi\chi}\frac{d}{dx}(r\sin\theta) &= U^{-1/2} V \,,
\end{align}
with $V = U^{1/2}r^2{\sin}^2\theta\,\sin\phi$ and $U=Z_1 Z_5$.
The second and third equations are redundant and imply
\begin{align}
  \cF_{tx} &= \frac{d}{dx}(r\cos\theta) \,,
\end{align}
while the fourth implies
\begin{align}
  \cF_{\phi\chi} (r\sin\theta)' = r^2{\sin}^2\theta\,\sin\phi \,.
\end{align}
Taking $\cF_{\phi\chi}$ to be $\SU(2)$-invariant and quantized,
\begin{align}
  \cF_{\phi\chi} &= \pi\alpha' q\, \sin\phi \,,
  \qquad\qquad
  q\in\Z \,,
\end{align}
gives
\begin{align}
  \frac{d}{dx}(r \sin\theta) &= \frac{(r\sin\theta)^2}{\pi\alpha' q} \,.
\label{eq:y'(x)}
\end{align}
This relation guarantees that the first equation, \eqref{eq: firstConstraint}, is automatically satisfied.
Choosing the interface to lie at $x=0$, it integrates to
\begin{align}
  x &= -\frac{\pi\alpha' q}{r\sin\theta} \,.
\end{align}

We now take the near-horizon limit, which corresponds to setting $U = g_s\alpha' N_5/r^2$.
In the limit the RR 2-form potential is
\begin{align}
  C^{(2)} &= A(r) dt\wedge dx + B(\theta) d\phi\wedge d\chi \,, &
  B(\theta) &= \theta - \sin\theta\,\cos\theta \,.
\end{align}
As a result, the quantization condition \eqref{eq:D3 quantization} associated to $\cF_{tx}$ takes the form
\begin{align}
  p = q \frac{Z_5}{g_s} \frac{d}{dx}(r\cos\theta) + \frac{N_5}{\pi} B(\theta) \,.
\label{eq:Ftx condition}
\end{align}
Solving \eqref{eq:y'(x)} and \eqref{eq:Ftx condition} for $\frac{dr}{d\theta}$ we can conclude that
\begin{align}
  \frac{d\log r}{d\theta} &= - \frac{\sin\theta + (\theta_p-\theta)\cos\theta}{(\theta_p-\theta)\sin\theta} \,,
\end{align}
with $\theta_p = \pi p/N_5$, from which we obtain
\begin{align}
  r(\theta) &= r_0 \frac{\theta_p - \theta}{\sin\theta} \,.
  \label{eq:r(theta)}
\end{align}
The $r$-$\theta$ relation \eqref{eq:r(theta)} coincides in fact with that appearing in \cite{Camino:2001at}, in spite of the non-trivial $x$ profile appearing in our case.
This, combined with \eqref{eq:y'(x)}, gives the full brane solution for the interface RG flow.

\subsubsection*{IR brane configuration}

We are also interesting in the flow endpoint of the D3-brane in Janus coordinates.
To make connection with the $(p,q)$-string computation given above, we give this computation in the F1/NS5 frame.
In this frame the D3-brane action gets contributions only from the DBI piece:
\begin{align}
  L_\text{D3} &= - T_3 e^{-\hat\phi} \sqrt{-\det(\hat g+\hat B-2\pi\alpha'F)} \,.
\end{align}
We write for convenience
\begin{align}
  F &= \frac{N_5}{2\pi}( \psi_q\,\omega_{\ads_2} + \theta_p\,\omega_{\S^2} ) \,.
\end{align}
In Janus coordinates the fixed point lies at constant $\psi$ and $\theta$, so we may drop the kinetic contributions.
We thus need to extremize
\begin{align}
  L_\text{D3} &= \frac{N_5^2}{8\pi^3} e^{-\hat\phi} \sqrt{({\cosh}^4\psi - \cE^2)({\sin}^4\theta - \cB^2)} \sqrt{-g_{\ads_2}} \sqrt{g_{\S^2}}
\end{align}
with
\begin{align}
  \cE &= \psi_q - \psi - \tfrac{1}{2}\sinh 2\psi \,, &
  \cB &= \theta_p - \theta + \tfrac{1}{2}\sin 2\theta \,,
\end{align}
which is accomplished with $\psi=\psi_q$ and $\theta=\theta_p$.
The D1-brane charge $p$ and F1-charge $q$ (we retain the labelling from the D1/D5 frame) are given by
\begin{align}
  p &= \int_{\S^2} F_{\phi\chi} = \frac{N_5\theta_p}{\pi} , &
  q &= -\int_{\S^2} \frac{\p L}{\p F_{tz}}
  = \bigl( p\, e^{-\hat\dil}{\textstyle \frac{\sin\theta_p}{\theta_p} } \bigr) \sinh\psi_q  ,
\end{align}
in terms of which the on-shell Lagrangian takes the form
\begin{align}
  L_\text{D3}^\text{o.s.} &= - \frac{N_5}{8\pi^2} \sqrt{q^2 + \bigl( p\, e^{-\hat\dil}{\textstyle \frac{\sin\theta_p}{\theta_p} } \bigr)^2 } \sqrt{-g_{\ads_2}} \sqrt{g_{\S^2}} .
  \label{eq:D3-on-shell}
\end{align}

\section{Backreacted supergravity dual of the interface fixed points}
\label{sec:sugra}
We have seen that, in the probe brane approximation, the configuration dual to our interface flow is described by a F1/D1 bound state that puffs up inside $S^3$, growing from a point into the $S^2$ located at polar angle $\theta_p = \pi p/N_5$.
While the probe brane computation is sufficient to determine the leading interaction between the interface and the CFT, it does not capture the effect of the interface on CFT observables.
These contributions are instead encoded in the backreaction of the geometry in response to the probe branes.
Thus, we turn our attention to the study of backreacted supergravity solutions dual to the UV and IR interfaces of previous sections.
Some further applications of backreacted supergravity solutions include the study of interactions between interfaces and the computation of the interface's reflection and transmission coefficients, although we leave these matters for future work.

The goal of this section is to write down the fully backreacted configurations dual to the RG fixed points.
Our analysis relies on the work of \cite{ChiodaroliOriginal, ChiodaroliJunctions}, which provides a general class of asymptotically $\ads_3\times \S^3\times M_4$ BPS solutions to type IIB supergravity, where $M_4$ is as usual $T^4$ or $K_3$.
These solutions are foliated by $\ads_2\times \S^2$ slices, and thus possess $SO(2,1)\times SO(3)$ symmetry in addition to eight supersymmetries -- all appropriate for our interfaces.

We begin with a review of the solutions of \cite{ChiodaroliOriginal} in \secref{sec: SugraReview}, and our modifications to them:
in order to write down the backreacted duals of our interface fixed points, we must relax the regularity conditions imposed in \cite{ChiodaroliOriginal}.
We determine in \secref{sec: SugraDefectSolutions} the configurations dual to both the UV and IR interfaces of section \secref{sec:interfaces}.
Finally, in \secref{sec: chargeMatching} we identify these interface solutions as fixed points of our Kondo-like RG flow, confirm the value of the polar angle, and discuss the interesting case of ``critical screening'' that occurs at $\theta_p=\pi$.

\subsection{Supergravity duals of conformal interfaces in \texorpdfstring{CFT$_2$}{CFT2}}
\label{sec: SugraReview}

As in \cite{ChiodaroliOriginal, ChiodaroliJunctions}, we consider solutions with the symmetries mentioned above, together with the constraints that all fields are constant on $T^4$, and that all moduli of $T^4$ are constant excepting the volume.
These constraints guarantee that the geometry dual to the conformal interfaces takes the form
\begin{equation} \label{Ansatz}
 ds_{10}^2 =
 f_1^2ds_{\ads_2}^2+f_2^2ds_{\S^2}^2+f_3^2ds_{T^4}^2+\rho^2\, dz\,d\bar{z} \,,
\end{equation}
where the $\ads_2$ and $\S^2$ metrics have unit radius.
We take $T^4$ to have volume $(2\pi)^4 \alpha'{}^2$, and where $z$ is a holomophic coordinate on a Riemann surface with boundary, $\Sigma$.
The assumptions listed imply that all fields depend only on $(z,\bar{z})$.
The complex coordinate $z$ should not be confused with the $\ads$ radial coordinate of previous sections.

One of the main results of \cite{ChiodaroliOriginal} was to show that the general local solution to the Killing spinor equations satisfying the above constraints is determined by four holomorphic functions: $\holA(z)$, $\holB(z)$, $\holU(z)$, and $\holV(z)$.
The solutions themselves are most easily written using the eight harmonic functions
\begin{align}
  \label{harmonics}
  \hola &= \holA+\bar{\holA}, &
  \holb &= \holB+\bar{\holB}, &
  \holu &= \holU+\bar{\holU}, &
  \holv &=\holV+\bar{\holV}, \\
  \label{dualHarmonics}
  \tilde{\hola} &= -i(\holA-\bar{\holA}), &
  \tilde{\holb} &= -i(\holB-\bar{\holB}), &
  \tilde{\holu} &= -i(\holU-\bar{\holU}), &
  \tilde{\holv} &= -i(\holV-\bar{\holV}).
\end{align}
In terms of these, the metric becomes
\begin{subequations} \label{metricFactors}
\begin{align}
 f_1^2&=\frac{e^{\dil}}{2f_3^2}\frac{|\holv|}{\holu}(\hola\,\holu+\tilde{\holb}^2),\label{metricFactorAdS2}\\
 f_2^2&=\frac{e^{\dil}}{2f_3^2}\frac{|\holv|}{\holu}(\hola\,\holu-\holb^2),\label{metricFactorS2}\\
 f_3^4&=e^{-\dil}\frac{\holu}{\hola},\label{metricFactorT4}\\
 \rho^4&=4e^{-\dil}\biggl|\frac{\p_z\holv}{\holB}\biggr|^4\frac{\hola\,\holu}{\holv^2}
 \label{metricFactorSigma}
 \,,
\end{align}
\end{subequations}
while the dilaton, axion $\chi$, and RR four-form%
\footnote{\cite{ChiodaroliOriginal} uses supergravity conventions in which the RR four-form is $1/4$ of its value in the most common convention.
In order to rescale to conventional string theory conventions, we express their function $\textrm{Hol}(\hat{h})=\holU/4$ and shift $B\rightarrow\holB/2$.
Since we use capital letters for meromorphic functions, we depart from their notation by writing $\holV=\textrm{Hol}(H)$.
\label{footnote: conventions}}
are
\begin{subequations}\label{SugraFields}
\begin{align}
 e^{-2\dil}&=\frac{1}{4\holu^2}(\hola\,\holu-\holb^2)(\hola\,\holu+\tilde{\holb}^2),\label{DilatonFormula}\\
 \chi&=\frac{1}{2\holu}(\holb\,\tilde{\holb}-\tilde{\hola}\,\holu),\label{AxionFormula}\\
 C_K&=\frac{1}{2\hola}(\holb\,\tilde{\holb}-\hola\,\tilde{\holu})\label{FourFormFormula}.
\end{align}
\end{subequations}
Here, $C_K$ denotes the coefficient of $\omega_{T^4}$ in $C^{(4)}$.
The ansatz of \cite{ChiodaroliOriginal} allows for one more component along $\ads_2\times S^2$,
$C^{(4)}
= C_K \, \omega_{T^4}
+ C_{\ads_2\times\S^2} \, \omega_{\ads_2} \wedge \omega_{\S^2}$,
which is determined by $C_K$ via the self-duality of $F_{(5)}$.
Each $\omega$ denotes the volume form induced by the corresponding ``unit'' fiber metric discussed above.

To match dual geometries to RG flow endpoints, it will suffice to match the charges carried by appropriate singularities in the $\Sigma$ plane.
It therefore behooves us to use the appropriate notion of charge.%
\footnote{A useful introduction to all notions of charge occuring in supergravity is presented in \cite{Marolf:2000cb}.}
For our purposes this is the Page charge, which is conserved, localized and quantized.
The first two properties enable us to identify the charges localized at a point on $\Sigma$, while the third allows us to match these charges with the quantum numbers defining the CFT and interface.
Note that the Page charges are not guaranteed to be invariant under large gauge transformations, which will be of some relevance to us later.

The Page charges differ from the commonly used Maxwell charges.
For instance, the D1-brane Page charge reads
\begin{equation}\label{PageChargeExample}
 \Pone=-\int_{\Sigma_7}\biggl( e^\dil\star\bigl(dC_{(2)}-\chi H_{(3)}\bigr)  -  C_{(4)}\wedge H_{(3)}\biggr) ,
\end{equation}
where $\Sigma_7$ is a 7-manifold enclosing the charge (but no other sources).
Due to the background ansatz \eqref{Ansatz}, the 3-forms take the form
\begin{subequations}
 \begin{align}
  H_{(3)}&=dB_{(2)}=(\p_a b^{(1)})d\zeta^a \wedge\omega_{\ads_2}
    + (\p_a b^{(2)}) d\zeta^a\wedge\omega_{\S^2} \\
  F_{(3)}&=dC_{(2)}=(\p_a c^{(1)})d\zeta^a\wedge\omega_{\ads_2}
    + (\p_a c^{(2)}) d\zeta^a \wedge\omega_{\S^2} \,,
 \end{align}
\end{subequations}
with $\zeta^a = (z,\bar z)$.
The expressions for $b^{(i)}$ and $c^{(i)}$ in terms of the holomorphic functions may be found in equation \eqref{2form} of the appendix.

The form of the Page charges used in this paper were derived in detail in the appendix of \cite{ChiodaroliJunctions};
here we only collect the final expressions.
The 1-brane Page charges are
\begin{subequations}\label{PageOneCharges}
\begin{align}
 \Pone=4\pi\biggl[&\int_\cC\frac{u}{a}\frac{a\,u-b^2}{a\,u+\tilde{b}^2}\,i(\p_zc^{(1)}-\chi\p_zb^{(1)})dz
 +\int_\cC C_K\p_zb^{(2)}dz\biggr]+ \cc \label{QD1}\\
 \Qone=4\pi\biggl[&\int_\cC\frac{(a\,u-b^2)^2}{4a\,u}i\p_zb^{(1)}dz-\int C_K\p_zc^{(2)}dz\notag\\
 -&\int_\cC\frac{u}{a}\frac{a\,u-b^2}{a\,u+\tilde{b}^2}\,\chi\, i(\p_zc^{(1)}-\chi\p_zb^{(1)})dz\biggr] + \cc \label{QF1}
\end{align}
\end{subequations}
The integration contour $\cC$ is a semicirle ending on the boundary $\Im{z}=0$ of $\Sigma$, and stems from integrating \eqref{PageChargeExample} over the 7-manifold given by the $T^4\times\S^2$ fibration of over $\cC$.
The Page 5-brane charges are%
\footnote{While the 1-brane charges are electric charges, the 5-brane charges are magnetic.}
\begin{subequations}\label{PageFiveCharges}
\begin{align}
 \Qfive &= 4\pi\int_\cC dz\,\p_zb^{(2)} + \cc \label{QF5}\\
 \Pfive &= 4\pi\int_\cC dz\,\p_zc^{(2)} + \cc \label{QD5}.
\end{align}
\end{subequations}
The contour $\cC$ is again a semicircle ending on $\p\Sigma$ and enclosing a pole on $\p\Sigma$.
Including the $\S^2$ fiber over $\cC$ yields an $\S^3$ threading the 5-brane.
The D3-brane Page charge is given analogously by
\begin{equation}\label{QD3}
  \Pthree = \oint_\cC \p_zC_K\,dz + \cc \,,
\end{equation}
where now the integration 5-manifold is $\cC$ together with the $T^4$ fiber.
In contrast to the previous charges, this contour encloses a point in the interior of $\Sigma$.

This concludes our summary of local $\frac{1}{2}$-BPS solutions of the form \eqref{Ansatz}.
When discussing concrete solutions below it is advantageous to work in terms of the charges, \eqref{PageFiveCharges} and \eqref{PageOneCharges}, evaluated at the asymptotic regions; see \appref{app: asymptoticRegions} for details.
Here we only remark that it is convenient to use the rescaled charges
\begin{equation}
 \pfive=\Pfive/(8\pi^2),  \quad  \qfive=\Qfive/(8\pi^2),  \quad  \pone=\Pone/(8\pi^2),  \quad  \qone=\Qone/(8\pi^2),
\end{equation}
which are found in \eqref{asymptoticCharges}.

\subsubsection{Regularity Constraints}
\label{sec:Regularity Constraints}
To go from the local solutions reviewed above to a global geometry requires imposing regularity conditions, which was done for non-singular bulk geometries in \cite{ChiodaroliOriginal}.
Unfortunately, to include brane backreactions requires allowing certain mild singularities, requiring us to relax these constraints slightly.

The holomorphic functions $\holA,\,\holB,\,\holU,\,\holV$ were assumed in \cite{ChiodaroliOriginal} only to have poles of order one, and we adopt this restriction, with the proviso that we allow certain logarithmic singularities.
We further impose the following conditions:
\begin{itemize}
 \item The $\ads_2$ metric factor $f_1$ is finite and non-zero everywhere except at most at isolated singular points.
 A pole corresponds to an asymptotic $\ads_3 \times \S^3 \times T^4$ region.
 \item The $\S^2$ metric factor $f_2$ is finite in the interior of $\Sigma$ and vanishes on its boundary, except at most at isolated singularities.
 \item The metric factor $f_3$ of the $T^4$ and the dilaton are finite and non-zero \textit{except at isolated points} in $\Sigma$.
\end{itemize}
The first two assumptions are the same as in \cite{ChiodaroliOriginal}.
The third assumption, however, differs: \cite{ChiodaroliOriginal} required both $f_3$ and $e^\dil$ to be everywhere finite and non-vanishing, a condition which excludes the brane solutions of interest to us.
Our modifications and their consequences are discussed in detail in what follows.

\medskip

We now review three important consequences of the above constraints discussed in \cite{ChiodaroliOriginal}, and specialize them to our particular case.

\subsubsection*{Functions vanishing on the boundary}
The two requirements $\res{f_2}{\p\Sigma}=0$ and $\res{f_1}{\p\Sigma}\neq0$ imply that
\begin{equation} \label{harmonicBoundaryVanishing}
 \res{\hola}{\p\Sigma} = \res{\holb}{\p\Sigma} = \res{\holu}{\p\Sigma} = \res{\holv}{\p\Sigma} = 0.
\end{equation}

The solutions we are interested in are obtained by modifying the vacuum solution, which is discussed in \secref{sec: SugraVacuum}.
Equation \eqref{harmonicBoundaryVanishing} represents an important constraint on possible modifications, and will be particularly important when discussing the D3 solution.

\subsubsection*{Spurious singularities}
It turns out to be possible for $\hola$ and $\holu$ to be singular without generating any singularities in the supergravity solution, provided certain constraints are satisfied.
We will call such singularities in description spurious singularities.
One constraints is that $f_3$ must be finite.
A look at \eqref{metricFactorT4} reveals that any singularities in $\hola$ and $\holu$ must be of the same order, and \eqref{metricFactorS2} then implies that $\holb$ must be singular, as well.
A second constraint is that $f_2$ must be everywhere non-negative,%
\footnote{For all metric factors \eqref{metricFactors} to be non-negative requires also that $a\geq0$ and $u\geq0$.}
which implies that
\begin{equation}\label{f2nonneg}
 \hola\,\holu-\holb^2 \geq 0 \,.
\end{equation}
Let $z_* \in \Sigma$ be the location of a spurious singularity.
Expanding
\begin{align}
 \holA(z)=i\frac{\hola_*}{z-z_*}+\cO(1),\qquad
 \holB(z)=i\frac{\holb_*}{z-z_*}+\cO(1),\qquad
 \holU(z)=i\frac{\holu_*}{z-z_*}+\cO(1),
\end{align}
we obtain the constraint
\begin{equation}\label{nonBranePoles}
 \hola_*\holu_* = \holb_*^2 .
\end{equation}
In the solutions of \cite{ChiodaroliOriginal}, which had finite, non-zero dilaton and $f_3$ everywhere, all the functions $\holA$, $\holB$, $\holU$ had singularities at the same points.
Our defect solutions, however, require additional singularities in $\holU$, and this singularity will not be shared by $\holA$ and $\holB$.
These singularities mark the defect loci and, while \eqref{f2nonneg} is still satisfied, equation \eqref{nonBranePoles} does not apply there.

\subsubsection*{Zeroes of $\holB$ and $\p_z\holV$}
The authors of \cite{ChiodaroliOriginal} considered solutions for which the curvature scalar
\begin{equation}
R_\Sigma=-2\frac{\p_z\p_{\bar{z}}\log\rho^2}{\rho^2}
\end{equation}
is non-singular everywhere on $\Sigma$, which forces $\p_z\holV$ and $\holB$ to have common zeroes.
In order to generate our defects below, we will modify only the functions $\holU(z)$ and $\bar{\holU}(\bar{z})$, which does not affect this relation between $\p_z\holV$ and $\holB$.

\subsubsection{Trivial interface: the D1/D5 geometry}\label{sec: SugraVacuum}
The simplest solution to these constraints is the trivial interface: the D1/D5 geometry.
We refer to this solution as the vacuum as it is dual to the vacuum state of the D1/D5 CFT.

We restrict ourselves in this paper to solutions with two asymptotic regions, which we place at $z=0$ and $z=\infty$ in $\Sigma$.
It can be readily checked using equation \eqref{asymptoticCharges} of the appendix that the choice
\begin{subequations}\label{vacFuns}
\begin{align}
 \holA(z)&=i\vaca\frac{z}{z^2-1},\qquad \hola=\frac{2\vaca\Im(z)}{|1-z^2|^2}(1+|z|^2)\\
 \holB(z)&=i\vacb\frac{z^2+1}{z^2-1},\qquad \holb=\frac{8\vacb\Im(z)\Re(z)}{|1-z^2|^2}\\
 \holU_0(z)&=i\vacu\frac{z}{z^2-1},\qquad \holu_0=\frac{2\vacu\Im(z)}{|1-z^2|^2}(1+|z|^2)\label{U0}\\
 \holV(z)&=i\vacv\biggl(\frac{1}{z}-z\biggr),\quad \holv=2\vacv\Im(z)(|z|^{-2}+1)
\end{align}
\end{subequations}
yields a geometry carrying only D1 and D5 charge, localized at the points $z=0$ and $z=\infty$.
This is the vacuum configuration.

The coefficients $\vaca,\,\vacb,\,\vacu,\,\vacv$ are chosen to be real.
The constraint \eqref{nonBranePoles} implies the relation  $4\vacb^2=\vaca\vacu$.
Due to the positivity of the metric scale factors \eqref{metricFactors}, $\vaca$ and $\vacu$ must not only have the same sign, but in fact both be positive.
We may also take $\vacv$ to be positive.
Without loss of generality we also choose $\vacb>0$, rendering charges at $z=0$ positive, and those at $z\rightarrow\infty$ negative.
The subscript of the meromorphic function $\holU_0$ anticipates the fact that we will later modify it from its vacuum value.

It is useful to express the three parameters $(\vaca,\vacu,\vacv)$ in terms of the asymptotic charges and the dilaton,
\begin{equation}\label{vacSolution}
\vacv=\frac{1}{2}\sqrt{\pfive\,\pone},\qquad
\vaca=2e^{-\dil(0)},\qquad
\vacu=2\frac{\pone}{\pfive}e^{-\dil(0)}.
\end{equation}
Both asymptotic regions have the same dilaton, while their asymptotic charges differ only in sign; see \appref{app: SugraVacuum} for details.

Given the functions \eqref{vacFuns}, the metric factors \eqref{metricFactors} realize $\AdS_3\times \S^3\times T^4$:
in coordinates $z=\exp(\psi+i\theta)$, the Einstein frame metric derived from \eqref{vacSolution} reads
\begin{equation}\label{D1D5vacuumMetric}
 ds_{10}^2=\adsL^2\biggl(d\psi^2+\cosh^2\psi\, ds_{\AdS_2}^2+d\theta^2+\sin^2\theta ds^2_{\S^2}\biggr)+\sqrt{\frac{\pone}{\pfive}e^{-\dil(0)}}ds^2_{T^4} ,
\end{equation}
and has $\AdS$ radius $\adsL^2=2|\pfive|e^{\dil(0)/2}$.
By comparison of \eqref{D1D5vacuumMetric} with the D1/D5 Einstein frame solution \eqref{eq:D1/D5 dual}, we can relate the Page charges to the integer charges $N_1$ and $N_5$ as follows:
\begin{equation}\label{D1D5Normalization}
 \Pfive=8\pi^2\pfive=(2\pi)^2\alpha' N_5 ,\qquad \Pone=8\pi^2\pone=(2\pi)^6\alpha'^3 N_1 ,
\end{equation}
and similarly for $\Qfive$ and $\Qone$.
Plugging this into the central charge \eqref{centralCharge} and using $4\pi\kappa_{10}^2=(2\pi)^8\alpha'^4$, we recover the well-known value $\sfc=6N_1N_5$.

\subsection{Backreacted defect solutions}
\label{sec: SugraDefectSolutions}
We now turn to our primary interest: modifying of the vacuum solution by the addition of new singularities to produce interface solutions.
We discuss F1/D1 interfaces in \secref{sec: F1defect} and D3 interfaces in \secref{sec: D3defect}.
Both families of solutions are found by
\begin{enumerate*}
  \item imposing the appropriate brane charges at each asymptotic boundary, and
  \item disallowing gravitational singularities except at the brane locus, where their charges and order of divergence must match that of the corresponding flat space solution.
\end{enumerate*}
We finally make contact with the probe brane RG flows in \secref{sec: chargeMatching} by identifying the particular solutions dual to either end of the RG flows.

Before we begin, let us make some general remarks for all interface solutions in this paper.
Due to the presence of interfaces, the asymptotic regions will generally have different D1-brane charges.
We distinguish these two D1-brane charges by superscripts $(0)$ and $(\infty)$ indicating their asymptotic regions.
Charge conservation then implies the relations
\begin{subequations}\label{DefectCharges}
\begin{align}
 \qone^{\cD} &= |\qone^{(\infty)}|-\qone^{(0)} \\
 \pone^{\cD} &= |\pone^{(\infty)}|-\pone^{(0)} .
\end{align}
\end{subequations}
We anticipate that our defect solutions all have $\qone^{(0)} = 0$ so that the asymptotic region at $z\rightarrow\infty$ mirrors the defect's F1 charge, $\pone^\cD=|\pone^{(\infty)}|$.
We also find it convenient to define
\begin{align}\label{D1mean}
 \ol{{\pone}}&\equiv\frac{|\pone^{(\infty)}|+\pone^{(0)}}{2} ,
\end{align}
as most of our results are simplest when written in terms of $\pone^{\cD}$ and $\ol{\pone}$.
As the interfaces we consider carry no D5 charge, we will not dress $\pfive$ with superscripts in the main text.

For later reference we introduce
\begin{equation}\label{bareD1Charge}
  \pbare = \frac{\vacv\vacu}{\vacb}.
\end{equation}
When there is no defect, \eqref{bareD1Charge} is the D1 charge (cf. \eqref{vacD1Charge} in the appendix), while in the presence of an interface it is only a notational convenience and does not represent a Page charge.
Even though we have used the same greek letters in \eqref{bareD1Charge} as in \eqref{vacFuns}, we emphasize that they will not take the same values as in \eqref{vacSolution}.

\subsubsection{F1/D1 interface (UV)}\label{sec: F1defect}

We begin by inserting into the D1/D5 geometry fundamental strings smeared over the $T^4$ directions.%
\footnote{In the true supergravity solution, the fundamental strings would be localized on $T^4$, but the ansatz of \cite{ChiodaroliOriginal} is not general enough to accomodate this situation. Nonetheless, in the weakly-coupled regime where $T^4$ is on the string scale while the $S^3$ is large, the true results will not differ appreciatively from those arising from this assumption.}
This is done by augmenting the functions \eqref{vacFuns} by terms carrying only the local charges and singularities sourced by the appropriate probe branes.

In order to embed a fundamental string (or F1-brane) in the D1/D5 background, one has to introduce a monodromy in the integrand of \eqref{QF1} at the boundary of $\Sigma$.
Due to the complicated form of the F1 charge it is simpler to S-dualize and then perform four T-dualities along the $T^4$ directions in order to turn the string into a D5-brane.
It is clear from \eqref{QD5} that one has to introduce a pole in $\holA$ at some point $\defx\in\p\Sigma$ in order to create D5 charge.
For positive values of $\defx$ the defect lies at the north pole of $\S^3$, while negative $\defx$ localizes it at the south pole (in our coordinate system $z=re^{i\theta}$).
We may choose $\defx>0$ without any particular loss of generality.

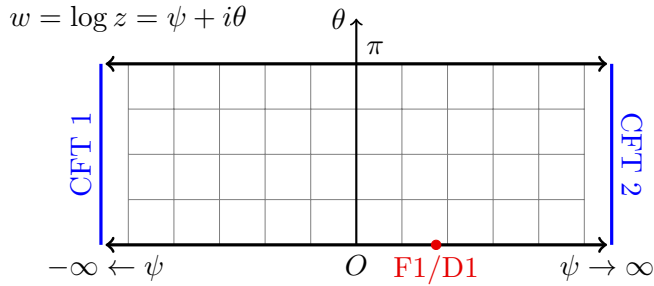
\begin{figure}[t]
 \tikzstyle{D0}=[circle, draw=red!90!black, fill=red!90!black, minimum size=3.5pt, inner sep=0pt]
  \centering
   \begin{tikzpicture}[scale=.6]
 \draw[step=1cm, gray, very thin] (-5,0) grid (5,4);
 \draw[very thick, <->] (-5.5,0) -- (5.5,0); 
 \draw[thick, ->] (0,0) -- (0,5);
 \draw[very thick,<->] (-5.5,4) -- (5.5,4);
 \node[anchor=south west] at (0,4) {$\pi$};
 \node[left] at (0,5){$\theta$};
 \node[anchor=north] at (5.5,0) {$\psi\rightarrow\infty$};
 \node[anchor=north] at (-5.5,0) { $-\infty\leftarrow\psi$};
 \node[below] (0.0) {$O$};
 \node at (-5,5){$w=\log z=\psi+i\theta$};
 \draw[very thick, blue] (-5.6,0) -- (-5.6,4);
 \node [rotate=90,anchor=south, blue] at (-5.6,2){CFT 1};
 \draw[very thick, blue] (5.6,0) -- (5.6,4);
 \node [rotate=270,anchor=south, blue] at (5.6,2){CFT 2};
 \node[D0] (D1) at (1.75,0){};
 \node[color=red!90!black, below] at (D1){ F1/D1};
\end{tikzpicture}
\caption[F1/D1 interface on open Riemann surface $\Sigma$ in strip coordinates]{In strip coordinates $w=\psi+i\theta$ the two asymptotic regions, depicted by blue bars, lie at $\psi\rightarrow\pm\infty$. Each harbors a CFT which differ due to the presence of the F1/D1 defect located at $\log \defx=\defxi\in\p\Sigma$ depicted as red dot. The lower boundary, $\theta=0$, corresponds to the north pole of the $\S^3$, while the upper boundary, $\theta=\pi$, corresponds to the southpole. }
\label{fig: F1defectSigma}
 \end{figure}

Now we trace back what this implies for our F1 interface by reversing these duality transformations step by step using the results of \cite{ChiodaroliOriginal}.
The four T-dualities are realized by the exchange $\holA(z)\leftrightarrow \holU(z)$, so that a D1 charge is introduced in place of the D5 charge provided we add a pole, not to $\holA$, but to $\holU$.
S-duality acts on the meromorphic functions by
\begin{equation}
\holA\rightarrow\frac{1}{\holA},\qquad \holB\rightarrow i\frac{\holB}{\holA},\qquad \holU\rightarrow \holU-\frac{\holB^2}{\holA}\,.
\end{equation}
We see from this transformation that the new pole in $U$ is preserved by S-duality.
Therefore, we choose to modify $U_0$ in \eqref{vacFuns} by
\begin{equation}\label{UF1}
 \holU(z)=\holU_0+\delta U^{F1/D1}, \qquad \delta \holU^{F1/D1}=i\defc\,\frac{z}{z-\defx},\qquad \delta \holu^{F1/D1}=\frac{2\defc\,\defx\,\Im(z)}{|z-\defx|^2} \,.
\end{equation}
This is depicted in strip coordinates $w=\log z=\psi+i\theta$ (with $\defx=\exp{\defxi}$) in \figref{fig: F1defectSigma}.
The real number $\defc$ will be evaluated in terms of the interface's F1 charge below.
As will become evident below, this modification gives rise not to a pure F1-string, but rather to a $(p,q)$ string -- hence the label F1/D1 -- with the interface's D1 charge set by the value of $\defx$.
The particular value $\defx=1$ results in a pure F1-string interface.

While this solution carries F1 charge, we must ask whether it has the appropriate singularities to be the backreacted F1 geometry.
We demonstrate that this is indeed so in the case $\defx=1$ ($\defxi=0$ and no D1 charge).
Set $z=1+i\epsilon\in\Sigma$, which is a radial distance $\epsilon$ from the brane, and plug $\holA,\,\holB,\,\holV$ of \eqref{vacFuns} and $U$ of \eqref{UF1} into \eqref{metricFactorAdS2}-\eqref{metricFactorSigma}.
We can approach the brane by sending $\epsilon$ to zero.
This results in the leading behavior
\begin{equation}\label{F1DefectAsymptotics}
f_1^2\simeq \epsilon^{3/2}, \qquad f_2^2\simeq \epsilon^{3/2},\qquad f_3^2\simeq \epsilon^{-1/2},\qquad e^\dil\simeq \epsilon.
\end{equation}
This singular behavior is the same as that of smeared fundamental strings in a flat background, confirming our claim.%
\footnote{There are a couple of simple ways to obtain the geometry produced by smeared F1 strings in a flat background. One is by taking the number of NS5 branes in the familiar F1/NS5 solution to vanish. Another is by inserting the harmonic function of the NS5 solution into the F1 solution. In both cases the near-brane behavior takes the shape \eqref{F1DefectAsymptotics}.}

\begin{figure}[t]
  \tikzstyle{D0}=[circle, draw=red!90!black, fill=red!90!black, minimum size=3.5pt, inner sep=0pt]
  \centering
  \begin{tikzpicture}[scale=1.2]
 \def\R{2 };
 \coordinate (O) at (0,0);
 \def\m{.5 };
 \def\n{.5 };
 \coordinate [label=above:\large $\ads_3$] (U) at (\m,\n);
 \def\RA{315 };
 \def\LA{210}
 \coordinate (A) at (\R,0);
 \coordinate (D) at (-\R,0);
 \coordinate (B) at ($(U)+(A)$);
 \coordinate (C) at ($(U)-(A)$);
 \filldraw[fill=blue!10!white, draw=blue] (A) -- (B) -- (C) -- (D) -- (A);
 \draw[violet!90!white, very thick] (O) -- (U);
 \coordinate [label=below:D1/F1] (E) at ($(O)+(\RA:\R)$);
 \coordinate (F) at ($(E)+(U)$);
 \coordinate (G) at (intersection of U--F  and A--D);
 \shadedraw[top color=red!40!white, bottom color=red!50!black] (O) -- (E) -- (F) -- (G) -- (O);
 \draw[dashed] (U) -- (G);
 \coordinate [label=below:$AdS_2$] (H) at ($(O)+(\LA:\R)$);
 \coordinate (I) at ($(H)+(U)$);
 \coordinate (J) at (intersection of U--I  and A--D);
 \shadedraw[top color=blue!10!white, bottom color=blue!40!black] (O) -- (H) -- (I) -- (J) -- (O);
 \draw[dashed] (U) -- (J);
 \draw (O) -- (H) node (rhoEnd) [midway, right] {};
 \draw (O) -- (E) node (rhoBegin) [midway, left] {};
 \draw[->, very thick] (rhoBegin) to [bend left=45] node (rhoMid) [midway, below]{\large $\psi$} (rhoEnd);
 \coordinate (K) at ($(D)- .45*(\R,0)$);
 \coordinate (L) at ($(C)- .45*(\R,0)$);
 \coordinate (M) at ($(D)- .15*(\R,0)$);
 \draw[->] (K) to node [midway, left]{$t$} (L);
 \draw[->] (K) to node [midway, below]{$x$} (M);

\def\Radius{.65*\R };
\def\BraneAngle{35 };
 \coordinate (BallCenter) at ($2.8*(A)$);
 \shade[ball color = blue!40, opacity = 0.4] (BallCenter) circle (\Radius);
 \draw (BallCenter) circle (\Radius);
 \coordinate[red, ultra thick] (North) at ($(BallCenter)+(0,\Radius)$);
 \node[D0] at (North){};
 \node[above] at (North){\large $\S^3$};
 \coordinate (Equator) at ($(BallCenter)+(\Radius,0)$);
 \draw (Equator) arc (0:-180:\Radius and 0.45);
 \draw[dashed] (Equator) arc (0:180:\Radius and 0.45);
 \coordinate (D2ArcBegin) at ($(BallCenter)+(\BraneAngle:\Radius)$);
 \draw[dashed] (BallCenter) --  (North) node (AngleBegin) [midway]{};
 \draw[dashed] (BallCenter) -- (D2ArcBegin)
			     node (AngleEnd) [midway]{};
 \draw[thick] (AngleBegin) to [bend left=45] node (AngleMid) [midway]{} (AngleEnd);
 \node (theta) at ($(BallCenter)!0.5!(AngleMid)$) {\large $\theta$};
\end{tikzpicture}
\caption{In coordinates $z=\exp(\psi+i\theta)$ $\AdS_3$ is foliated by $\AdS_2$ sheets shaded in dark blue and labelled by $\psi$.
	 A D1/F1 string, shaded in red, is embedded into $\AdS_3\times \S^3$ at $\defx=\exp\psi_\defx$, i.e. it sits at the north pole
	 of $\S^3$.
	 The boundary of $\AdS_3$, shaded in light blue, harbors the CFT and its intersection with the brane is the wordline of the
	 field theory defect, colored in violet.}
\label{fig:F1Defect}
\end{figure}

The geometry is depicted in \figref{fig:F1Defect}.
Since the defect lies on $\p\Sigma$, it sits at a pole of $\S^3$ and occupies one $\ads_2$ slice.
A pure F1-string corresponds in \figref{fig:F1Defect} to an interface, which intersects the CFT spacetime orthogonally (i.e. $\defx=1\Leftrightarrow\defxi=0$).

Since the new pole in \eqref{UF1} represents a genuine singularity, the constraint \eqref{nonBranePoles} does not apply at $\defx$.
Because the $\hola$ is positive, the requirement \eqref{f2nonneg} implies
\begin{equation}
\hola\,\holu_0 -\holb^2 + a \, \delta\holu^{F1/D1} \geq 0
\qquad\Rightarrow\qquad
\defc\,\defx \geq 0 .
\end{equation}
Since we are building on the vacuum solution \eqref{vacFuns}, the sum of the first two terms is positive by itself, cf. \eqref{vacuumf2nonneg}. Our choice $\defx>0$ then renders $\defc$ positive.

Our next step is to express all parameters of the solution through the available charges and the dilaton.
Let us note that through \eqref{UF1} we have added two new parameters, $\defc$ and $\defx$, to the system and obtained two new independent charges \eqref{DefectCharges}.
Our goal is then to solve the variables $(\vaca,\vacu,\vacv,\defc,\defx)$ for $(\dil(0),\pfive,\ol{\pone},\pone^\cD,\qone^\cD)$.
The ratio
\begin{align}
 \kappa = \frac{\ol{{\pone}}}{\pbare}.\label{F1SolutionD1mean}
\end{align}
quantifies the departure of the geometric quantity $\pbare$ from its vacuum value $\pone$,
and is useful for expressing the properties of our solutions.
For details the reader is referred to \appref{app:OneBraneDefect};
here we only present the solution:
\begin{subequations}\label{F1D1Solution}
\begin{align}
\vacv=\frac{1}{2}\sqrt{\frac{\pfive\,\ol{\pone}}{\kappa}},\quad
\vaca=2e^{-\dil(0)}&\sqrt{\kappa\,\frac{2\ol{\pone}-\pone^\cD}{2\ol{\pone}}},\quad
\vacu=2\sqrt{\frac{1}{\kappa}\frac{\ol{\pone}\,(2\ol{\pone}-\pone^\cD)}{2\pfive^2}}e^{-\dil(0)} ,
\label{F1D1SolutionAlphaEtaUpsilon}\\[.2cm]
\sinh\defxi&=e^{-\dil(0)}\sqrt{\kappa\,\frac{2\ol{\pone}-\pone^\cD}{2\ol{\pone}}}\frac{\pone^\cD}{\qone^\cD} ,
\label{F1D1SolutionXasCharges}\\[.25cm]
&\quad\defc=\frac{\qone^\cD}{\pfive} .
\label{F1D1SolutionC}
\end{align}
\end{subequations}
The proportionality factor $\kappa$ is best expressed in terms of the Einstein frame $(p,q)$-string tension,
\begin{equation}\label{tensionEinstein}
T_{(\qone,\pone)}=\frac{1}{2\pi\alpha'}\sqrt{e^{\dil(0)}\qone^2+e^{-\dil(0)}\pone^2}\,.
\end{equation}
Then we get
\begin{subequations}\label{kappa}
\begin{align}
\kappa = \kappa(\qone^\cD,\,\pone^\cD) &=
\frac{T^2_{\bigl(\qone^\cD,\, 4\sqrt{ \scriptscriptstyle{ \ol{\pone}\pone^{(0)} } }\bigr)}-T^2_{\bigl(\qone^\cD,\,0\bigr)}}
{\biggl(\sqrt{
  \sigma^2(\pone^\cD)
    + T^2_{\bigl(\qone^\cD,\,4\sqrt{\scriptscriptstyle{\ol{\pone}\pone^{(0)}}}\bigr)}} - T_{\bigl(\qone^\cD,\, 0\bigr)}\biggr)^2-\sigma^2\bigl(\pone^\cD\bigr)}\\[.3cm]
&\qquad\qquad\sigma(\pone^\cD)=\frac{T_{\bigl(0,\,4\pone^{(0)}\bigr)}}{\,T_{\bigl(\qone^\cD,\,0\bigr)}}T_{\bigl(0,\, \pone^\cD\bigl)}
\end{align}
\end{subequations}
The solution \eqref{F1D1Solution} exhibits a natural bound $|\pone^\cD|<2\ol{\pone}$ for the interface's D1 charge;
otherwise, the D1-brane charges have opposite signs on either side of the interface, and no BPS solution exists.
Note that $\kappa>0$ and that, due to \eqref{F1D1SolutionXasCharges}, we have $\sgn(\pone^\cD)=\sgn(\defxi)$.

\subsubsection*{Pure F1 defect, $\pone^\cD\rightarrow0$}

Let us now conider the pure fundamental string defect, which lives on the $\AdS_2$ sheet of smallest size
\begin{equation}
\sinh\defxi=0\quad\Leftrightarrow\quad \defx=1,
\end{equation}
In this case we have $\sigma(0)=0$, and since the D1 charges coincide at both asymptotic we drop the superscripts: $\pone\equiv\pone^{(0)}=-\pone^{(\infty)}=\ol{\pone}$.
The solution \eqref{F1D1Solution} then becomes
\begin{equation}\label{F1Solution}
  \vacv=\frac{1}{2}\sqrt{\frac{\pfive\,\pone}{\kappa_0}},\qquad
  \vaca=2\sqrt{\kappa_0}e^{-\dil(0)},\qquad
  \vacu=\frac{2}{\sqrt{\kappa_0}}\frac{\pone}{\pfive}e^{-\dil(0)},\qquad
  \defc=\frac{\qone^\cD}{\pfive},
\end{equation}
where $\kappa$ reduces to
\begin{equation}\label{kappaPureF1}
  \kappa_0\equiv\kappa(\qone^\cD,\,0)=
    \frac{T_{(\qone^\cD,\,4\pone)}+T_{(\qone^\cD,\,0)}}{T_{(\qone^\cD,\,4\pone)}-T_{(\qone^\cD,\,0)}} \,.
\end{equation}
In the limit $\qone^{\cD} \propto \defc\rightarrow 0$ the interface disappears, so that $\kappa_0 \rightarrow 1$ and the triple $(\vaca,\vacu,\vacv)$ reduces to the vacuum expressions \eqref{vacSolution}.
In the appendix \appref{app:OneBraneDefect} we also discuss the case $\qone^{\cD}\to 0$ with finite $\pone^\cD$.

\subsubsection{D3 interface (IR)}\label{sec: D3defect}

\begin{figure}
 \tikzstyle{D0}=[circle, draw=red!90!black, fill=red!90!black, minimum size=3.5pt, inner sep=0pt]
 \tikzstyle{Mirror}=[circle, draw=violet!90!black, fill=violet!90!black, minimum size=3.5pt, inner sep=0pt]
  \centering
  \begin{tikzpicture}[scale=.6]
 \draw[step=1cm, gray, very thin] (-5,0) grid (5,4);
 \draw[very thick, <->] (-5.5,0) -- (5.5,0); 
 \draw[thick, ->] (0,0) -- (0,5);
 \draw[very thick,<->] (-5.5,4) -- (5.5,4);
 \node[anchor=south west] at (0,4) { $\pi$};
 \node[left] at (0,5){$\theta$};
 \node[anchor=north] at (5.5,0) { $\psi\rightarrow\infty$};
 \node[anchor=north] at (-5.5,0) { $-\infty\leftarrow\psi$};
 \node[below] (0.0) { $O$};
 \node at (-5,5){ $w=\log z=\psi+i\theta$};
 \draw[very thick, blue] (-5.6,0) -- (-5.6,4);
 \node [rotate=90,anchor=south, blue] at (-5.6,2){ CFT 1};
 \draw[very thick, blue] (5.6,0) -- (5.6,4);
 \node [rotate=270,anchor=south, blue] at (5.6,2){ CFT 2};
 \node (D1) at (1.75,0){};
 \node[D0]	(D3)	at (3.2,1.8){};
 \node[Mirror]  (Mirror) at (3.2,-1.8){};
 \node[color=red!90!black, below] at (D3){ D3};
 \node[color=violet!90!black, above] at (Mirror){Mirror Charge};
\end{tikzpicture}
\caption[D3 interface in open Riemann surface $\Sigma$ in strip coordinates]{In contrast to the 1-brane interface of the previous section, the D3 brane interface is located in the interior of $\Sigma$ at $\log\defw=\defrho+i\Theta$, with $R=e^{\defrho}$. Hence the defect is no longer located at the poles of the $\S^3$, but wraps an $\S^2$ at some constant value $\Theta\in(0,\pi)$. It is depicted as a red dot. The corresponding mirror charge lies outside of $\Sigma$; here depicted as violet dot. }
\label{fig:D3DefectStrip}
\end{figure}
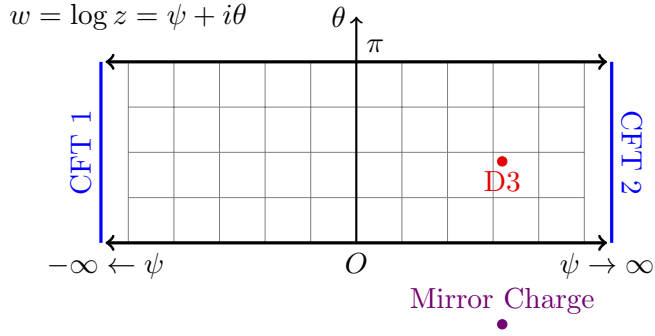

The other kind of interface relevant to us is dual to a D3-brane embedded in the D1/D5 background, whose probe brane description was given in \secref{sec:D3 probe brane}.
In terms of the description of \secref{sec: SugraReview}, a D3-brane is characterized by an additive monodromy in $C_K$ around some point $\defw\in\Sigma$.
As is evident from \eqref{FourFormFormula}, a monodromy in $C_K$ that does not affect any other charge should arise from $\tilde{\holu}$, since this harmonic function appears in no other charge or field.
The obvious way to get such a monodromy is to augment $U$ by a term of the form $\delta U\sim\log(z-\defw)$.
A logarithmic singularity can be made to respect \eqref{harmonicBoundaryVanishing} by employing the method of images, placing a mirror charge at $\bar{\defw}=R\,e^{-i\Theta}$ as depicted in \figref{fig:D3DefectStrip}.
It can be readily checked that the modification
\begin{align}\label{UD3}
	 \delta \holU^{D3}&=-\frac{\pthree^{\cD}}{2}\log\biggl(\frac{z/\defw-1}{z/\bar{\defw}-1}\biggr),\\[.25cm]
  \qquad \delta \holu^{D3}=-\frac{\pthree^{\cD}}{2}\log\biggl|\frac{z-\defw}{z-\bar{\defw}}\biggr|^2&,
  \qquad \delta \tilde{\holu}^{D3}=i\frac{\pthree^{\cD}}{2}\log\biggl[\biggl(\frac{\bar{\defw}}{\defw}\biggr)^2\frac{(z-\defw)(\bar{z}-\defw)}{(z-\bar{\defw})(\bar{z}-\bar{\defw})}\biggr]
\end{align}
produces
\begin{equation}\label{D3Normalization}
  \Pthree^\cD=\pi \,\pthree^\cD, \qquad \pthree^\cD=\frac{(2\pi)^4\alpha'^2}{\pi}\,N_3
\end{equation}
via \eqref{QD3}. The second equation introduces the integer valued D3-brane charge $N_3$. Since $w$ lies on the upper half plane $\Sigma$, we have $\Theta\in(0,\pi)$. Then the constraint $\hola\,\holu-\holb^2\geq0$ enforces positivity of $\pthree^\cD$. The geometry is depicted in \figref{fig:D3Defect} in coordinates $z=\exp(\psi+i\theta)$ with the fixed values $R=\exp\defrho$ and $\theta=\Theta$.

\begin{figure}[t]
  \tikzstyle{D0}=[circle, draw=red!90!black, fill=red!90!black, minimum size=3.5pt, inner sep=0pt]
  \centering
  \begin{tikzpicture}[scale=1.2]
 \def\R{2 };
 \coordinate (O) at (0,0);
 \def\m{.5 };
 \def\n{.5 };
 \coordinate [label=above:\large $\ads_3$] (U) at (\m,\n);
 \def\RA{315 };
 \def\LA{210}
 \coordinate (A) at (\R,0);
 \coordinate (D) at (-\R,0);
 \coordinate (B) at ($(U)+(A)$);
 \coordinate (C) at ($(U)-(A)$);
 \filldraw[fill=blue!20!white, draw=blue] (A) -- (B) -- (C) -- (D) -- (A);
 \draw[violet!90!white, very thick] (O) -- (U);
 \coordinate [label=below:D3] (E) at ($(O)+(\RA:\R)$);
 \coordinate (F) at ($(E)+(U)$);
 \coordinate (G) at (intersection of U--F  and A--D);
 \shadedraw[top color=red!40!white, bottom color=red!50!black] (O) -- (E) -- (F) -- (G) -- (O);
 \draw[dashed] (U) -- (G);
 \coordinate [label=below:$AdS_2$] (H) at ($(O)+(\LA:\R)$);
 \coordinate (I) at ($(H)+(U)$);
 \coordinate (J) at (intersection of U--I  and A--D);
 \shadedraw[top color=blue!20!white, bottom color=blue!40!black] (O) -- (H) -- (I) -- (J) -- (O);
 \draw[dashed] (U) -- (J);
 \draw (O) -- (H) node (rhoEnd) [midway, right] {};
 \draw (O) -- (E) node (rhoBegin) [midway, left] {};
 \draw[->, very thick] (rhoBegin) to [bend left=45] node (rhoMid) [midway, below]{\large $\psi$} (rhoEnd);
 \coordinate (K) at ($(D)- .45*(\R,0)$);
 \coordinate (L) at ($(C)- .45*(\R,0)$);
 \coordinate (M) at ($(D)- .15*(\R,0)$);
 \draw[->] (K) to node [midway, left]{$t$} (L);
 \draw[->] (K) to node [midway, below]{$x$} (M);

\def\Radius{.65*\R };
\def\BraneAngle{35 };
 \coordinate (BallCenter) at ($2.5*(A)$);
 \shade[ball color = blue!40, opacity = 0.4] (BallCenter) circle (\Radius);
 \draw (BallCenter) circle (\Radius);
 \coordinate (North) at ($(BallCenter)+(0,\Radius)$);
 \node[above] at (North){\large $\S^3$};
 \coordinate (Equator) at ($(BallCenter)+(\Radius,0)$);
 \draw (Equator) arc (0:-180:\Radius and 0.3);
 \draw[dashed] (Equator) arc (0:180:\Radius and 0.3);
 \coordinate (D2ArcBegin) at ($(BallCenter)+(\BraneAngle:\Radius)$);
 \draw[dashed, thick] (BallCenter) --  (North) node (AngleBegin) [midway]{};
 \draw[dashed, thick] (BallCenter) -- (D2ArcBegin)
                 node (D3anchor) [pos=1, right]{D3}
			     node (AngleEnd) [midway]{};
 \draw[thick] (AngleBegin) to [bend left=45] node (AngleMid) [midway]{} (AngleEnd);
 \node (theta) at ($(BallCenter)!0.5!(AngleMid)$) {\large $\Theta$};
 \draw[draw=red!75!black, thick] (D2ArcBegin) arc (0:-180:1.06 and 0.2);
 \draw[draw=red!75!black, dashed, thick] (D2ArcBegin) arc (0:180:1.06 and 0.2);
\end{tikzpicture}

\caption{In coordinates $z=\exp(\psi+i\theta)$ $\AdS_3$ is foliated by $\AdS_2$ sheets shaded in dark blue and labelled
	  by $\psi$. A D3 brane, shaded in red, is embedded into $\AdS_3\times \S^3$ at $\defw=\exp(\psi_R+i\Theta)$,
	  i.e. it wraps an $\S^2$ on the $\S^3$.
	 The boundary of $\AdS_3$, shaded in light blue, harbors the CFT and its intersection with the brane is the wordline of the
	 field theory defect, colored in violet.}
\label{fig:D3Defect}
\end{figure}

To go on, it is convenient to define an effective F1 charge
\begin{equation}\label{qTheta}
\qone^\Theta\equiv\qone^{\cD}\frac{\sin\Theta}{\Theta}
\end{equation}
and, similar to \eqref{F1SolutionD1mean}, define a ratio, which characterizes how much the D1 charge differs from the expression for the trivial interface \eqref{bareD1Charge},
\begin{align}
 \kappa^{(\Theta)}=\frac{\ol{\pone}}{\pbare}.\label{D3SolutionD1mean}
\end{align}
We may now express all parameters of the D3 solution in terms of the charges and the boundary dilaton (we leave the details to \appref{app:ThreeBraneDefect}),
\begin{subequations}\label{D3Solution}
\begin{align}
\vacv=\frac{1}{2}\sqrt{\frac{\pfive\,\ol{\pone}}{\kappa^{(\Theta)}}},\quad
\vaca=2e^{-\dil(0)}&\sqrt{\kappa^{(\Theta)}\,\frac{2\ol{\pone}-\pone^\cD}{2\ol{\pone}}}\quad
\vacu=2\sqrt{\frac{1}{\kappa^{(\Theta)}}\frac{\ol{\pone}\,(2\ol{\pone}-\pone^\cD)}{2\pfive^2}}e^{-\dil(0)},\label{D3SolutionAlphaEtaUpsilon}\\[.2cm]
\sinh\defrho=&e^{-\dil(0)}\sqrt{\kappa^{(\Theta)}\,\frac{2\ol{\pone}-\pone^\cD}{2\ol{\pone}}}\,\frac{\pone^\cD}{\qone^\Theta},\label{D3SolutionRasCharges}\\[.25cm]
&\quad\Theta=\frac{1}{\pfive}\frac{\qone^{\cD}}{\pthree^\cD}\label{thetaSolution}.
\end{align}
\end{subequations}
The proportionality factor $\kappa$ actually coincides with \eqref{kappa} upon making the replacement $\qone^\cD\rightarrow\qone^\Theta$:
\begin{subequations} \label{lambda}
\begin{align}
\kappa^{(\Theta)} \equiv \kappa(\qone^\Theta,\,\pone^\cD) =&
\frac{T^2_{\bigl(\qone^\Theta,\, 4\sqrt{ \scriptscriptstyle{ \ol{\pone}\pone^{(0)} } }\bigr)}-T^2_{\bigl(\qone^\Theta,\,0\bigr)}}
{\biggl(\sqrt{\sigma_\Theta^2\bigl(\pone^\cD\bigr)+T^2_{\bigl(\qone^\Theta,\, 4\sqrt{\scriptscriptstyle{\ol{\pone}\pone^{(0)}}}\bigr)}}-T_{\bigl(\qone^\Theta,\,0\bigr)}\biggr)^2-\sigma_\Theta^2\bigl(\pone^\cD\bigr)} \\[.3cm]
& \qquad\qquad
\sigma_\Theta(\pone^\cD)
= \frac{ T_{\bigl(0,\,4\pone^{(0)}\bigr)}} {\,T_{\bigl(\qone^\Theta,\, 0 \bigr)}} T_{\bigl(0,\,\pone^\cD\bigl)},
\end{align}
\end{subequations}
Clearly, $\kappa^{(\Theta=0)}=\kappa$ and $\kappa^{(\Theta)}>0$.
Analogously to before, \eqref{D3SolutionRasCharges} implies $\sgn(\pone^\cD)=\sgn(\defrho)$.
The system is described by the three defect charges $\pone^\cD$, $\qone^\cD$, and $\pthree^\cD$.
Instead of using $\pthree^\cD$ we will find it convenient to work in terms of the ratio $\qone^\cD/\pthree^\cD\propto\Theta$.

It is important in what follows that \eqref{thetaSolution}, which encodes the fundamental string charge per D3-brane, determines the value of the polar angle of the interface on the $\S^3$.
We will come back to this point in the next section when we discuss the RG flow.
Observe that the D3 solution reflects a natural bound on the dissolved fundamental string charge: when $\qone^\cD > \pi\pthree\pfive$, $\Theta>\pi$ and so $\defw\notin\Sigma$.
The interface's D1 charge is bounded as before by $|\pone^\cD|<2\ol{\pone}$;
note that it has no influence on the polar angle $\Theta$.

In the limit where the singularity's D3-brane charge is of the same order as its F1 charge, the brane approaches the boundary of $\Sigma$ ($\Theta\rightarrow0$), or equivalently $\qone^\Theta\rightarrow\qone^\cD$.
This implies that all expressions in \eqref{D3Solution} and \eqref{lambda} reduce to those of the 1-brane interface, \eqref{F1Solution} and \eqref{kappa} respectively.
Hence, as in the probe brane computation there is a regime where the D3 interface is realized as a 1-brane defect.
We will comment on this in the next section when making contact with the Kondo effect.
For future reference we study the limit of vanishing defect D1 charge. The limit of vanishing F1 charge is relegated to \appref{app:ThreeBraneDefect}.

\subsubsection*{Pure F1 defect, $\pone^\cD\rightarrow0$}
As with the 1-brane defect we drop all subscripts on the D1 charge, $\pone\equiv\pone^{(0)}=-\pone^{(\infty)}=\ol{\pone}$.
The solution then reduces to
\begin{subequations}\label{D3SolutionPureF1}
\begin{align}
\vacv=\frac{1}{2}\sqrt{\frac{\pfive\,\pone}{\kappa^{(\Theta)}_0}},\qquad
\vaca=2&\sqrt{\kappa^{(\Theta)}_0}e^{-\dil(0)},\qquad
\vacu=\frac{2}{\sqrt{\kappa^{(\Theta)}_0}}\frac{\pone}{\pfive}e^{-\dil(0)}\\
\Theta&=\frac{1}{\pfive}\frac{\qone^{\cD}}{\pthree^\cD}
\end{align}
\end{subequations}
with the considerable simplification,
\begin{equation}\label{lambdaPureF1}
\kappa^{(\Theta)}_0\equiv\kappa(\qone^\Theta,\,0)=
\frac{T_{(\qone^\Theta,\,4\pone)}+T_{(\qone^\Theta,\,0)}}{T_{(\qone^\Theta,\, 4\pone)}-T_{(\qone^\Theta,\,0)}}\,.
\end{equation}

\subsection{Matching to the RG flow's fixed points}\label{sec: chargeMatching}
So far we have described \onehalf-BPS asymptotically $\AdS_3\times\S^3\times T^4$ supergravity solutions containing D1/F1 and D3-branes dual to CFT interfaces.
In this section we identify particular solutions as endpoints of an RG flow, in which a D1/F1 interface flows in the IR to a D3 interface.

From the field theory point of view we are considering interface RG flows, meaning that the ambient CFTs themselves are unaltered by the flow.
As a result, the charges and supergravity fields characterizing the CFT should remain unchanged under the flow:
\begin{equation}
  Q_{brane}^{\IR} \overset{!}{=} Q^{\UV}_{brane},\qquad \dil^{\IR}\overset{!}{=}\dil^{\UV},
\end{equation}
where these expressions refer to the values in either asymptotic region.
The goal of this matching is to relate the individual parameters in the 1-brane solution \eqref{F1D1Solution} to those in the 3-brane solution \eqref{D3Solution}.

Let us start with the F1 charge.
Equating the value of $\qone^\cD$ in \eqref{F1D1SolutionC} and \eqref{thetaSolution} yields the relation
\begin{equation}
\pthree^{\cD}\Theta=\defc.
\end{equation}
Using this, it is readily verified that, for small $\Theta$, the 3-brane modification \eqref{UD3} reduces to the 1-brane modification \eqref{UF1}
provided we fix $\defx=R$, which means that the interface remains on the same $\AdS_2$ sheet throughout the flow.
This holds true only when $\Theta$ is small, or when the interface carries no D1 charge.
Recall that when $\pone^\cD=0$, the 1-brane and the 3-brane interface both occupy the $\AdS_2$ sheet of smallest size, $\defx=1$ ($\defxi=0$) and $R=1$ ($\defrho=0$).
This property does not hold true when the interface is stabilized by extra D1 charge on a non-minimal $\AdS_2$ slice.
Indeed, comparing \eqref{F1D1SolutionXasCharges} and \eqref{D3SolutionRasCharges} we conclude that the interface shifts to a new $\ads_2$ slice,
\begin{equation}\label{MatchingRandX}
\sinh\defrho=\sqrt{\frac{\kappa^{(\Theta)}}{\kappa}}\,\frac{\Theta}{\sin\Theta}\,\sinh\defxi \,,
\end{equation}
in agreement with the probe brane computation.
The expressions \eqref{kappa} and \eqref{lambda} together imply the bound
\begin{equation}\label{kappaLambdaComparison}
1\,\leq\,\frac{\kappa}{\kappa^{(\Theta)}}\,\leq\,\biggl(\frac{\Theta}{\sin\Theta}\biggr)^2 ,
\end{equation}
which is saturated at $\Theta=0$.
Plugged into \eqref{MatchingRandX}, this implies that under the RG flow the interface is pushed towards the boundary of $\AdS_3$, $|\defrho|\geq|\defxi|$ when $\pone^\cD\neq0$.
Their sign matches that of the interface's D1 charge, $\sgn(\defxi) = \sgn(\defrho) = \sgn(\pone^\cD)$.

Using \eqref{F1D1Solution} and \eqref{D3Solution}, we can express the triple $(\vaca^{IR},\vacu^{IR},\vacv^{IR})$ in terms of their UV values via
\begin{equation}
\vacv^{\IR}=\sqrt{\frac{\kappa}{\kappa^{(\Theta)}}}\,\vacv^{\UV},\qquad
\vaca^{\IR}=\sqrt{\frac{\kappa^{(\Theta)}}{\kappa}}\,\vaca^{\UV}, \qquad
\vacu^{\IR}=\sqrt{\frac{\kappa}{\kappa^{(\Theta)}}}\,\vacu^{\UV}.
\end{equation}

Recall from our probe brane discussion that the angle $\Theta$ indicates the endpoint of the flow, and so \eqref{thetaSolution} now indicates its relation to the UV configuration,
\begin{equation}\label{ThetaFlow}
\Theta=\frac{1}{\pfive}\frac{\qone^{\cD,\UV}}{\pthree^{\cD,\IR}}=\frac{\pi}{N_5}\frac{p}{N_3}=\theta_p.
\end{equation}
We have employed the integer valued charges of \eqref{D3Normalization} and \eqref{D1D5Normalization}.
The last equality matches the probe brane computation, where we considered the special case $N_3=1$.
The charge of the UV 1-brane is dissolved in the D3-brane in the UR.
Moreover, the more units of F1 charge are dissolved into a single D3-brane, the further the D3-branes slide down the $\S^3$.
This process is described by the $\SU(2)$ WZW model appearing in the original Kondo effect, where the UV number of branes at the north pole of the $\S^3$ determines the $\affine{\su(2)}$ representation defining the final boundary state in the IR.

\subsubsection*{Critical Screening}
We now turn to the critical case $\Theta=\pi$, which occurs when $p/N_3=N_5$.  We take for simplicity $\pone^\cD=0$ and $N_3=1$.
Observe that the modification to $U$ in \eqref{UD3} vanishes when $\defw\in\p\Sigma$, i.e. $\Theta=\pi$ and $\Theta=0$.
We therefore expect to obtain a pure D1/D5 geometry in this limit.
Using $\qone^{\Theta=\pi}=0$ and thus $\kappa_0^{(\Theta=\pi)}=1$, it is easy to check from \eqref{D3SolutionPureF1} that this is indeed the case.

This would be the end of the story if it weren't for the RR 4-form potential, which according to \eqref{D3SolutionCK} has a finite jump across the interface:
\begin{equation}\label{criticalScreeningCK}
  C_K(0)=0 \quad \text{vs.} \quad C_K(\infty)=-(2\pi)^4\alpha'^2 \,.
\end{equation}
This puzzle has its origin in the fact that the modification \eqref{UD3} is not globally defined, because going once around the 3-brane causes a finite shift in the value of $C_K$.
In order to return it to its original value we must perform a large gauge transformation, which in fact realizes a duality transformation.

\begin{figure}[t]
  \newcommand\ContourOneColor{blue}
  \newcommand\ContourTwoColor{orange}
  \tikzset{->-/.style={decoration={
  markings,
  mark=at position #1 with {\arrow{>}}},postaction={decorate}}}
  \tikzstyle{D0}=[circle, draw=red!90!black, fill=red!90!black, minimum size=3.5pt, inner sep=0pt]
    \centering
    \begin{tikzpicture}[scale=.5]
    \def\GridWidth{ 3};
    \def\GridHeight{ 4};
    \def\DefectX{ 1.2};
    \def\MidPanel{ 11};
    \def\RightPanel{ 22};

   \coordinate (XAxis) at (\GridWidth,0);
   \coordinate (RightXaxis) at ($(XAxis)+(.5,0)$);
   \coordinate (LeftXaxis) at ($-1*(RightXaxis)$);
   \coordinate (Height) at (0,\GridHeight);
   \coordinate (RightXaxisTop) at ($(RightXaxis)+(Height)$);
   \coordinate (LeftXaxisTop) at ($(LeftXaxis)+(Height)$);
   \coordinate (YAxis) at ($(Height)+(0,1)$);
   \coordinate (RightCFTx) at ($(RightXaxis)+(0,0.1)$);
   \coordinate (LeftCFTx) at ($-1*(RightCFTx)+(0,0.1)$);
   \coordinate (RightCFTxTop) at ($(RightCFTx)+(Height)$);
   \coordinate (LeftCFTxTop) at ($(LeftCFTx)+(Height)$);
   \coordinate (RightCFTxMid) at ($(RightCFTx)+.5*(Height)$);
   \coordinate (LeftCFTxMid) at ($(LeftCFTx)+.5*(Height)$);
   \coordinate (F1) at (\DefectX,0);
   \coordinate (F1arrow) at ($(F1)+.3*(Height)$);
   \coordinate (ContourL) at ($.48*(Height)+(LeftXaxis)$);
   \coordinate (ContourR) at ($.48*(Height)+(RightXaxis)$);
   \coordinate (Control2) at ($(F1)+.55*(Height)$);
   \coordinate (Control3) at ($(F1)+.15*(Height)$);
   \coordinate (Mid) at (\MidPanel,0);
   \coordinate (Right) at (\RightPanel,0);

   \draw[step=2cm, gray, very thin, dashed] (-\GridWidth,0) grid (\GridWidth,\GridHeight);
   \draw[very thick, <->] (LeftXaxis) -- (RightXaxis); 
   \draw[thick, ->] (0,0) -- (YAxis);
   \draw[very thick,<->] (LeftXaxisTop) -- (RightXaxisTop);
   \node[anchor=south west] at (Height) { $\pi$};
   \node[left] at (YAxis){$\theta$};
   \node[anchor=north] at (RightXaxis) { $\psi\rightarrow\infty$};
   \node[] at ($(LeftXaxisTop)+(0,1)$){$w=\psi+i\theta$};
   \node[anchor=north] at (LeftXaxis) { $-\infty\leftarrow\psi$};
   \draw[very thick, blue] (LeftCFTx) -- (LeftCFTxTop);
   \node [rotate=90,anchor=south, \ContourOneColor] at (LeftCFTxMid) { $(N_5,N_1,0)$};
   \draw[very thick, blue] (RightCFTx) -- (RightCFTxTop);
   \node [rotate=270,anchor=south, \ContourOneColor] at (RightCFTxMid){ $(N_5,N_1,p)$};
   \node[D0]	at (F1){};
   \node[color=red!90!black, below] at (F1){$p$ F1};
   \draw[red, ->, dashed] (F1) -- (F1arrow);

   \draw[rounded corners, \ContourOneColor, thick, ->-=.35] (ContourL) 
              -- node [midway, above] {$\cC_1$}(Control2) -- (ContourR);

   \draw[step=2cm, gray, very thin, dashed, xshift=\MidPanel cm] (-\GridWidth,0) grid (\GridWidth,\GridHeight);
   \draw[very thick, <->] ($(LeftXaxis)+(Mid)$) -- ($(RightXaxis)+(Mid)$); 
   \draw[thick, ->] (Mid) -- ($(YAxis)+(Mid)$);
   \draw[very thick,<->] ($(LeftXaxisTop)+(Mid)$) -- ($(RightXaxisTop)+(Mid)$);

   \node[anchor=south west] at ($(Height)+(Mid)$) { $\pi$};
   \node[left] at ($(YAxis)+(Mid)$){$\theta$};
   \node[anchor=north] at ($(LeftXaxis)+(Mid)$) { $-\infty\leftarrow\psi$};
   \draw[very thick, blue] ($(LeftCFTx)+(Mid)$) -- ($(LeftCFTxTop)+(Mid)$);
   \node [rotate=90,anchor=south, \ContourOneColor] at ($(LeftCFTxMid)+(Mid)$) { $(N_5,N_1,0)$};
   \draw[very thick, blue] ($(RightCFTx)+(Mid)$) -- ($(RightCFTxTop)+(Mid)$);
   \node [rotate=270,anchor=south, \ContourOneColor] at ($(RightCFTxMid)+(Mid)$){ $(N_5,N_1,p)$\textcolor{black}{, }\textcolor{\ContourTwoColor}{$(N_5,N_1,p - N_5)$}};

   \node[D0]	at ($(F1arrow)+(Mid)+.1*(Height)$){};
   \node[color=red!90!black, below] at ($(F1arrow)+(Mid)+.1*(Height)$){D3};
   \draw[red, dashed, ->] ($(F1arrow)+(Mid)+.1*(Height)$) -- ($(F1arrow)+(Mid)+.4*(Height)$);

   \draw[rounded corners, \ContourOneColor, thick, ->-=.35] ($(ContourL)+(Mid)$) 
              -- node [midway, above] {$\cC_1$}($(Control2)+(Mid)+.2*(Height)$) -- ($(ContourR)+(Mid)$);
   \draw[rounded corners, \ContourTwoColor, thick, ->-=.35] ($(ContourL)+(Mid)$) 
              -- node [midway, below] {$\cC_2$}($(Control3)+(Mid)$) -- ($(ContourR)+(Mid)$);

   \draw[step=2cm, gray, very thin, dashed, xshift=\RightPanel cm] (-\GridWidth,0) grid (\GridWidth,\GridHeight);
   \draw[very thick, <->] ($(LeftXaxis)+(Right)$) -- ($(RightXaxis)+(Right)$); 
   \draw[thick, ->] (Right) -- ($(YAxis)+(Right)$);
   \draw[very thick,<->] ($(LeftXaxisTop)+(Right)$) -- ($(RightXaxisTop)+(Right)$);

   \node[left] at ($(YAxis)+(Right)$){$\theta$};
   \node[anchor=north] at ($(RightXaxis)+(Right)$) { $\psi\rightarrow\infty$};
   \node[anchor=north] at ($(LeftXaxis)+(Right)$) { $-\infty\leftarrow\psi$};

   \draw[very thick, blue] ($(LeftCFTx)+(Right)$) -- ($(LeftCFTxTop)+(Right)$);
   \node [rotate=90,anchor=south, \ContourOneColor] at ($(LeftCFTxMid)+(Right)$) { $(N_5,N_1,0)$};
   \draw[very thick, blue] ($(RightCFTx)+(Right)$) -- ($(RightCFTxTop)+(Right)$);
   \node [rotate=270,anchor=south, \ContourTwoColor] at ($(RightCFTxMid)+(Right)$){ $(N_5,N_1,0)$};

   \node[D0]	at ($(F1)+(Right)+(Height)$){};
   \node[color=red!90!black, above] at ($(F1)+(Right)+(Height)$){\small nothing};
   \draw[rounded corners, \ContourTwoColor, thick,->-=.35] ($(ContourL)+(Right)$) 
              -- node [midway, below] {$\cC_2$}($(Control3)+(Right)+.25*(Height)$) -- ($(ContourR)+(Right)$);
  \end{tikzpicture}
  \caption[]{Left: When comparing fields between the two asymptotic regions we follow contour $\cC_1$. Note that the defect charge is absorbed by the CFT on the right.
            Middle: Contour $\cC_2$ defines another duality frame for the fields for which the right CFT carries no F1 charge.
            Right: In the frame $\cC_2$ we can have the defect move all the way to the south pole, where it turns into the trivial interface.}
  \label{fig: criticalScreening}
\end{figure}
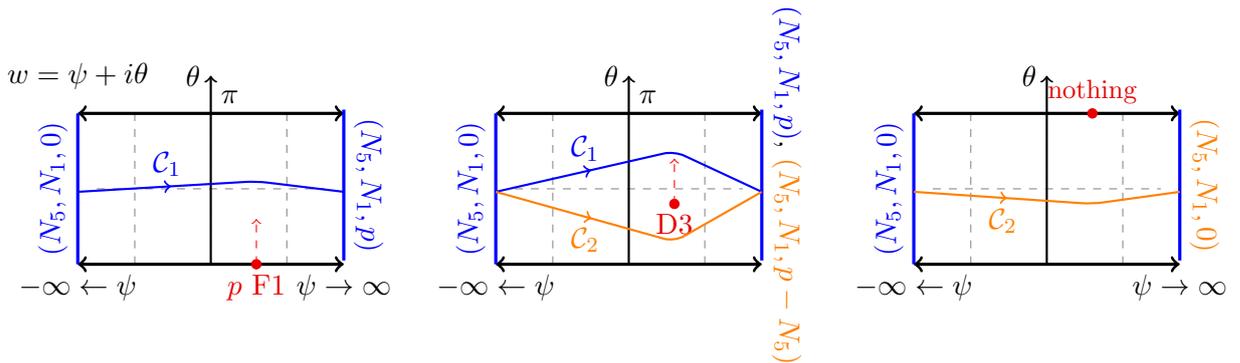

Let us see where the tension with \eqref{D3SolutionCK} arises.
When we compare field values in the two asymptotic regions we follow their values along a contour stretching from one asymptotic region to the other, as illustrated for the UV solution in the left-hand diagram of \figref{fig: criticalScreening}.
Observe that the asymptotic value of $C_K$, and thus the coupling $\theta_5$ of the CFT on the right, is determined by the interface's F1 charge $p$.
In contrast to the D3 solution, the F1 solution \eqref{UF1} features no monodromies and so is globally defined.
This means that any other choice of $\cC_1$ yields the same description.

When we trip the RG flow, the brane puffs into a D3-brane, which is characterized by the logarithms of \eqref{UD3}.
The monodromy means that a different choice of contours can result in a different description of the CFTs in each asymptotic region;
this is shown in the center picture of \figref{fig: criticalScreening}.
While $\cC_1$ is as before, $\cC_2$ gives a description in which the CFT at its end carries F1 charge $p - N_5$.

These considerations are valid for any $p$.
We now set $p=N_5$.
Then $\Theta=\pi$, implying that under the flow the interface moves all the way to the upper boundary of $\Sigma$, squeezing $\cC_1$ in the process.
The result \eqref{criticalScreeningCK} used $\cC_1$ and is precisely the shift in $C_K$ obtained under one of the parabolic generators of the extended U-duality group of IIB on $T^4$.
Working in this frame, the interface starts out as a non-trivial interface joining two CFT's that happen to be dual, but flows in the infrared to a duality interface.
In contrast, if we choose $\cC_2$, the interface can approach the upper boundary unhindered.
Since the CFT on the right carries no F1 charge, we find that in this description the interface is the trivial interface.
This is depicted in the right-hand diagram of \figref{fig: criticalScreening}.

In the WZW description of the original Kondo model, the limit $\Theta = \pi$ is known as \textit{critical screening}.
Physically, it describes a situation where the defect is screened completely by the conduction electrons.

\section{Interface entropy}
\label{sec:entropy}

The \emph{boundary entropy}, or \emph{$g$-factor}, has several equivalent definitions.
The original definition is as follows~\cite{AffleckGFactor, PhysRevB.48.7297}.
Place the CFT on a cylinder of radius $\beta$ and length $\ell$, with boundary states $A$ and $B$ at either end.
In the limit $\ell\gg\beta$ of a unitary BCFT, the partition function has an expansion
\begin{equation}
  \log Z = \frac{\pi c}{6} \frac{\ell}{\beta} + (s_A + s_B) + O(\beta/\ell)
\end{equation}
where $s_A$ and $s_B$ depend only on the choice of $A$ and $B$, respectively.
The $g$-factor corresponding to boundary $A$ is then $g_A=e^{s_A}$.
We can obtain the same quantity using only a single boundary state $A$ by performing a conformal transformation to the annulus, and then plugging a hole to produce a disk.
In this case, the disk partition function becomes simply%
\footnote{Technically speaking, $Z$ can be multiplied by an arbitrary constant by including a conformally invariant counterterm. This can be eliminated by comparing $Z_{D_2}^2$ to $Z_{\S^2}$, which is independent of renormalization scheme.}
\begin{equation}
Z = g_A \,.
\end{equation}

It was proved in \cite{Calabrese:2009qy} that the $g$-factor for a boundary conformal field theory is also encoded in the entanglement entropy.
Let the entangling region be an interval of length $\zeta_0$ starting at the boundary.
Then
\begin{equation}\label{EntanglementEntropy}
\cS_{\zeta_0}=
  \frac{\sfc}{6}\log\frac{\zeta_0}{\epsilon}
  + s_A
  + O(\epsilon) \,,
\end{equation}
with central charge $\sfc$, UV-cutoff $\epsilon$, and conformal boundary condition $A$.
In the case of an interface, we may fold the system along the interface (called the folded picture), and can think of our interface theory as a BCFT with central charge $\sfc=\sfc^{(0)}+\sfc^{(\infty)}$.
If we now unfold the system again, the entanglement interval lifts to an interval of length $2\zeta_0$ centered on the interface.

In \secref{sec: gFactorProbe} we derive the interface entropy in the probe brane description and thereafter, in \secref{sec: gFactorSugra}, we compute the interface entropies using the supergravity solutions of \secref{sec:sugra}.
Finally, we consider the probe brane limit of the supergravity solutions, and compare the result with the probe brane computation.

\subsection{Probe brane computation}\label{sec: gFactorProbe}

We reviewed the static probe brane configurations in \secref{sec: pqStringProbeBrane}.
Consider a $(p,q)$ string in the D1/D5 background.
In the S-dual frame this was described by a $(q,p)$ string on a background with $H^{(3)}$ flux.
On shell, its Lorentzian action took the form
\begin{align}
  L_{(q,p)} &= - \alpha' N_5 T_{(q,p)} \sqrt{-g_{\ads_2}} \,.
\end{align}
After analytic continuation, the probe brane Euclidean action thus takes the form
\begin{align}
  L^E_{(q,p)} &= -\alpha' N_5 T_{(q,p)} \sqrt{g_{H^2}} \,,
\end{align}
where $H^2$ denotes hyperbolic 2-space (i.e. Euclidean $\ads_2$).
In global coordinates, the bulk Euclidean metric takes the form
\begin{align}
  ds^2_{H^3} &= \alpha' N_5(d\rho^2 + {\cosh}^2\rho\, d\tau^2 + {\sinh}^2\rho\, d\theta^2) \,,
\end{align}
where $\tau$ is the coordinate running along the cylinder;
the defect is located at $\tau=0$.
The defect entropy is evaluated by computing the action in the conformal frame defined by the family of cutoff surfaces of the form $\rho=\rho_*$.
Placing a brane of tension $T_{(q,p)}$ (recall we are in the S-dual frame, with dilaton $e^{-\hat\dil}=e^{\dil}$) at $\tau=0$ yields the regularized Euclidean on-shell action
\begin{align}
  S^*_E = \alpha' N_5 T_{(q,p)} \int_0^{\rho_*} d\rho \int_0^{2\pi} d\theta\, \sinh\rho = 2\pi \alpha' N_5 T_{(q,p)} (\cosh\rho_* - 1) \,,
\end{align}
which is $\alpha'N_5 T_{(q,p)}$ times the regularized volume of Euclidean unit $\ads_2$.
Holographic renormalization of the probe brane demands \cite{Rey:1998ik,Maldacena:1998im,Karch:2005ms} that we add the covariant counterterm
\begin{align}
  S_\text{c.t.} &= - T_{(q,p)} \alpha' N_5 \int_0^{2\pi}d\theta \sqrt{h}\vert_{\rho=\rho_*}
   = -2\pi \, T_{(q,p)} \alpha' N_5 \sinh\rho_* \,,
\end{align}
where $h$ is the induced (unit radius) metric on the boundary surface.
Taking the limit gives the on-shell Euclidean action
\begin{align}\label{probeStringEntropy}
  \log g_\text{p.b.} = -S_{E}^{\text{ren}} &= -\alpha' N_5 T_{(q,p)} \Vol(H^2)_\text{ren} = 2\pi \alpha' N_5 T_{(q,p)} = N_5 \sqrt{q^2 + p^2 e^{2\dil}} \,,
\end{align}
where the final expression is written in terms of the dilaton in the original D1/D5 frame.
When $(p,q)=(1,0)$ this reduces to
\begin{align}
  \log g &= N_5 e^\dil = 2 \pi T_\text{1} L^2 \,,
\end{align}
which coincides with the obvious computation for the fundamental string in the original D1/D5 geometry, and this is the classical $g$-factor in the supergravity limit.

On the other hand, when $q\ne 0$, $g_\text{p.b.}$, it is impossible to work entirely in the probe brane approximation.
This is because the brane causes a jump in the $\ads$ radius, leading to large contributions to the action from the asymptotic regions.
To get the first non-trivial term in the $\pone^\cD / \overline{\pone}$ expansion, holographic renormalization must be performed taking into account the differing values of the $\ads$ radius.
While it is not difficult to perform a schematic version of this computation by truncating to $d=3$ and working through the holographic renormalization as in \cite{Bak:2016rpn}, it is not immediately clear without understanding the $\ads_3\times S^3$ backreaction whether this captures all effects at this order.
It is therefore desirable to take into account the complete backreaction; this is done in the next section.

The process of evaluating the $g$-factor for the IR D3-brane probe is essentially identical.
Integrating the Euclidean version of the on-shell action \eqref{eq:D3-on-shell} over $\ads_2\times\S^2$ and renormalizing the $\ads_2$ volume, we find
\begin{align}
  \log g &= - S_{E}^\text{ren} = N_5 \sqrt{q^2 + \bigl( p\, e^{\dil}{\textstyle \frac{\sin\theta_p}{\theta_p}} \bigr)^2 } \,.
  \label{probeD3Entropy}
\end{align}

\subsection{Supergravity computation of the interface entropy}\label{sec: gFactorSugra}
A detailed derivation of the interface entanglement entropy for the solutions of \cite{ChiodaroliOriginal} was given in \cite{ChiodaroliEntropy}.
We begin by reviewing their procedure and then apply it to the supergravity solutions of \secref{sec:sugra}.
This enables us to verify that the $g$-theorem applies to the RG flow endpoints.
Lastly, we take the probe brane limit and compare with the results of the previous section.
In this section we work exclusively in the D1/D5 frame.

\subsubsection*{Interface entropy of asymptotically $\ads_3\times \S^3$ geometries}
If $\cA$ is some $(d-1)$-dimensional subregion of some spatial slice in a $d$-dimensional QFT, the entanglement entropy $\cS_\cA$ of that region quantifies the degree of entanglement between $\cA$ and its complement in the quantum state on that slice.
The Ryu-Takayanagi (RT) proposal \cite{Ryu} posits that, at leading order in the $1/c$ expansion, the vacuum entanglement entropy of a $d$-dimensional holographic QFT is given by the area of the codimension~2 submanifold $\gamma_\cA$ of the bulk geometry bounded by $\p\cA$:
\begin{equation}
\cS_\cA = \frac{\textrm{Area}(\gamma_\cA)}{4G_N} ,
\end{equation}
with $G_N$ the bulk Newton's constant.
In the original construction the bulk spacetime was $\ads_{d+1}$ and $\gamma_\cA$ was $(d-1)$-dimensional.
When the dual is a superstring theory, however, the vacuum geometry has compact internal factors, and $\gamma_\cA$ has codimension~2 in the entire bulk spacetime.
Thus, in our case $\gamma_\cA$ is 8-dimensional.

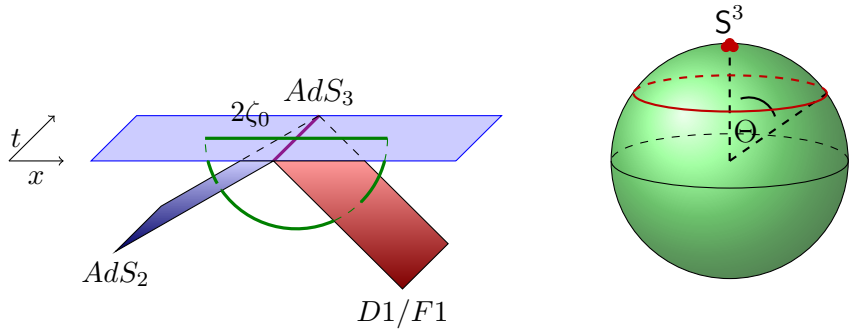
\begin{figure}[t]
  \tikzstyle{D0}=[circle, draw=red!90!black, fill=red!90!black, minimum size=3.5pt, inner sep=0pt]
  \centering
  \begin{tikzpicture}[scale=1.2]
\tikzstyle{D0}=[circle, draw=red!75!black, fill=red!75!black, minimum size=3.5pt, inner sep=0pt]
 \def\R{2 }; 
 \def\ZETA{1 }
 \coordinate (O) at (0,0);
 \def\m{.5 };
 \def\n{.5 };
 \coordinate [label=above:\large $AdS_3$] (U) at (\m,\n);
 \def\RA{315 };
 \def\LA{210}
 \coordinate (A) at (\R,0);
 \coordinate (D) at (-\R,0);
 \coordinate (B) at ($(U)+(A)$);
 \coordinate (C) at ($(U)-(A)$);
 \filldraw[fill=blue!20!white, draw=blue, name path=CFTspacetime] (A) -- (B) -- (C) -- (D) -- (A); 
 \draw[violet!90!white, very thick] (O) -- (U);
 \coordinate [label=below:$D1/F1$] (E) at ($(O)+(\RA:\R)$);
 \coordinate (F) at ($(E)+(U)$);
 \coordinate (G) at (intersection of U--F  and A--D);
 \shadedraw[top color=red!40!white, bottom color=red!50!black, name path=Defect] (O) -- (E) -- (F) -- (G) -- (O);
 \draw[dashed] (U) -- (G);
 \coordinate [label=below:$AdS_2$] (H) at ($(O)+(\LA:\R)$);
 \coordinate (I) at ($(H)+(U)$);
 \coordinate (J) at (intersection of U--I  and A--D);
 \shadedraw[top color=blue!20!white, bottom color=blue!40!black, name path=AdS2Sheet] (O) -- (H) -- (I) -- (J) -- (O);
 \draw[dashed] (U) -- (J);
 \coordinate (K) at ($(D)- .45*(\R,0)$);
 \coordinate (L) at ($(C)- .45*(\R,0)$);
 \coordinate (M) at ($(D)- .15*(\R,0)$);
 \draw[->] (K) to node [midway, left]{$t$} (L);
 \draw[->] (K) to node [midway, below]{$x$} (M);

 \coordinate (Y) at ($.5*(\m,\n)$);
 \coordinate (Z) at ($(Y)+(\ZETA,0)$);
 \coordinate (W) at ($(Y)-(\ZETA,0)$);
 \draw[very thick, green!50!black] (W) -- (Z)
		    node [pos=0.25, above] {$\textcolor{black}{2\zeta_0}$}; 
 \path [name path=Geodesic, draw, dashed, green!50!black](Z)
     arc[start angle=0, end angle=-180,radius=\ZETA];
 \path [name intersections={of = Geodesic and CFTspacetime}];
  \coordinate (N)  at (intersection-1);
  \coordinate (P)  at (intersection-2);
 \coordinate (Q) at ($(E)+(Y)$);
 \coordinate (S) at ($(H)+(Y)$);
 \path [name path=CenterOfSheets, draw=none] (S) -- (Y) -- (Q);
 \path [name intersections={of = Geodesic and CenterOfSheets}];
  \coordinate (piercingDefect)  at (intersection-1);
  \coordinate (piercingSheet)  at (intersection-2);
 \path [name intersections={of = Geodesic and Defect}];
  \coordinate (T)  at (intersection-1);
  \path [name intersections={of = Geodesic and AdS2Sheet}];
  \coordinate (V)  at (intersection-1);
 \pic [draw, angle radius=1.2*\ZETA cm, very thick, green!50!black] {angle=piercingDefect--Y--N};
 \pic [draw, angle radius=1.2*\ZETA cm, very thick, green!50!black] {angle=P--Y--piercingSheet};
 \pic [draw, angle radius=1.2*\ZETA cm, very thick, green!50!black] {angle=V--Y--T};

\def\Radius{.65*\R };
\def\BraneAngle{35 };
 \coordinate (BallCenter) at ($2.5*(A)$);
 \shade[ball color = green!60, opacity = 0.8] (BallCenter) circle (\Radius);
 \draw (BallCenter) circle (\Radius);
 \coordinate (North) at ($(BallCenter)+(0,\Radius)$);
 \node[above] at (North){\large $\S^3$};
 \coordinate (Equator) at ($(BallCenter)+(\Radius,0)$);
 \draw (Equator) arc (0:-180:\Radius and 0.3);
 \draw[dashed] (Equator) arc (0:180:\Radius and 0.3);
 \coordinate (D2ArcBegin) at ($(BallCenter)+(\BraneAngle:\Radius)$);
 \draw[dashed, thick] (BallCenter) --  (North) node (AngleBegin) [midway]{};
 \draw[dashed, thick] (BallCenter) -- (D2ArcBegin)
			     node (AngleEnd) [midway]{};
 \draw[thick] (AngleBegin) to [bend left=45] node (AngleMid) [midway]{} (AngleEnd);
 \node (theta) at ($(BallCenter)!0.5!(AngleMid)$) {\large $\Theta$};
 \draw[draw=red!75!black, thick] (D2ArcBegin) arc (0:-180:1.06 and 0.2);
 \draw[draw=red!75!black, dashed, thick] (D2ArcBegin) arc (0:180:1.06 and 0.2);
 \node[D0] at (North){};
 \node[D0, below right] at (North){};
 \node[D0, below left] at (North){};
\end{tikzpicture}

\caption{Entanglement minimal surface (depicted in green) wraps all of $\S^3$ (and $T^4$) and is a geodesic inside $\AdS_3$ anchored
at a CFT space interval of size $2\zeta_0$}
\label{fig: RT}
\end{figure}

Choose Poincar\'e patch coordinates $(t,\zeta)$ on the $\AdS_2$ fiber.
In the asymptotic region $r = e^\psi \to 0$, we may expand the metric in powers and logarithms of $r$.
In the notation of \eqref{asymptoticMetric},
\begin{equation}
ds_{\AdS_3}^2=\adsL^2\biggl(\frac{dr^2}{r^2}+\frac{\mu}{4}\frac{1}{r^2}\frac{d\zeta^2-dt^2}{\zeta^2}\biggr) + O(\log r) ,
\qquad\qquad
r=e^\psi .
\end{equation}
Sending $r\to 0$, the $\ads_2$ sheet approaches a half-space of the boundary CFT, and $(t,\zeta)$ can be identified with its Minkowski CFT coordinates.
We place one boundary of the entanglement interval at a distance $\zeta_0$ from the interface at $\zeta = 0$.
The entanglement interval extends an equal distance to the other side of the interface, giving $\cA = [-\zeta_0,\zeta_0]$ (see \figref{fig: RT}).

The fiber structure of the spacetime allows us to immediately identify an extremal 8-dimensional submanifold: the one defined by $(t,\zeta)=\text{const}$.
This surface has the important property that if it starts at a distance $\zeta_0$ to the right of the interface, it ends a distance $\zeta_0$ to the left of the interface, making it precisely the surface relevant to computing the interface entropy discussed above.
Thus, $\gamma_\cA$ lies at a constant point in the $\ads_2$ slice and wraps the remaining directions, which are a fibration of $\S^2\times T^4$ over $\Sigma$.
Computing the area of $\gamma_\cA$ then yields the entropy
\begin{equation}\label{RT}
  S_\cA=\frac{1}{4G_N^{(10)}}\int_{\S^2}d\Omega_2\int_{T^4}d\Omega_4\int_\Sigma\rho^2\,f_2^2\,f_3^4,
\end{equation}
where $d\Omega_2$ and $d\Omega_4$ denote the volume elements of $\S^2$and $T^4$, respectively, with unit radii.
Recall that the metric functions depend only on the coordinates of $\Sigma$.
Given the general form of the metric factors \eqref{metricFactorAdS2}-\eqref{metricFactorSigma} we deduce
\begin{equation}\label{EE}
S_\cA=\frac{\Vol(\S^2)}{4G_N^{(10)}}\int_\Sigma\,(a\,u-b^2)\biggl|\frac{\p_zV}{B}\biggr|^2.
\end{equation}
This area stretches to the boundary and is of course divergent.
In coordinates $z=re^{i\theta}$, the cutoffs at $r\rightarrow0$ and $r\rightarrow\infty$ are related to the UV cutoff $\epsilon$ in the CFT by
\begin{equation}\label{cutoff}
  r_{\infty}=\frac{2\zeta_0}{\epsilon\sqrt{\mu^{(\infty)}}}, \qquad r_{0}^{-1}=\frac{2\zeta_0}{\epsilon\sqrt{\mu^{(0)}}},
\end{equation}
where the scale factors $\mu^{(i)}$ are given in \eqref{scaleFactor}. In the F1/D1 solution this yields $\mu^{(i)}=\ol{\pone}/(|\pone^{(i)}|\kappa)$, while in the D3 solution we replace $\kappa\rightarrow \kappa^{(\Theta)}$.

\subsubsection*{D1/F1 and D3 interface entropy}

We now compute the interface entropy for the D1/F1 and D3 solutions of sections \secref{sec: F1defect} and \secref{sec: D3defect}, respectively.
Since in both cases the only modification is $\holu = \holu_0 + \delta\holu$, the entanglement entropy \eqref{EE} splits into two pieces.
First, we have the vacuum contribution
\begin{equation}\label{Ivac}
  \cI_{\textrm{0}}=\Vol(\S^2)\int_\Sigma\,(\hola\,\holu_{0}-\holb^2)\biggl|\frac{\p_z\holV}{\holB}\biggr|^2 ,
\end{equation}
together with the effect of the deformation \eqref{UF1} in the UV (F1/D1), or \eqref{UD3} for the IR (D3) interface:
\begin{equation}
 \cI_{\UV,\,\IR}=\Vol(\S^2)\int_\Sigma\,\hola\,\delta \holu^{F1/D1,\,D3}\,\biggl|\frac{\p_z\holV}{\holB}\biggr|^2 .
\end{equation}
We find
\begin{subequations}
\begin{align}
 \cI_{0}=&4\Vol(\S^3)\,\pfive\,\pbare\,\log\frac{r_\infty}{r_0},\\[.2cm]
 \cI_{\UV}=&4\Vol(\S^3)\,\pfive
	\biggl[\frac{\qone^\cD}{\vaca^{\scriptscriptstyle \UV}}
	      \biggl(\defx\,\log r_{\infty}+\frac{1}{\defx}\log\frac{1}{r_0}\biggr)
	      +\ol{\pone}-\pbare{}^{,\,\scriptscriptstyle \UV}-\pone^\cD\log|\defx|\biggr]\\[.25cm]
 \cI_{\IR}=&4\Vol(\S^3)\,\pfive
 \biggl[\frac{\qone^\cD}{\vaca^{\scriptscriptstyle \IR}}
	      \biggl(R\,\log r_{\infty}+\frac{1}{R}\log\frac{1}{r_0}\biggr)
	      +\ol{\pone}-\pbare{}^{,\,\scriptscriptstyle \IR}-\pone^\cD\log R\biggr].
\end{align}
\end{subequations}
The divergent pieces contain the two summands of the asymptotic D1 charges \eqref{F1SolutionD1Charge}, \eqref{D3SolutionD1Charge}. When adding the integrals it is convenient to repackage them using the central charges \eqref{AdSRasCharges}, giving
\begin{align}
 S^{\UV,\,\IR}_{2\zeta_0}=\frac{\cI_{0}+\cI_{\UV,\,\IR}}{4G_N^{(10)}}
		       =\frac{\sfc^{(\infty)} +\sfc^{(0)}}{6}\log\frac{2\zeta_0}{\epsilon}+\log \sfg^{\UV,\,\IR}
\end{align}
It is reassuring that the entanglement entropy assumes the expected form \eqref{EntanglementEntropy} for an interface CFT in the folded picture.
The sought-after g-factors are
\begin{subequations}\label{gFactors}
 \begin{align}
 s_{\UV}=\log \sfg^{\UV}&=\frac{\sfc^{(\infty)}+\sfc^{(0)}}{12}\biggl(\log\kappa+1-\frac{1}{\kappa}-\frac{\pone^\cD}{\ol{\pone}}\,\defxi\biggr)+ \sum_{i=0,\infty}\frac{c^{(i)}}{12}\log{\left(\frac{|\pone^{(i)}|}{\ol{\pone}}\right)}\\[.25cm]
 s_{\IR}=\log \sfg^{\IR}&=\frac{\sfc^{(\infty)}+\sfc^{(0)}}{12}\biggl(\log\kappa^{(\Theta)}+1-\frac{1}{\kappa^{(\Theta)}}-\frac{\pone^\cD}{\ol{\pone}}\,\defrho\biggr)+ \sum_{i=0,\infty} \frac{c^{(i)}}{12} \log{\left(\frac{|\pone^{(i)}|}{\ol{\pone}}\right)},\label{gFactorsIR}
\end{align}
\end{subequations}
where we employed \eqref{F1SolutionD1mean} and \eqref{D3SolutionD1mean}, and wrote the brane locations in terms of the coordinates $\defxi = \log\defx$ and $\defrho = \log R$.

One crucial property of any boundary RG flow in a 2d CFT is the $g$-theorem: the $g$-factor in the IR of a boundary RG flow is always smaller than in the UV \cite{Friedan:2003yc, Casini:2016fgb}.
In our case this means that, if an RG flow connects our UV and IR solutions, then the quantity
\begin{equation}\label{gTheorem}
\log\frac{\sfg^{UV}}{\sfg^{IR}}=\frac{\sfc^{(\infty)}+\sfc^{(0)}}{12}\biggl(\log\frac{\kappa}{\kappa^{(\Theta)}}+\frac{1}{\kappa^{(\Theta)}}-\frac{1}                           {\kappa}+\frac{\pone^\cD}{\ol{\pone}}\bigl(\defrho-\defxi\bigr)\biggr)\,.
\end{equation}
must be positive.
The lower bound in \eqref{kappaLambdaComparison} establishes this for all but the last term.
That term is also never negative:
indeed, from \eqref{F1D1SolutionXasCharges} we see that
$|\defrho| - |\defxi| \geq 0$, while \eqref{F1D1SolutionXasCharges} and \eqref{D3SolutionRasCharges} show that the signs of $\defrho$ and $\defxi$ both match that of $\pone^\cD$.
This guarantees that the last term is never negative.

That \eqref{gTheorem} has the positivity required of an interface RG flow by the $g$-theorem is an important check that these solutions are indeed those connected by our probe brane RG flows.

Observe that \eqref{gTheorem} vanishes for $\Theta\ll1$, since $\kappa^{(\Theta\rightarrow0)}=\kappa$ and $R=\defx$, which confirms that in this case the D3 interface reduces to the stack of 1-brane interfaces. For the other extreme, $\Theta=\pi$, we again restrict to pure F1 interfaces with $\pone^\cD=0\Leftrightarrow|\pone^{(i)}|=\ol{\pone}$. Recall $\qone^{\Theta=\pi}=0$ and from \eqref{lambdaPureF1} that $\kappa_0^{(\Theta=\pi)}=1$.
This limit corresponds to critical screening as discussed above. Indeed, we confirm that the defect degrees of freedom disappear --- they are being screened by the ambient degrees of freedom --- since the interface entropy \eqref{gFactorsIR} vanishes. This draws a beautiful parallel to the original Kondo model.

\subsubsection*{Comparison with probe brane calculation}
To compare with the probe brane computations we consider the limit of small string coupling, $e^\dil \ll 1$.
Recall that if the interface carries D1-brane charge in the D1/D5 frame, it is sent off to infinity as $e^\dil\to 0$, so to make sense of this computation we must set $q\propto\pone^\cD=0$.
In this case $\pone^{(0)}=-\pone^{(\infty)}$, meaning that both asymptotic regions have the same central charge, and that $\kappa$ reduces to the simpler forms \eqref{kappaPureF1} and \eqref{lambdaPureF1}.
Expanding \eqref{gFactors} in powers of $e^\dil$ then yields
\begin{subequations}\label{gFactorsProbeLimit}
 \begin{align}
 \log \sfg^{UV}&=\frac{\Pfive\Qone^\cD}{4\pi\kappa^2_{10}}e^{\dil}+\cO(e^{2\dil})=N_5\,p\,e^{\dil}+\cO(e^{2\dil}),\\[.25cm]
 \log \sfg^{IR}&=\frac{\Pfive\Qone^\cD}{4\pi\kappa^2_{10}}\frac{\sin\Theta}{\Theta}e^{\dil}+\cO(e^{2\dil})=N_5\,p\,\frac{\sin\Theta}{\Theta}e^{\dil}+\cO(e^{2\dil}).
\end{align}
\end{subequations}
These coincide with \eqref{probeStringEntropy} and \eqref{probeD3Entropy} upon setting $q=0$.
In this limit it is easy to see that the $g$-theorem is satisfied:
\begin{equation}\label{gTheoremProbeLimit}
\log\frac{\sfg^{UV}}{\sfg^{IR}}=N_5\,p\,\biggl(1-\frac{\sin\Theta}{\Theta}\biggr)e^{\dil}+\dots,
\end{equation}
which is manifestly positive.

\section{Discussion}
\label{sec:conclusions}

In this paper, we used holography to study a class of supersymmetric interface RG flows in the D1/D5 CFT that share important characteristics with the Kondo model.
The simplest of these interfaces arise in the brane construction as multiple fundamental strings attached to the D1/D5 system.
In the gauged linear sigma model description these interfaces arise from localized charged fermions and join CFTs with different $\U(N_5)$ theta angle, while in the NLSM description of the CFT they arise from fermions coupled to the $\U(N_5)$ connection induced on the target space by the ADHM construction, and join D1/D5 CFTs with different values of $B$-field cycles on the target manifold.
Generalizations of these interfaces allow D1-branes to be peeled off from the D1/D5 system in the presence of fundamental string charge;
these interfaces join CFTs with different values of the central charge.

The gravitational dual to these interfaces require embedding branes into the dual geometry.
In the regime where the DBI-CS description is reliable, the UV interface fixed point is dual to the $(p,q)$-string configuration first described in \cite{Bachas:2000fr}.
This string extends along an $\ads_2$ sheet inside $\ads_3$ and lies at a pole of $\S^3$.
This configuration has $\SO(2,1)\times\SU(2)$ bosonic symmetry;
the first factor is the $(0+1)$-dimensional conformal group, while the second factor is suitable for mimicking aspects of Kondo physics.

These $(p,q)$-string interfaces provide the UV fixed point for the aforementioned holographic RG flows.
We showed, in the probe brane approximation, that there exists a marginally relevant perturbation for which the embedding coordinates of the interface on $\S^3$ exhibit non-abelian polarization via the Myers effect.
This is in fact a consequence of the WZW description of the Kondo effect, in which a stack of $p$ D0 branes on $\SU(2)\simeq \S^3$ condenses into a single spherical D2-brane stabilized at constant polar angle $\theta\propto p$.
As a result, the holographic $(p,q)$-string interfaces puff up in $\S^3$ from a point into $\S^2$.

The dual of the IR interface is described by a D3 with $\ads_2\times \S^2$ geometry and carrying $p$ units of fundamental string charge and $q$ units of D1-brane charge.
We studied its evolution along RG time as parametrized by the radial coordinate of $\ads_3$.
When the $\S^2$ is above the string scale, this process is governed by the D3-brane DBI-CS action \eqref{D3 DBI CS action}.
Using $\kappa$ symmetry, we derive exact $\nicefrac{1}{2}$-BPS solutions that reliably describe the entire flow (except very near the UV fixed point).
These flow solutions indeed exhibit a non-trivial IR fixed point at the polar angle $\theta=\theta_p=\pi p/N_5$;
when the interface carries non-zero D1-brane charge, the brane locus also moves further away from the minimal area $\ads_2$ slice as we flow to the IR.

Thus far, the analysis had been limited to the probe brane description.
As a next step, we studied the gravitational backreaction of these branes.
Including backreaction along the entire flow is a difficult task, so we limited ourselves in this paper to the more tractable problem of identifying the Type IIB supergravity duals to the interface fixed points.
We construct these solutions using the general class of junction configurations constructed in \cite{ChiodaroliOriginal}.
By relaxing their regularity conditions, we can obtain exact backreacted supergravity backgrounds with localized brane sources containing either a $(p,q)$-string interface for the UV fixed point, or a D3-brane interface with $(p,q)$ units of 1-brane charge.
Both solutions preserve $\SO(2,1)\times\SU(2)$ symmetry and are $\nicefrac{1}{2}$-BPS.
The D3-brane solution features an upper bound on interface F1 charge, $p\leq N_3N_5$.

In order to relate these solutions to our RG flows, we must solve for the parameters defining the D3-brane solution in terms of those yielding the $(p,q)$-string solution.
This is accomplished by demanding that the ambient CFTs are themselves unaffected by the interface RG flow.
Doing so, we reproduce the value of the polar angle $\theta_p=\pi p/(N_5N_3)$ at which the D3 stabilizes, and also that the $\ads_2$ sheet of the interface is pushed toward the boundary of $\ads_3$ when the interface carries D1 charge $q$.

One immediate advantage of the supergravity description over the probe brane analysis is that it is valid for parameters that are not reliable in the probe brane description.
In particular, it is easy to dial up the fundamental string charge to  $\theta_p=\pi$.
In this case, we saw that the IR geometry approached the D1/D5 vacuum geometry;
however, describing the dual geometry globally required applying a duality transformation in one of the asymptotic regions relative to the UV.
In terms of the UV frame, the interface flows to a duality interface.
We call this process critical screening, in analogy with the Kondo effect.
Note that the remaining supersymmetric flows with $\theta_p<\pi$ ($p<N_5$) correspond in Kondo physics to ``overscreened'' impurities. Overscreened impurities occur in the multi-channel Kondo model when the number of conduction electron flavors (or channels) outweighs the impurity spin $p$. In our case the analogue of electron flavors is set by $N_5$, the number of background D5-branes.

We finally applied the gravitational description of our interfaces to compute their $g$-factors, which in some sense quantify the interface degrees of freedom.
In the probe brane description, the interface entropy is given by the brane's free energy.
This is computed via holographic renormalization of the brane's Euclidean action.
Equation \eqref{probeStringEntropy}, quantifies the interface degrees of freedom via the tension of the $(p,q)$ string.

We then considered the same problem including gravitational backreaction.
This was made possible by the framework of \cite{ChiodaroliEntropy}, which described how to compute the entropy of interfaces dual to the regular solutions of \cite{ChiodaroliOriginal} (and by trivial extension, the singular brane solutions we considered).
Using this we could verify that the $g$-theorem is satisfied by our purported flows, i.e. that the $g$-factor of the IR interface must be smaller than that in the UV.
Finally, we showed that the supergravity result reduces to the probe brane computation in the appropriate limit.

\subsection*{Future directions}
Based on the results presented, there are several natural directions for future research.

For example on the field theory side we left the description of the deformation generating the interface flows for future work.
It is also interesting to compute the exact field-theory values of the interface entropies, which should be possible using the localization results of \cite{Hori:2013ika}.

In view of applications to condensed matter physics, it is interesting to obtain the finite-temperature gravity solution for the model considered here. Moreover, while in the previous work \cite{Erdmenger:2013dpa}, the electrons were taken to be non-propagating away from the defect, the construction of the present paper allows to calculate electronic conductivities in the ambient CFT. Generalizations of the present model to finite temperature will in particular allow to evaluate the temperature dependence of the conductivity from a backreacted gravity solution, beyond probe-limit results presented in \cite{Padhi:2017uxc}. In particular, in the model presented here, we have full control of the boundary behaviour of the gauge field dual to the conserved current, both at and away from the interface.

A different direction would make further application of the gravitational dual solutions described in \secref{sec:sugra}.
While the probe brane description gives the leading behavior of the defect itself, if we want to understand its effect on CFT correlation functions or the structure of bulk-boundary fusion, we must take backreaction into account.
In particular, the CFT one-point functions can easily be derived from our supergravity solutions, and a study of fluctuations on this background may make it possible also to derive two-point functions.

\acknowledgments{We are grateful to Juan Maldacena for discussions that initiated this project.
Moreover, we are grateful to Mario Flory, Dilyn Fullerton, Kevin Grosvenor, Carlos Hoyos, Darya Krym, Ren\'e Meyer, Nina Miekley, Andy O'Bannon, Brandon Robinson, Ronnie Rodgers and Kostas Skenderis for useful discussions.
C.M.T. would further like to thank Davide Gaiotto for helpful discussions.
The work of C.M.T. was supported through a research fellowship from the Alexander von Humboldt foundation.
C.N. acknowledges financial support through the W\"{u}rzburg-Dresden Cluster of Excellence on Complexity and Topology in Quantum Matter -- ct.qmat (EXC 2147, Project-id No. 39085490).}

\appendix

\section{Conventions and background}
\label{app:conventions}

\subsection{Spacetime and spinors}
The metric is mostly plus, $\eta_{MN}=\text{diag}(-+\cdots+)$.
Our Clifford algebra convention is $\{\Gamma_M,\Gamma_N\} = 2\eta_{MN}$.
Multi-index $\Gamma$ matrices are antisymmetrized products as usual, e.g.
$\Gamma_{MN}=\frac{1}{2}[\Gamma_M,\Gamma_N]$.
We take complex conjugation of Grassmann scalars to reverse order, $(\psi\eta)^* = \eta^* \psi^*$.

The invariant bilinear form associated to any complex doublet index $A$ is the antisymmetric tensor, which in the standard basis has the form
\begin{align}
  (C^{AB}) &= (C_{AB}) = \begin{pmatrix} 0 & 1 \\ -1 & 0 \end{pmatrix} .
\end{align}
Indices are raised and lowered using the $C$ tensor according to the NW-SE convention:
\begin{align}
  \psi^A &= C^{AB}\psi_B &
  \psi_B &= \psi^A C_{AB} \,.
\end{align}

\subsection{Type IIB supergravity}
\label{app:IIB}
Local supersymmetry parameters in Type IIB form a doublet of Majorana-Weyl spinors with the same chirality.
The doublet index is acted on by 3 matrices $I$, $J$, and $K$ which generate $\SL(2,\R)$:
\begin{align}
  I^2 &= -1 & J^2 &= 1 & K^2 &= 1 \\
  IJ &= K & KI &= J & KJ &= I \,.
\end{align}
In this paper we only utilize bosonic fields supergravity fields.
In addition to the metric $g$, dilaton $\phi$, and Kalb-Ramond 2-form potential $B$, there are the RR $p$-form potentials $C^{(p)}$: $C^{(0)}$, $C^{(2)}$, and $C^{(4)}$.
Their differentials $F^{(p+1)} = dC^{(p)}$ are not fully gauge-invariant;
the gauge-invariant field strengths are $F^{(1)}$, $\tilde F^{(3)}=F^{(3)}-C^{(0)}H$, and $\tilde F^{(5)}=F^{(5)}-\frac{1}{2}C^{(2)}\wedge H+\frac{1}{2}B\wedge F^{(3)}$.

In this paper we absorb the string coupling into the value of the dilaton, so that the 10d Newton's constant is $\frac{1}{\kappa_{10}^2}=\frac{2\pi}{(2\pi\ell_s)^8}$.
When we do this, we must also absorb a factor of $g_s^{-1}$ into the RR potentials relative to the most prevalent convention.

The bulk bosonic action decomposes as $S_\mathrm{bos} = S_{0} + S_{CS}$:
\begin{align}
S_0 &=
\begin{aligned}[t]
 \frac{1}{2\kappa_{10}^2}\int d^{10}x \Bigl(
  & e^{-2\Phi} (R+4(\nabla\Phi)^2-\frac{1}{2}|H|^2) \\
  & -\frac{1}{2}|F^{(1)}|^2-\frac{1}{2}|\tilde F^{(3)}|^2-\frac{1}{4}|\tilde F^{(5)}|^2 \Bigr) ,
\end{aligned} \\
S_{CS} &= -\frac{1}{2\kappa_{10}^2}\frac{1}{2}\int B\wedge F^{(3)}\wedge F^{(5)} .
\end{align}
The resulting equations need to be supplemented by the condition $*\tilde F^{(5)} = \tilde F^{(5)}$.

\subsection{Gauge theory}
In a gauge theory with gauge group $G$ and real Lie algebra $\g$, we write the
covariant derivative in the form $D_\mu=\p_\mu - i\dA_\mu$, where $\dA_\mu(x)$ is a matrix valued in $i\g$.
In particular, in any Hermitian representation of the gauge group, $\dA$ is a Hermitian matrix.
The structure constants in such a basis take the form
$[t_a,t_b] = i f_{ab}{}^c t_c$.
If $\psi$ is a matter field and $\Omega$ is a $G$-valued local parameter, gauge transformations act as
\begin{align}
\tilde\psi &= \Omega\,\psi &
\tilde{A}_\mu &= \Omega(i\p_\mu + A_\mu)\Omega^{-1} \,.
\end{align}
Finally, we take traces to be in the fundamental representation of $\U(N)$.

\subsection{\texorpdfstring{$\cN=(4,4)$ gauge theory}{N=(4,4) gauge theory}}
\label{app:vectorMultiplet}
When working with 2-dimensional $\cN=(4,4)$ supersymmetric field theories we encounter three complex doublet indices $i$, $\alpha$, and $\dot\alpha$.
In the conformal limit, $\alpha$ and $\dot\alpha$ become the doublet indices for the left-moving R-symmetry $\SU(2)_-$ and for the right-moving R-symmetry $\SU(2)_+$, respectively.
There are 4 left-moving supercharges $Q_\alpha$ and four right-moving supercharges $Q_{\dot\alpha}$, complemented in the conformal limit by the superconformal charges.

In the main text we require several details of the $\cN=(4,4)$ non-abelian vector multiplet $(A_\mu, A_I, \lambda_{+i\alpha}, \lambda_{-i\dot\alpha})$, together with a symmetric doublet of auxiliary fields $D_{(ij)}$.
We omit these for hypermultiplets as their explicit form is not required in the text.
Recall from the main text that $\mu=(01)$, $I=(6789)$, $(\alpha,\dot\alpha)$ denote doublet indices for the $\SU(2)_-\times\SU(2)_+$ R-symmetry, while $i$ is a doublet index for the $\SU(2)$ rotating the three complex structures on the hypermultiplet target space.

Denote by $\tau_{I\alpha}{}^{\dot\alpha}$ and $\btau_{I\dot\alpha}{}^{\alpha}$ the $\Spin(4)=\SU(2)_-\times\SU(2)_+$ $\sigma$ matrices,
\begin{align}
  (\tau_{(I}\btau_{J)})_\alpha{}^\beta &= \delta_{IJ} \delta_\alpha{}^\beta &
  (\btau_{(I}\tau_{J)})_{\dot\alpha}{}^{\dot\beta} &= \delta_{IJ} \delta_{\dot\alpha}{}^{\dot\beta} \,,
\end{align}
which give an embedding of the basis quaternions ($\tau_I$) and their quaternion conjugates ($\btau_I$) into $M_2(\C)$.
They further satisfy
\begin{align}
  (\tau_{I\alpha}{}^{\dot\alpha})^* &= \btau_{I\dot\alpha}{}^{\alpha} = -\tau_I{}^\alpha{}_{\dot\alpha} &
  (\tau_{I\alpha\dot\alpha})^* &= \tau_I{}^{\alpha\dot\alpha} &
  \tau_{I\alpha\dot\alpha} &= - \btau_{I\dot\alpha\alpha}
  \,.
\end{align}
In an appropriate basis they take the form
\begin{align}
  \tau_I &= (i\vec\sigma, \id) &
  \btau_I &= (-i\vec\sigma, \id) \,.
\end{align}
Rotations are generated using the matrices
\begin{align}
  \tau^{IJ}{}_\alpha{}^\beta &=
  	(\tau^{[I}\btau^{J]})_\alpha{}^\beta &
  \btau^{IJ}{}_{\dot\alpha}{}^{\dot\beta} &=
  	(\btau^{[I}\tau^{J]})_{\dot\alpha}{}^{\dot\beta} \,,
\end{align}
which satisfy the properties
\begin{subequations}
\begin{align}
  \tau^I \btau^J &= \delta^{IJ} \id + \tau^{IJ} &
  \btau^I \tau^J &= \delta^{IJ} \id + \btau^{IJ} \\
  \tau^{IJ}{}_{\alpha\beta} &= \tau^{IJ}{}_{\beta\alpha} &
  \btau^{IJ}{}_{\dot\alpha\dot\beta} &= \btau^{IJ}{}_{\dot\beta\dot\alpha} \\
  \tau^I{}_{\alpha\dot\alpha} \tau_{I\beta\dot\beta}
  	&= 2 C_{\alpha\beta} C_{\dot\alpha\dot\beta} &
  \tau^I{}_{\alpha}{}^{\dot\alpha} \btau_{I\dot\beta}{}^\beta
    &= 2 \delta_\alpha{}^\beta \delta_{\dot\beta}{}^{\dot\alpha}
\end{align}
\end{subequations}
and
\begin{subequations}
\begin{align}
  \tau^{IJ}\tau^K
  	&= \tau^I \delta^{JK} - \tau^J \delta^{IK} - \epsilon^{IJKL} \tau_L \,, &
  \tau^I \btau^{JK}
  	&= \delta^{IJ} \tau^K - \delta^{IK} \tau^J - \epsilon^{IJKL}\tau_L \,, \\
  \btau^{IJ} \btau^K
  	&= \btau^I \delta^{JK} - \btau^J \delta^{IK} + \epsilon^{IJKL} \btau_L \,, &
  \btau^I \tau^{JK}
  	&= \delta^{IJ} \btau^K - \delta^{IK} \btau^J + \epsilon^{IJKL} \btau_L \,.
\end{align}
\end{subequations}
The gauginos $\lambda_\pm$ obey a reality relation  of ``symplectic Majorana'' type,
\begin{align}
  (\lambda_{-i\dot\alpha})^* &= \lambda_-^{i\dot\alpha} &
  (\lambda_{+i\alpha})^* &= \lambda_+^{i\alpha} \,.
  \label{eq:reality}
\end{align}
By abbreviating
\begin{subequations}
\begin{align}
  \lambda_- &= \lambda_{-i\dot\alpha} &
  \lambda_+ &= \lambda_{+i\alpha} \\
  \tau_I &= \tau_{I\alpha}{}^{\dot\alpha} &
  \btau_I &= \btau_{I\dot\alpha}{}^{\alpha} \\
  (\epsilon_- \lambda_-) &= \epsilon_-^{i\dot\alpha} \lambda_{-i\dot\alpha} &
  (\epsilon_+ \lambda_+) &= \epsilon_+^{i\alpha} \lambda_{+i\alpha}
\end{align}
\end{subequations}
we write the Lagrangian of a $\U(N)$ vector multiplet
\begin{align}
  L = \frac{1}{g^2}\Tr\biggl(&
    \frac{1}{2} (F_{01})^2
    - \frac{i}{2} (\lambda_- D_+ \lambda_{-})
    - \frac{i}{2} (\lambda_+ D_- \lambda_{+})\notag \\
    &
    + \frac{1}{2} D_+ A^I D_- A_I
    + \frac{1}{4} [A_I,A_J]^2
    + (\lambda_- \btau^I [A_I, \lambda_+])
  \biggr) \,.
\end{align}
The trace is in the fundamental representation of $\U(N)$ and $D_\pm=D_0\pm D_1$. We have omitted the auxiliary fields $D_{ij}$ here, because we are interested in on-shell configurations. Straightforward computations show that this Lagrangian is invariant under the supersymmetry variations
\begin{subequations}\label{vectorVariations}
\begin{align}
  \delta A_+ &= 2 i (\epsilon_+ \lambda_+) &
  \delta F_{01} &= i \cd_- (\epsilon_+\lambda_+) - i \cd_+ (\epsilon_- \lambda_-) \\
  \delta A_- &= 2 i (\epsilon_- \lambda_-) &
  \delta \lambda_- &= \cd_- A_I (\btau^I \epsilon_+)
    - \frac{i}{2} [A_I,A_J] (\btau^{IJ}\epsilon_-) - F_{01} \epsilon_- \\
  \delta A_I &= i (\epsilon_+ \tau \lambda_-) + i (\epsilon_- \btau \lambda_+) &
  \delta \lambda_+ &= \cd_+ A_I (\tau^I \epsilon_-)
    - \frac{i}{2} [A_I, A_J] (\tau^{IJ} \epsilon_+) + F_{01} \epsilon_+ \,.
\end{align}
\end{subequations}
In the main text we combine the gaugino variations of \eqref{vectorVariations} with \eqref{defectSUSYs} to yield \eqref{vanishingGauginoVariations}. Specifically we demand $\delta(\lambda_+\pm \tau^9\lambda_-)=0$.

\section{Derivation of the jump in theta angle}
\label{app:theta}
An important feature of the interfaces we are studying is that the theta angle of the D5 gauge theory differs on either side.
The theta angle descends from the Chern-Simons coupling to $C^{(4)}$,
\begin{align}
  S_\text{CS} &\supset
  T_5\int_{\R^2\times M_4} C^{(4)}\wedge 2\pi\alpha' \Tr F = \frac{\theta_5}{2\pi} \int_{\R^2}\! \Tr F &
  \theta_5 &= 4\pi^2\alpha' T_5 \int_{M_4}C^{(4)} \,.
\end{align}
We will now derive the jump in $\theta_5$.

It is simplest to use the truncation of IIB to $\ads_3$, which accurately describes physics at distances much larger than the $\ads_3$ radius.
This can be done at the level of the equations of motion by taking the ansatz
\begin{align}
  F^{(3)} &= \frac{2}{L}e^{-\dil}(\omega_{\ads}+\omega_{\S^3}) \\
  B^{(2)} &= B_2 &
  H^{(3)} &= H_3 = dB_2 \\
  C^{(4)} &= A_0 \wedge \omega_{M_4} + A_1 \wedge \omega_{\S^3} &
  G_1 &= dA_0 \\
  \tilde F^{(5)} &= G_1 \wedge \omega_{M_4} + \tilde G_2 \wedge \omega_{\S^3} &
  \tilde G_2 &= dA_1 + \mu B_2
\end{align}
where $A_0$, $A_1$, and $B_2$ are constant on $\S^3\times M_4$, and $\mu = \frac{2}{L}e^{-\dil}$.
On this ansatz, the equations of motion
\begin{align}
  d(e^{-2\dil}*H^{(3)}) &= F^{(1)}\wedge *\tilde F^{(3)} + \tilde F^{(3)}\wedge\tilde F^{(5)} \\
  d\tilde F^{(5)} &= H_3\wedge\tilde F^{(3)} \\
  \tilde F^{(5)} &= {*_{10}}\tilde F^{(5)} \\
\end{align}
take the form
\begin{align}
  d(e^{-2\dil}{*_3}H_3) &= \mu G_1 + J_1  \\
  dG_1 &= d({*_3}\tilde G_2) = 0 &
  d\tilde G_2 &= -d({*_3}G_1) = \mu H_3
\end{align}
where $J_1$ is a source.

We now introduce $p$ fundamental strings at the locus $\Sigma=\{\psi=0\}$ of $\ads_3$,
\begin{align}
  S_\text{F1} &= \frac{p}{2\pi\alpha'} \int B_2 &
  \implies J_1 &= -\frac{2\kappa_3^2}{2\pi\alpha'}p\, d(\Theta(\psi)) \,,
\end{align}
with $\Theta$ the Heaviside theta function, and
\begin{align}
  \frac{1}{2\kappa_3^2} = \frac{1}{2\kappa_{10}^2}(2\pi^2L^3)(\ell_s^4N_1 N_5^{-1})
\end{align}
the 3d gravitational coupling.
Eliminating $H_3$ from the equations of motion and integrating, we obtain
\begin{align}
  *d*dA_0 &= -\nabla^2 A_0 = - \mu^2 (A_0 - \alpha) + \frac{2\kappa_3^2}{2\pi\alpha'}p\,\mu\,\Theta(\psi) \,,
\end{align}
with $\alpha$ a constant of integration.
$A_0$ has mass $\mu$, meaning $dA_0$ vanishes in the asymptotic region.
The shift in $A_0$ in the asymptotic region is therefore given by the $\Theta$ contribution,
\begin{align}
  \Delta A_0 &= \frac{2\kappa_3^2}{2\pi\alpha'}\frac{p}{\mu} = \frac{p}{N_1} \,.
\end{align}
The corresponding jump in $\theta_5$ across the interface is then
\begin{align}
  \Delta\theta_5 &= \ell_s^2 T_5 \Delta A_0 \int_{M_4}\omega_{M_4}
  = \frac{2\pi p}{N_5} \,.
\end{align}

\section{Supergravity defect solutions}
\label{app:SugraDefects}
This appendix provides technical details on the supergravity solutions presented in \secref{sec:sugra}. In particular, we give the explicit expressions for the fields and charges at the asymptotic regions required to obtain the solutions \eqref{F1D1Solution} and \eqref{D3Solution}.

Before we begin, we fill in the missing two-form potentials needed to compute the Page charges \eqref{PageFiveCharges} and \eqref{PageOneCharges},
\begin{subequations}
\begin{align}\label{2form}
 b^{(1)}&=-\frac{2\holv\,\holb}{\hola\,\holu-\holb^2}-h_1,\qquad& h_1=\int\frac{\p_z\holv}{\holB}+c.c.,\\
 b^{(2)}&=\frac{2\holv\,\tilde{\holb}}{\hola\,\holu+\tilde{\holb}^2}+\tilde{h}_1,\qquad& \tilde{h}_1=\frac{1}{i}\int\frac{\p_z\holv}{\holB}+c.c.,\\
 c^{(1)}&=-\holv\frac{\hola\,\tilde{\holb}-\tilde{\hola}\,\holb}{\hola\,\holu-\holb^2}+\tilde{h}_2,\qquad& \tilde{h}_2=\frac{1}{i}\int \holA\frac{\p_z\holv}{\holB}+c.c.,\\
 c^{(2)}&=-\holv\frac{\hola\,\holb+\tilde{\hola}\,\tilde{\holb}}{\hola\,\holu+\tilde{\holb}^2}+h_2,\qquad& h_2=\int \holA\frac{\p_z\holv}{\holB}+c.c.
\end{align}
\end{subequations}

\subsection{Asymptotic Regions}\label{app: asymptoticRegions}
Our goal is to characterize the interface geometry in terms of expressions associated with the CFTs, which live at the asymptotic regions, $z=0$ and $z\rightarrow\infty$. In this subsection we present all the charges \eqref{PageFiveCharges}, \eqref{PageOneCharges} and fields \eqref{SugraFields} evaluated at the asymptotic regions.
Even though, our intterface solutions are not included within the set of solutions studied in \cite{ChiodaroliOriginal, ChiodaroliJunctions}
the explicit form of all required expressions at the asymptotic regions remains unchanged (up to the different scalings, \autoref{footnote: conventions}). This becomes evident when expanding our modifications \eqref{UF1} and \eqref{UD3} at the asymptotic regions. Thus we cite all charges, fields and metric factors as computed in \cite{ChiodaroliJunctions}.

Singularities in $f_1$ designate asymptotic $\AdS_3\times \S^3\times T^4$ regions. In terms of the holomorphic functions the asymptotic regions are singled out as poles of $\holV$. We are interested in solutions with two asymptotic regions, which we place at $z=0$ and $z\rightarrow\infty$ in $\Sigma$. These regions are interchanged via inversion $z\rightarrow-1/z$. In the vicinity of $z=0$ the meromorphic functions assume the form
\begin{subequations}
\begin{align}
 \holV(z)&=i\holv_{-1}z^{-1}+i\holv_1z+\dots\\
 \holA(z)&=i\hola_0+i\hola_1z+\dots\\
 \holB(z)&=i\holb_0+i\holb_1z+\dots\\
 \holU(z)&=i\holu_0+i\holu_1z+\dots
\end{align}
\end{subequations}
All coefficients $\holv_j,\,\holu_j,\,\hola_j,\,\holb_j$ in these expansions are real. The coefficient $\holu_0$ is not to be confused with the harmonic function \eqref{U0} characterizing the trivial interface. Which one is used will always be clear from context. Switching coordinates to $z=re^{i\theta}$ we find expressions for the dilaton,
axion and RR four-form potential (see \eqref{DilatonFormula}-\eqref{FourFormFormula})
\begin{subequations}\label{asymptoticFields}
\begin{align}
  e^{-2\dil}&=\frac{\holb_0^2}{\holu_1^2}(\hola_1\holu_1-\holb_1^2)+\cO(r)\label{asymptoticDilaton},\\
  \chi&=\frac{\holb_0\holb_1}{\holu_1}-\hola_0+\cO(r)\label{asymptoticAxion},\\
  C_K&=\frac{\holb_0\holb_1}{\hola_1}-\holu_0+\cO(r)\label{asymptoticFourForm}.
\end{align}
\end{subequations}
Similarly, the metric factors \eqref{metricFactors} become
\begin{subequations}\label{asymptoticMetricFactors}
\begin{align}
 f_1^4&=\frac{1}{r^4}\frac{4\hola_1\holb_0\holv_{-1}^2}{(\hola_1\holu_1-\holb_1^2)^{3/2}}+\cO(r^{-3}),\\
 f_2^4&=\sin^4\theta\frac{4a_1v_{-1}^2}{b_0^3}\sqrt{a_1u_1-b_1^2}+\cO(r),\\
 f_3^4&=\frac{b_0}{a_1}\sqrt{a_1u_1-b_1^2}+\cO(r),\\
 \rho^4&=\frac{1}{r^4}\frac{4a_1v_{-1}^2}{b_0^3}\sqrt{a_1u_1-b_1^2}+\cO(r^{-3}).
\end{align}
\end{subequations}
Using coordinates $z=\exp(\psi+i\theta)$ the metric assumes the form ($\psi\rightarrow-\infty, r\rightarrow 0$)
\begin{equation}\label{asymptoticMetric}
 ds_{10}^2=\adsL^2\biggl(d\psi^2+\frac{\mu}{4}e^{-2\psi}ds_{\AdS_2}^2+d\theta^2+\sin^2\theta ds^2_{\S^2}\biggr)+\sqrt{\frac{\holu_1}{\hola_1}e^{-\dil}}ds^2_{T^4}.
\end{equation}
Here, we defined the ten-dimensional $\AdS$ radius $\adsL$ and a scale factor $\mu$, which becomes important when choosing a cutoff for $\AdS_3$,
\begin{equation}\label{Radius10}
\adsL^2=2\sqrt{\frac{\hola_1\holv_{-1}^2\holu_1}{\holb_0^4}}\,e^{-\frac{1}{2}\dil},\qquad \mu=4\frac{\holb_{0}^4}{\holu_{1}^2}\,e^{2\dil}.
\end{equation}
The six-dimensional $\AdS$ radius $\adsR=\adsL f_3$ is useful and appears in the scale factor,
\begin{equation}\label{scaleFactor}
\mu=\frac{(4\holv_{-1})^2}{\adsR^4}.
\end{equation}
The asymptotic 5-brane Page charges \eqref{PageFiveCharges} and the asymptotic 1-brane Page charges \eqref{PageOneCharges} are expressed through
\begin{subequations}\label{asymptoticCharges}
\begin{align}
 \pfive\equiv&\frac{Q_{D5}}{8\pi^2}\;=\holv_{-1}\frac{\hola_1\holb_0-\hola_0\holb_1}{\holb_0^2}\label{qD5},\\
 \qfive\equiv&\frac{Q_{F5}}{8\pi^2}\;=\holv_{-1}\frac{\holb_1}{\holb_0^2}\label{qF5},\\
 \pone\equiv&\frac{Q_{D1}}{8\pi^2}=-\holv_{-1}\frac{\holb_1\holu_0-\holb_0\holu_1}{\holb_0^2}\label{qD1},\\
 \qone\equiv&\frac{Q_{F1}}{8\pi^2}=-\holv_{-1}\frac{\holb_0^2\holb_1+a_0\holb_1\holu_0-a_1\holb_0\holu_0-a_0\holb_0\holu_1}{\holb_0^2}\label{qF1}.
\end{align}
\end{subequations}
The ten-dimensional gravitational constant and the ten-dimensional Newton constant are
\begin{equation}
  \kappa_{10}^2=8\pi G_N^{(10)}, \qquad G_N^{(10)}=G_N^{(3)}\Vol(S_\adsL^3)\Vol(T_{f_3}^4)=G_N^{(3)}\,2\pi^2\adsL^3\,f_3^4,
\end{equation}
where the subscripts in the volumes denote the respective radii\footnote{Whenever we omit the radius in Volume expressions it implies unit radius, i.e. $\Vol(\S^3)=2\pi^2$}. The Brown-Henneaux formula then provides the central charge of the CFT at the asymptotic region,
\begin{equation}\label{centralCharge}
  \sfc=\frac{3\adsL}{2G_N^{(3)}}=\frac{6}{4\pi\kappa_{10}^2}\biggl(Q_{D5}\,Q_{D1}+Q_{F5}\,Q_{F1}\biggr).
\end{equation}
Lastly, the observation
\begin{equation}\label{AdSRasCharges}
\adsR^4=4\biggl(\pfive\,\pone+\qfive\,\qone\biggr)=\frac{4G_N^{(10)}}{\Vol(\S^3)}\frac{\sfc}{6}
\end{equation}
is convenient. Keep in mind that all expressions in this section hold only at one asymptotic region.

\subsection{Trivial interface in the D1/D5 CFT: vacuum \texorpdfstring{$\ads_3$}{AdS3}}\label{app: SugraVacuum}
Here we present details for the vacuum solution in \secref{sec: SugraVacuum}. The necessary coefficients for \eqref{asymptoticCharges} are found after expanding \eqref{vacFuns} at $z=0$ and $z\rightarrow\infty$. It is readily seen that $\qfive=0=\qone$ at both asymptotic regions. The remaining charges in \eqref{asymptoticCharges} and the dilaton \eqref{asymptoticDilaton} on both sides are not independent,
\begin{subequations}\label{vacCharges}
\begin{align}
\pfive^{(0)}&=\frac{\vaca\,\vacv}{\vacb}=-\pfive^{(\infty)},\label{vacD5charge}\\
\pone^{(0)}&=\frac{\vacu\,\vacv}{\vacb}=-\pone^{(\infty)},\label{vacD1Charge}\\
e^{-2\dil(0)}&=\frac{\vaca\,\vacb^2}{\vacu}=e^{-2\dil(\infty)}.
\end{align}
\end{subequations}
Superscripts are used to indicate where this charge is evaluated, $z=0$ or $z\rightarrow\infty$. Obviously, the first two equations are simply the expected charge conservation. The axion and the RR four-form, equation \eqref{asymptoticAxion} and \eqref{asymptoticFourForm} respectively, vanish. We see that the sign of $\vacb$ determines the signs of the charges. The signs of both, D1 and D5 charges, coincide at one asymptotic region. In what follows we choose without loss of generality $\vacb>0$.

The harmonics $\hola,\,\holb,\,\holu_0,\,\holv$ vanish on the boundary $\p\Sigma$, \eqref{harmonicBoundaryVanishing} and the meromorphics $\holA,\,\holB,\,\holU$ share their singularites, \eqref{nonBranePoles}, at $z=\pm1$. The requirements \eqref{nonBranePoles} at these loci give rise to the same constraint and reduce the number of independent parameters in \eqref{vacFuns},
\begin{equation}\label{betaSquared}
\vacb^2=\frac{\vaca\vacu}{4}.
\end{equation}
This identification will persist\footnote{In what follows we will sometimes keep the parameter $\vacb$ to avoid clutter in equations. Unless otherwise stated it will be determined by \eqref{betaSquared}.}
through any modification that we will employ in order to give rise to interfaces later on. Also, $\holB$ and $\p_z\holV$ share their zeroes at $z=\pm i$ as required by the last point in \secref{sec:Regularity Constraints}. Moreover, it can be checked that
\begin{equation}\label{vacuumf2nonneg}
\hola\,\holu_0-\holb^2=\frac{4\vaca\vacu\Im^2(z)}{|1-z^2|^4}\bigl((1+|z|^2)^2-4\Re^2(z)\bigr)\geq0,
\end{equation}
as desired by \eqref{nonBranePoles}.
It is useful to replace the three parameters $(\vaca,\,\vacb,\,\vacu)$ by the physically meaningful charges and the dilaton
Therefore, we invert the system of equations \eqref{vacCharges},
\begin{equation}\label{vacSolutionAppendix}
\vacv=\frac{1}{2}\sqrt{\pfive^{(0)}\,\pone^{(0)}},\qquad
\vaca=2e^{-\dil(0)},\qquad
\vacu=2\frac{\pone^{(0)}}{\pfive^{(0)}}e^{-\dil(0)}.
\end{equation}
In the main text, \eqref{vacSolution}, we omitted dressing the charges by a label indicating its asymptotic region, because the charges at infinity differ only in sign.

\subsection{D1/F1 defect (UV)}\label{app:OneBraneDefect}
In the main text we convinced ourselves that the addition \eqref{UF1} to \eqref{vacFuns} generates a 1-brane interface embedded into the D1/D5 geometry. In this appendix we present the details leading up to \eqref{F1D1Solution}. This is achieved following the same philosophy as for the trivial interface, the only difference being that we now also have defect charges. Even though we use the same symbols $\vacv,\,\vaca,\,\vacu$ here as in the vacuum solution \eqref{vacFuns}, their values will differ due to the presence of the defect as we will see below. Nevertheless, they will reduce to their vacuum pendants \eqref{vacSolution} once the interface is removed.

Straightforward computation of the asymptotic charges \eqref{asymptoticCharges} gives
\begin{subequations}\label{F1SolutionCharges}
\begin{align}
 \pfive^{(0)}=&\frac{\vaca\vacv}{\vacb},\qquad& \pfive^{(\infty)}&=-\frac{\vaca\vacv}{\vacb},\label{F1SolutionD5Charge}\\
 \pone^{(0)}=&\frac{\vacv}{\vacb}\biggl(\vacu+\frac{\defc}{\defx}\biggr),\qquad& \pone^{(\infty)}&=-\frac{\vacv}{\vacb}\biggl(\vacu+\defc\,\defx\biggr),\\
 \qone^{(0)}=&0,\qquad& \qone^{(\infty)}&=-\defc\frac{\vaca\vacv}{\vacb}=\defc\,\pfive^{(\infty)},\label{F1solutionF1Charge}
\end{align}
\end{subequations}
while $\qfive$ still vanishes at both asymptotic regions. The parameters here still satisfy \eqref{betaSquared}. Indeed, in the vacuum solution the requirements \eqref{nonBranePoles} at $z=\pm1$ both gave rise to the same constraint. While at $z=1$ we do not require regularity anymore due to the possibility $\defx=1$, at $z=-1$ the constraint remains untouched yielding again $4\vacb=\vaca\vacu$.

The defect's F1 and D1 charges are
\begin{subequations}\label{F1DefectChargesAppendix}
\begin{align}
 \qone^{\cD}&\equiv-\qone^{(0)}-\qone^{(\infty)}=\defc\,\pfive^{(0)},\label{F1defectF1Charge}\\
 \pone^{\cD}&\equiv-\pone^{(0)}-\pone^{(\infty)}=c\frac{\vacv}{\vacb}\biggl(\defx-\frac{1}{\defx}\biggr)\label{F1defectD1Charge}.
\end{align}
\end{subequations}
For the particular value $\defx=1$ corresponding to the $\AdS_2$ sheet of smallest size, $\pone^\cD$ vanishes. In this case the D1 charges at both asymptotic regions differ only in sign.

The asymptotic values of the fields \eqref{asymptoticFields} are
\begin{subequations}\label{F1SolutionFields}
\begin{align}
 e^{-2\dil(0)}&=\frac{\vacb^2\vaca}{\vacu+\defc/\defx}=\vacb^2\frac{\pfive^{(0)}}{\pone^{(0)}},
 \qquad&
 e^{-2\dil(\infty)}&=\frac{\vacb^2\vaca}{\vacu+\defc\,\defx}=\vacb^2\frac{\pfive^{(\infty)}}{\pone^{(\infty)}}\\
 C_K(0)&=0,\qquad& C_K(\infty)&=-\defc=-\frac{\qone^\cD}{\pfive^{(0)}},
\end{align}
\end{subequations}
while the axion $\chi$ still vanishes at both regions. This configuration features a jump in the dilaton, which is controlled by the discrepancy in D1 charge at the asymptotic regions,
\begin{equation}\label{DilatonJump}
e^{2\dil(\infty)}=-e^{2\dil(0)}\frac{\pone^{(\infty)}}{\pone^{(0)}}\,.
\end{equation}
The dilaton jump is therefore not independent.

For the remainder of this article we drop the superscript on the D5 charge, $\pfive\equiv \pfive^{(0)}=-\pfive^{(\infty)}$. Without loss of generality we choose $\vacb>0$, which renders all charges at zero and $\qone^\cD$ positive, while all charges at infinity are then negative. For future reference we rewrite the D1 charges in \eqref{F1SolutionCharges} in the more suggestive form
\begin{equation}\label{F1SolutionD1Charge}
 \pone^{(0)}=\frac{\vacv\vacu}{\vacb}+\frac{1}{\vaca \defx}\qone^\cD,\qquad \pone^{(\infty)}=-\frac{\vacv\vacu}{\vacb}-\frac{ \defx}{\vaca}\qone^\cD,
\end{equation}
which elicits that we recover the vacuum expression \eqref{vacD1Charge} when $\qone^\cD$ tends to zero. Then the defect D1 charge and the D1 arithmetic mean read
\begin{subequations}\label{F1SolutionChargeCombinations}
\begin{align}
 \pone^{\cD}&=|\pone^{(\infty)}|-\pone^{(0)}=\frac{\qone^\cD}{\vaca}2\sinh\defxi,\\
 \ol{{\pone}}&=\frac{|\pone^{(\infty)}|+\pone^{(0)}}{2}=\frac{\vacv\vacu}{\vacb}+\frac{\qone^\cD}{\vaca}\cosh\defxi=\kappa\frac{\vacv\vacu}{\vacb} .
 \label{F1SolutionD1meanAppendix}
\end{align}
\end{subequations}
Here, we have expressed the locus of the defect through its Janus coordinate $x=\exp\defxi$. The last equality uses \eqref{F1SolutionD1mean}, which quantifies how much the D1 charge differs from the vacuum case, \eqref{vacD1Charge}.

Overall we have added two new parameters, $\defc$ and $\defx$ to the system and obtained two new independent charges \eqref{F1DefectChargesAppendix}. Our next step is to express the variables $(\vaca,\vacu,\vacv,\defc,\defx)$ in terms of the charges and the dilaton $(\dil(0),\pfive,\ol{\pone},\pone^\cD,\qone^\cD)$. To that end we invert the set of equations \eqref{F1SolutionCharges}, \eqref{F1DefectChargesAppendix} and \eqref{F1SolutionFields}. The result is presented in \eqref{F1D1Solution} alongside \eqref{kappa}.

In the main text we discussed the pure F1 defect. Here we also present the pure D1 defect.

\subsubsection*{Pure D1 defect, $\qone^\cD\rightarrow0$}

When there is no F1 charge on the defect it is pushed to the boundary
\begin{equation}
\sinh\defxi\rightarrow\,\sgn{\pone^\cD}\times\infty,
\end{equation}
where it merges with the CFT. Indeed, the triple in \eqref{F1D1SolutionAlphaEtaUpsilon} reduces to the vacuum expressions \eqref{vacSolution} with modified D1 charge,
\begin{subequations}\label{F1D1SolutionPureD1}
\begin{align}
 \pone^\cD>0:&\quad   \vacv=\frac{1}{2}\sqrt{\pfive\,\frac{2\ol{\pone}-\pone^\cD}{2}},\quad
                     \vaca=2e^{-\dil(0)},\quad
                     \vacu=2e^{-\dil(0)}\frac{(2\ol{\pone}-\pone^\cD)}{2\,\pfive},\\[.28cm]
 \pone^\cD<0:&\quad   \vacv=\frac{1}{2}\sqrt{\pfive\,\frac{2\ol{\pone}+\pone^\cD}{2}},\quad
                     \vaca=2e^{-\dil(\infty)},\quad
                     \vacu=2e^{-\dil(\infty)}\frac{(2\ol{\pone}+\pone^\cD)}{2\,\pfive},
\end{align}
\end{subequations}
where we employed \eqref{DilatonJump}. In both cases the defect's charge is added to the $D1$ charge $\ol{\pone}$ of the pure D1/D5 solution. Equivalently, this could have been written via the asymptotic charges $2\pone^{(0)}=2\ol{\pone}-\pone^\cD$ or $2|\pone^{(\infty)}|=2\ol{\pone}+\pone^\cD$, which justifies the dilaton being evaluated either at zero or infinity. Obviously, reducing $\pone^\cD=0$ leads exactly to the vacuum expressions \eqref{vacSolution}. 

\subsection{D3 defect (IR)}\label{app:ThreeBraneDefect}
In the main text we have convinced ourselves that the addition \eqref{UD3} to \eqref{vacFuns} generates a D3 interface inside the D1/D5 geometry. In this appendix we present the details leading up to \eqref{D3Solution}. Again we use the same symbols $\vacv,\,\vaca,\,\vacu$ and as before their dependence on the charges and the dilaton differs from the on-brane defect and the trivial interface. Nevertheless, they reduce to their pendants \eqref{vacSolution} and \eqref{F1D1Solution} in the appropriate limits.

Straightforward computation of the asymptotic charges \eqref{asymptoticCharges} gives
\begin{subequations}\label{D3SolutionCharges}
\begin{align}
 \pfive=&\frac{\vaca\vacv}{\vacb},\qquad& \pfive^{(\infty)}&=-\frac{\vaca\vacv}{\vacb},\label{D3SolutionD5Charge}\\
 \pone^{(0)}=&\frac{\vacv}{\vacb}\biggl(\vacu+\frac{\pthree^\cD}{R}\sin\Theta\biggr),\qquad& \pone^{(\infty)}&=-\frac{\vacv}{\vacb}\biggl(\vacu+\pthree^\cD R\sin\Theta\biggr),\\
 \qone^{(0)}=&0,\qquad& \qone^{(\infty)}&=-\pthree^\cD\,\Theta\frac{\vacv\vaca}{\vacb}=\pthree^\cD\Theta\,\pfive^{(\infty)}\label{D3SolutionF1Charge},
\end{align}
\end{subequations}
while $\qfive$ still vanishes at both asymptotic regions. Again, the defect carries $D1$ and $F1$ charge,
\begin{subequations}\label{D3DefectCharges}
\begin{align}
 \qone^{\cD}&\equiv-\qone^{(0)}-\qone^{(\infty)}=\pthree^\cD\Theta\,\pfive,\label{D3defectF1Charge}\\
 \pone^{\cD}&\equiv-\pone^{(0)}-\pone^{(\infty)}=\pthree^\cD\frac{\vacv}{\vacb}\biggl(R-\frac{1}{R}\biggr)\sin\Theta\label{D3defectD1Charge}.
\end{align}
\end{subequations}
The asymptotic values of the fields \eqref{asymptoticFields} are
\begin{align}\label{D3SolutionFields}
 e^{-2\dil(0)}&=\frac{\vacb^2\vaca}{\vacu+\pthree^\cD R^{-1}\sin\Theta}=\vacb^2\frac{\pfive}{\pone^{(0)}},
 \qquad&
 e^{-2\dil(\infty)}&=\frac{\vacb^2\vaca}{\vacu+\pthree^\cD R\sin\Theta }=\vacb^2\frac{\pfive^{(\infty)}}{\pone^{(\infty)}}\\[.22cm]
 C_K(0)&=0,\qquad& C_K(\infty)&=-\pthree^\cD\Theta=-\frac{\qone^\cD}{\pfive},\label{D3SolutionCK}
\end{align}
while the axion $\chi$ still vanishes at both regions. As before the jump in the dilaton is not independent, cf. \eqref{DilatonJump}.

Let us define an effective F1 charge
\begin{equation}\label{AppendixqTheta}
\qone^\Theta\equiv\qone^{\cD}\frac{\sin\Theta}{\Theta}
\end{equation}
and use it to rewrite the D1 charges in \eqref{D3SolutionCharges},
\begin{equation}\label{D3SolutionD1Charge}
\pone^{(0)}=\frac{\vacv\vacu}{\vacb}+\frac{1}{\vaca R}\qone^\Theta,\qquad \pone^{(\infty)}=-\frac{\vacv\vacu}{\vacb}-\frac{ R}{\vaca}\qone^\Theta.
\end{equation}
Their linear combinations are
\begin{subequations}\label{D3SolutionChargeCombinations}
\begin{align}
 \pone^{\cD}&=|\pone^{(\infty)}|-\pone^{(0)}=\frac{\qone^\Theta}{\vaca}2\sinh\defrho,\\
 \ol{{\pone}}&\equiv\frac{|\pone^{(\infty)}|+\pone^{(0)}}{2}=\frac{\vacv\vacu}{\vacb}+\frac{\qone^\Theta}{\vaca}\cosh\defrho\equiv\kappa^{(\Theta)}\frac{\vacv\vacu}{\vacb}.
 \label{D3SolutionD1meanAppendix}
\end{align}
\end{subequations}
In the second line we have again quantified the difference to the vacuum D1 charge \eqref{vacD1Charge} via $\kappa^{(\Theta)}$. Evidently, the D1 charges \eqref{D3SolutionD1Charge} and their linear combinations \eqref{D3SolutionChargeCombinations} look exactly like their counterparts \eqref{F1SolutionD1Charge} and \eqref{F1SolutionChargeCombinations}, respectively, with the replacements $\qone^\cD\rightarrow\qone^\Theta$ and $\defx\rightarrow R$ ($\defxi\rightarrow\defrho$). As the reader might have observed already the other relevant expressions, namely the D5 charge, \eqref{D3SolutionD5Charge}, and the dilaton in \eqref{D3SolutionFields} assume exactly the same form as their counterparts for the 1-brane interface \eqref{F1SolutionD5Charge} and \eqref{F1SolutionFields}. Therefore the result \eqref{F1D1Solution} of the F1/D1 interface carries over with the adjustments
 $\qone^\cD\rightarrow\qone^\Theta$, $\defxi\rightarrow \defrho$, yielding the result \eqref{D3Solution}.

In the main text we discussed the pure F1 defect.
Here we also present the pure D1 case.

\subsubsection*{Pure D1 case, $\qone^\Theta\rightarrow 0$}

Since the 3-brane defect cares only about the effective F1 charge \eqref{AppendixqTheta} we have two options to remove the effect of F1 charge. The first is as before $\qone^\cD\rightarrow0$.
The second is when $\Theta=\pi$, which happens at a large value of F1 charge $\qone^\cD=\pi\pfive\pthree^\cD$.
Of course, the defect is again pushed to the boundary of $\AdS_3$, $\sinh\defrho\rightarrow\,\sgn{\pone^\cD}\times\infty$ and the triple $(\vaca, \vacu,\vacv)$ behaves in the same way as before, \eqref{F1D1SolutionPureD1}.

\bibliographystyle{JHEP}
\bibliography{kondo}

\end{document}